\documentclass{aa}

\usepackage{graphicx}
\usepackage{txfonts}
\usepackage{natbib}
\usepackage[breaklinks=true, colorlinks=true, allcolors=blue]{hyperref}
\usepackage{amsmath}    
\usepackage{amssymb}    
\usepackage{multirow}
\usepackage{longtable}
\usepackage{threeparttable}

\usepackage{multicol}

\begin{document}

   \title{Inverted level populations of hydrogen atoms in ionized gas}


   \author{F.-Y. Zhu\inst{1,2},
          J. Z. Wang\inst{1,3},
          Q.-F. Zhu\inst{4},
          \and
          J.-S. Zhang\inst{5}
          }

   \institute{Shanghai Astronomical Observatory, Chinese Academy of Sciences,80 Nandan Road, Shanghai, 200030, PR China\\
              \email{junzhiwang@gxu.edu.cn, zhufy@zhejianglab.com}
         \and
             Research Center for Intelligent Computing Platforms, Zhejiang Laboratory, Hangzhou, 311100, PR China
         \and
             School of Physical Science and Technology, Guangxi University, Nanning 530004, PR China
         \and
             Department of Astronomy, University of Science and technology of China, Hefei, 230026, PR China
                \and
             Physics Department, Guangzhou University, Guangzhou, 510006, PR China
             }

 \date{Received xx; accepted xxx}

 \authorrunning{Zhu et al.}


  \abstract
   {Level population inversion of hydrogen atoms in ionized gas may lead to stimulated emission of hydrogen recombination lines, and the level populations can in turn be affected by powerful stimulated emissions.}
   {In this work the interaction of the radiation fields and the level population inversion of hydrogen atoms is studied. The effect of the stimulated emissions on the line profiles is also investigated.}
   {Our previous nl-model for calculating level populations of hydrogen atoms and hydrogen recombination lines is improved. The effects of line and continuum radiation fields on the level populations are considered in the improved model. By using this method the properties of  simulated hydrogen recombination lines and  level populations are used in analyses.}
   {The simulations show that hydrogen radio recombination lines are often emitted from the energy level with an inverted population. The widths of Hn$\alpha$ lines can be significantly narrowed by strong stimulated emissions to be even less than 10 km s$^{-1}$. The amplification of hydrogen recombination lines is more affected by the line optical depth than by the total optical depth. The influence of stimulated emission on the estimates of electron temperature and density of ionized gas is evaluated. We find that comparing multiple line-to-continuum ratios is a reliable method for estimating the electron temperature, while the effectiveness of the estimation of electron density is determined by the relative significance of stimulated emission.}
   {}


   \maketitle
%

\section{Introduction}

Ionized gas, which is  mainly composed of ionized hydrogen and electrons, is a common phase of    the interstellar medium (ISM). This gas  is usually formed by    the extreme-ultraviolet (h$\nu~\ge$ 13.6 eV) photons emitted from massive stars, post-asymptotic giant branch (post-AGB) stars, supernovas (SNs), or  active galactic nuclei (AGNs). Star formation activities and     the evolution of  massive stars in molecular clouds can be studied from     observations    toward     ionized gas caused by massive stars \citep{kim17,liu21}. Hydrogen radio recombination lines (RRLs) and     continuum emissions can be used    to derive  the electron    temperature and density of  ionized gas \citep{woo89}. These basic properties of  ionized gas are helpful    to determine    the effective    temperature and    the mass of    the ionizing star,    the chemical abundances of the ionized gas, and    the conditions of    the ambient medium \citep{pei79,sha83,aff94,zhu19}

The line-to-continuum ratio at centimeter and millimeter wavelengths is a common    tool used    to estimate    the electron    temperature of    the ionized gas with an assumption of local    thermodynamic equilibrium (LTE) conditions \citep{chu75,sha83}. However, in actual situations, the level populations of     hydrogen atoms do not perfectly satisfy    the LTE assumption \citep{sej69,bro70}. This could lead    to a significant difference between    the estimated and    the actual electron    temperatures \citep{sor65,hog65,gor02}. The excitation states of hydrogen can be classified into    the cases of population inversion, overheating,    thermalization, and overcooling \citep{str96}. In    the cases of population inversion and overheating,  line emission can be increased by stimulated emission and become higher than that expected under the LTE assumption. When    the hydrogen recombination lines are strongly amplified due    to    the inverted level populations,     line masers occur. The first  hydrogen recombination line maser source discovered was    the young stellar object (YSO) MWC 349A \citep{mar89}; hydrogen recombination line masers were then also found in other sources \citep{cox95,jim11,ale18}. The first extragalactic hydrogen recombination line maser was discovered in NGC 253 \citep{bae18}. For some H II regions  stimulated emissions are also important, although    the line masers cannot  be identified  \citep{aff94}. For these cases, the estimation method under    the LTE assumption is not applicable.

In order    to improve    the accuracy of    the estimation,    the non-LTE conditions of    the level populations of hydrogen atoms have been considered in some previous works \citep{aff94,bae13}. The accuracy of    the departure coefficients of    the level populations of hydrogen from    the populations under     LTE conditions is important for    the estimation of electron    temperature. The departure coefficients have been studied for more    than 50 years \citep{sej69}. In     early works, the departure coefficients were calculated via    the n-model in which only    the    transitions between different energy levels were considered \citep{bro70,wal90}. The nl-model, which  included angular momentum changing transitions, was used in \citet{sto95}, and the calculated values were then applied to specifying the conditions necessary for hydrogen recombination line masers \citep{str96}. \citet{pro18} improved    the nl-model with    the consideration of    the effect of    the continuum radiation. By using    the departure coefficients calculated by    the nl-model without    the effect of radiation fields, we provided a method    for  estimating    the electron    temperature and density of ionized gas from hydrogen RRLs in \citet{zhu19}. However,    the accuracy of    this method has not been checked for H II regions with high electron density ($n_e\ge10^4$ cm$^{-3}$) and high emission measure (EM), where the departure coefficients could be significantly affected by     line and continuum radiation \citep{pro18}.

The departure coefficients and the intensities of hydrogen recombination lines were recently calculated under  maser conditions by \citet{pro20}. Both the effects of radio free-free continuum and hydrogen recombination line radiation on the  level populations were considered in their work. However, in this model     the hydrogen recombination lines are all assumed    to have a box profile with    the same amplitude as    the Doppler profile. This simple assumption cannot reproduce the actual line profiles. Since the optical depth of the hydrogen recombination line is clearly affected by    the line width,    the fluxes of    the hydrogen recombination lines should also be influenced. The broadening of    the hydrogen recombination line is related to the velocity field, electron    temperature, and density \citep{pet12}. Conversely,     line masers can also influence    the  widths of    the hydrogen recombination lines \citep{thu95}. Moreover, the spectral simulation code CLOUDY \citep{fer17} can also treat hydrogen recombination lines under maser conditions.  The radiation fields and realistic line profiles are included in this code \citep{guz19}. However, the effects of the radiation fields and line profiles on hydrogen level populations have not been explored with CLOUDY in previous works.

In this work we improve    the nl-model produced in \citet{zhu19} with    the effects of    the radiation fields due    to    the hydrogen recombination lines and    the continuum emission. The effects of    the line and continuum radiation fields on    the departure coefficients are studied. The relation between    the line profile and    the emissions of    the hydrogen recombination lines is also investigated. The performance of    the estimation method    to derive    the electron    temperature from    the line-to-continuum ratio under     LTE assumptions for     ionized gas with a series of properties is presented. Finally, we    test    the reliability of    the estimated properties of H II regions by using hydrogen recombination lines and    the continuum emissions in multiple wavebands.


\section{Method}\label{sec:method}

\subsection{The calculation of departure coefficients in Case B}

The accuracy of  calculations of  level populations and  departure coefficients of hydrogen atoms is important for the simulation of    hydrogen recombination lines and continuum emission. Level population equations consists of a series of linear equations \citep{sal17,pro18}, and can be written in matrix form as 

\begin{equation}
\mathbf{A}\cdot\mathbf{b}=\mathbf{y}~~~,
\end{equation}where the elements of matrix \textbf{A} and vector $\mathbf{y}$ refer to the processes of different transitions from or to the state $nl$. The vector $\mathbf{b}$ is composed of the partial departure coefficients ($b_{nl}$). Then, by solving the matrix equation, $b_{nl}$ can be obtained. An iterative method for evaluating $b_{nl}$ is described in \citet{bro71}, and is also used in \citet{sto95} and \citet{sal17}. In that method, the values of $b_{nl}$ with the same principal quantum number $n$ are solved simultaneously with a $n\times n$ matrix, while the partial departure coefficients with the other principal quantum numbers are treated as known quantities in the equations. \citet{pro18} used a direct solver to calculate the departure coefficients, and provide an error estimate of the quality of the results derived from the norms of the matrix and vectors as

\begin{equation}
\epsilon \approx ||\mathbf{A}\cdot\mathbf{b}-\mathbf{y}||/||\mathbf{b}||~~~.
\label{eq:error}
\end{equation}This value can be used to estimate the convergence of the departure coefficients.

In our previous work, an algorithm    to calculate    the departure coefficients ($b_n$) was used in    the case    when    the radiation field does not have a significant effect on  level populations (Case B) \citep{zhu19}. The iterative method was used in the algorithm, and the calculations were terminated if the largest difference of departure coefficients between two iterations $\Delta b_n^t=|b_n^t-b_n^{t-1}|$ is less than $1\times10^{-6}$, where t means the sequence number of iterations. This criterion could be too simple. In this work  we directly solve the matrix equation. The error estimate is $\epsilon\sim 3\times10^{-5}$ in all the cases. The convergence should be reached with this value of the error estimate.

However, as    mentioned in \cite{zhu19}, the resulting departure coefficients of our method are closer to the results provided in \citet{sto95} than those in \citet{pro18}, although the rates of transition processes we used are similar to those  in \citet{pro18}. More checks were done, and they confirm the reliability of our results. The details are presented in Appendix \ref{sec:com}.

\begin{figure}
\begin{center}
\includegraphics[scale=0.4]{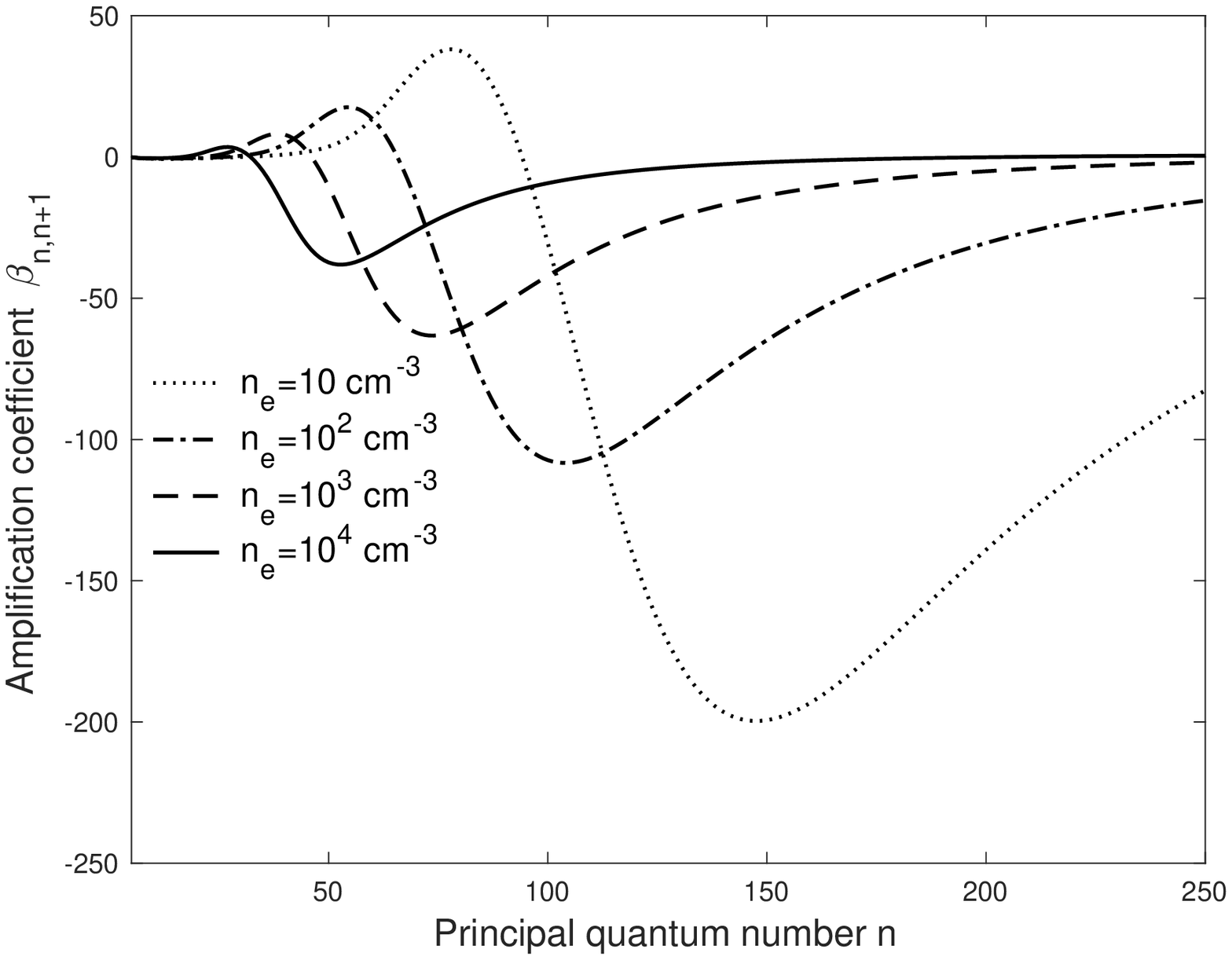}
\includegraphics[scale=0.4]{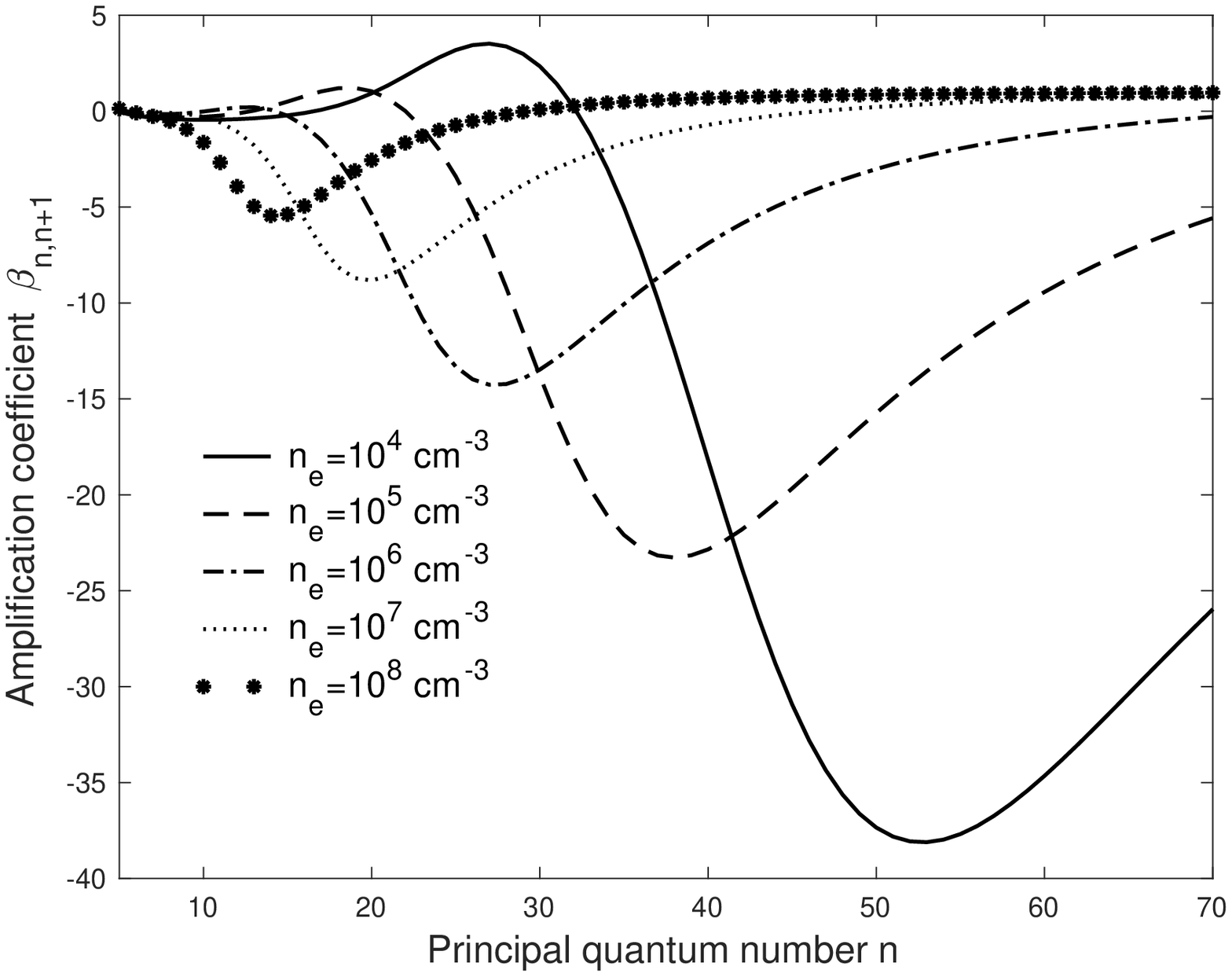}
\caption{Value of $\beta_{n,n+1}$ varies with principal quantum number $n$ for    $T_e$=10000 K and different electron densities $n_e$. }\label{fig:betalist}
\end{center}
\end{figure}

The amplification coefficients ($\beta_{n,n+1}$) with the principal quantum number $n$ calculated by this algorithm for different electron densities ($n_e$) are presented in Fig. \ref{fig:betalist}. The amplification coefficients represent the departure of the level population ratios from their LTE values \citep{str96,pro18}. Shown in Fig. \ref{fig:betalist}, the result is similar    to    those calculated in    the previous works in    the literature  \citep{sto95,str96}. The ranges of principal quantum number $n$ in    the cases of population inversion ($\beta_{n,n+1}<0$), overheating ($0<\beta_{n,n+1}<1$),    thermalization ($\beta_{n,n+1}=1$), and overcooling ($\beta_{n,n+1}>1$) are shown, while  the locations and    the widths of    these ranges changing with $n_e$ are also presented. The center of the valley of $\beta_{n,n+1}$ is closer to the left side of the low principal quantum number with  high electron density $n_e$, while the width and depth of the valley rises with  decreasing $n_e$. The result shows that population inversion is a common phenomenon for hydrogen radio recombination lines. This motivates us to study in detail the significance of the stimulation effect due to population inversion under different conditions. The level populations of hydrogen atoms could be influenced when stimulated emission is important. So in this work, the effects of    the line and continuum radiation fields on    the level populations and departure coefficients are both added to the new algorithm.

\subsection{Departure coefficients calculated with the effects of line and continuum radiation fields}

Hydrogen recombination lines, continuum emission, and the ``one-layer'' assumption are used when calculating departure coefficients.  Ionized gas is homogeneous and isothermal. In the calculation    the initial conditions include     electron    temperature    ($T_e$),     electron density ($n_e$), and emission measure (EM). In the beginning the departure coefficients are derived by directly solving the matrix equation of level populations with no radiation fields. In the second step these values of departure coefficients are used to calculate the mean intensities of incident radiation fields due to line and continuum emissions. Then the departure coefficients are calculated with the intensities of the incident radiation fields, while the new values of the intensities are estimated from the new departure coefficients. This process is operated iteratively until the values of the departure coefficients converge.

Because the radiation fields and the calculated departure coefficients are inter-related, the error estimate derived from the solution of the matrix equation cannot directly be used to evaluate the convergence. We replace $\mathbf{b}^{t}$ (the vector $\mathbf{b}$ composed of $b_{nl}$ in the current step $t$) with $\mathbf{b}^{t-1}$ in the last step $t-1$ to calculate the error estimate, and terminate the calculation when the two error estimates calculated with $\mathbf{b}^{t}$ and $\mathbf{b}^{t-1}$ are approximately equal. This is a reasonable stopping criterion since the values of $\mathbf{b}^{t}$ and $\mathbf{b}^{t}-1$ should be very close when the iterative calculation has converged. It is also found that the $\Delta b_n^t$ is always lower than $10^{-9}$ when the calculation converges.


In addition, the mean intensity of line and continuum radiation fields ($J_\nu$) need to be calculated in order to evaluate the effects of the radiation fields on the level populations. However, for a cloud of ionized gas, $J_\nu$ should be treated nonlocally  in principle with different quantities in every volume element. This would make the calculations very time consuming, so following \citet{keg79}, we assume that the electron temperature, density, absorption coefficient ($\kappa_\nu$), and source function ($S_\nu$) are all homogeneous, and then the mean intensity of the incident radiation field is

\begin{equation}
J_\nu=S_\nu(1-e^{-\tau_\nu})
,\end{equation}where $\tau_\nu$ is the optical depth at the frequency $\nu$  and $e^{-\tau_\nu}$ is also called the escape probability. However, the box line profile as a further simplification used in \citet{keg79}, \citet{koe80}, and \citet{rol99} is not assumed in this work. As defined in \citet{hum87}, the mean radiation field is

\begin{equation}
\bar{J}=\int J_\nu\phi_\nu d\nu~~~
\end{equation}where $\phi_\nu$ is the line profile function \citep{pet12}. In the calculation of departure coefficients, we replace the term $J_\nu$ in Eq. (5) of \citet{pro18} with $\bar{J}$. Moreover, for the transitions about $n=1$, the Lyman transitions are still assumed to be optically thick, and the collisional excitations from $n=1$ and $n=2$ states are also ignored, as in \citet{zhu19}.

\subsection{Radiative transfer}

In this work only    the hydrogen recombination lines and the continuum emission are considered to be    the radiation fields    that could affect    the level populations of hydrogen atoms. The radiation from     sources out of    the ionized gas such as the  stellar radiation field and    the cosmic microwave background,  and    the emission from dust are not included since    the influences of    these emissions are generally weak \citep{pro18}. In addition,    the gas    temperature is assumed    to be equal    to    the electron    temperature. For uniform ionized gas, the continuum absorption coefficient $\kappa_{\nu}$ is derived by \citet{ost61} and \citet{gor02} with the Rayleigh-Jeans approximation ($h\nu \ll kT_e$), which is evaluated numerically as

\begin{equation}
\kappa_{\nu,\textrm{C}}=9.770\times10^{-3}\frac{n_en_i}{\nu^2T_e^{1.5}}[17.72+\textrm{ln}\frac{T_e^{1.5}}{\nu}]~~~,
\end{equation}
where $n_e$ and $n_i$ are respectively   the number densities of electrons and ions in units of cm$^{-3}$, $T_e$   is    the electron    temperature in units of K, $\nu$ is    the frequency in Hz, and $\kappa_{\nu,\textrm{C}}$ is in units of cm$^{-1}$. The continuum optical depth is $\tau_{\nu,\textrm{C}}=\kappa_{\nu,\textrm{C}}D$ with    the line-of-sight (LOS) depth $D$. Then    the free-free continuum emissivity $j_{\nu,\textrm{C}}$ and    the intensity of    the continuum emission $I_{\nu,\textrm{C}}$ are

\begin{equation}
j_{\nu,\textrm{C}}=B_{\nu}(T_e)\kappa_{\nu,\textrm{C}}
\end{equation}

and

\begin{equation}
I_{\nu,\textrm{C}}=B_\nu(T_e)(1-e^{-\tau_{\nu,\textrm{C}}}) ~~~,
\end{equation}where $B_\nu(T_e)$ is    the intensity of a blackbody at        temperature $T_e$ and     frequency $\nu$. The line absorption coefficient at     frequency $\nu$ for    the    transition from lower level $n$ to higher level $m$ including    the processes of absorption and stimulated emission is

\begin{equation}
\begin{split}
\kappa_{\nu,\textrm{L}}&=\frac{h\nu}{4\pi}(N_nB_{nm}-N_mB_{mn})\phi_\nu \\
&=\frac{h\nu}{4\pi}\sum_{l=0}^{n-1}\sum_{l'=l\pm1}(N_{nl}B_{nlml'}-N_{ml'}B_{ml'nl})\phi_\nu
\end{split},
\end{equation}
where $N_{n}$ is    the    total number density of hydrogen atoms with    the principal quantum number $n$,  $N_{nl}$ is    the number density of     those of level nl, and l is the angular momentum number. Similarly, $B_{nm}$ and $B_{nlml'}$ are the Einstein coefficients for absorption and stimulated emissions of the corresponding transitions, and    they are derived from    the spontaneous Einstein coefficients $A_{mn}$ and $A_{ml'nl}$ \citep{bro71,pro20}. The parameter $\phi_\nu$ is the line profile function. Following \citet{gor02} and \citet{pet12}, when calculating the profiles of the hydrogen recombination lines, the line profile function is calculated by considering the thermal, velocity field, and pressure broadenings. The population of hydrogen atoms in level n under the LTE assumption is  \citep{pet12}

\begin{equation}
N_n^{\textrm{LTE}}=\frac{n_en_i}{T_e^{1.5}}\frac{n^2h^3}{(2\pi m_ek)^{1.5}}e^\frac{E_n}{kT_e}~~~
,\end{equation}
where $k$ is the Boltzmann constant,  $h$ the Planck constant,  $E_n$ is the energy of level $n$ below the continuum, and $m_e$ is the electron mass. The emission coefficient of a hydrogen recombination line is

\begin{equation}
j_{\nu,\textrm{L}}=\frac{h\nu}{4\pi}\phi_\nu N_mA_{mn}=b_m B_\nu(T_e)\kappa_{\nu,\textrm{L}}^\textrm{LTE}~~~,
\end{equation}where $\kappa_{\nu,\textrm{L}}^\textrm{LTE}$ is the line absorption coefficient calculated with the LTE level populations. The source function $S_\nu$ is provided by

\begin{equation}
S_\nu=\frac{j_{\nu,\textrm{L}}+j_{\nu,\textrm{C}}}{\kappa_{\nu,\textrm{L}}+\kappa_{\nu,\textrm{C}}}=B_\nu(T_e)\frac{\kappa_{\nu,\textrm{C}}+\kappa_{\nu,\textrm{L}}^{\textrm{LTE}}b_m}{\kappa_{\nu,\textrm{C}}+\kappa_{\nu,\textrm{L}}^\textrm{LTE}b_n\beta_{n,m}}~~~,
\label{eq:Sv}
\end{equation}where $\beta_{n,m}$ is    the amplification coefficient given by

\begin{equation}
\beta_{n,m}=\frac{1-(b_m/b_n)e^{-\frac{h\nu}{kT_e}}}{1-e^{-\frac{h\nu}{kT_e}}}~~~.
\end{equation}

The total intensity and the observed line intensity at the frequency $\nu$ for a uniform H II region are respectively given by

\begin{equation}
I_\nu=S_\nu(1-e^{-\tau_\nu})
\end{equation}

and

\begin{equation}
I_{\nu,\textrm{L}}=S_\nu(1-e^{-\tau_\nu})-B_\nu(T_e)(1-e^{-\tau_{\nu,\textrm{C}}})~~~,
\label{eq:Sl}
\end{equation}where    the    total optical depth is $\tau_\nu=\kappa_\nu D$. The    total absorption coefficient is $\kappa_\nu=\kappa_{\nu,\textrm{C}}+\kappa_{\nu,\textrm{L}}$, and    this value may be negative if    the contribution of    the stimulated emission is more    than    that of absorption. The line absorption coefficient can also be presented as $\kappa_{\nu,\textrm{L}}=b_n\beta_{n,m}\kappa_{\nu,\textrm{L}}^{\textrm{LTE}}$.




In this paper, the results of a series of simulations are presented. The effect of the radiation field on level populations of hydrogen atoms is considered in almost all of the simulations except for those especially indicated. On the contrary, the effect of the velocity field is only included in the simulations whose results shown in Fig. \ref{fig:profile} in Sect. \ref{sec:vel}. In addition, the one-layer model is used in all the simulations, except where a ``two-layer'' model is included in the calculations in Sect. \ref{sec:two}.

\section{Results and discussions}\label{sec:result}

\subsection{Variations in departure and amplification coefficients with radiation field} \label{sec:tem_losdepth}

\begin{figure}
\begin{center}
\includegraphics[scale=0.4]{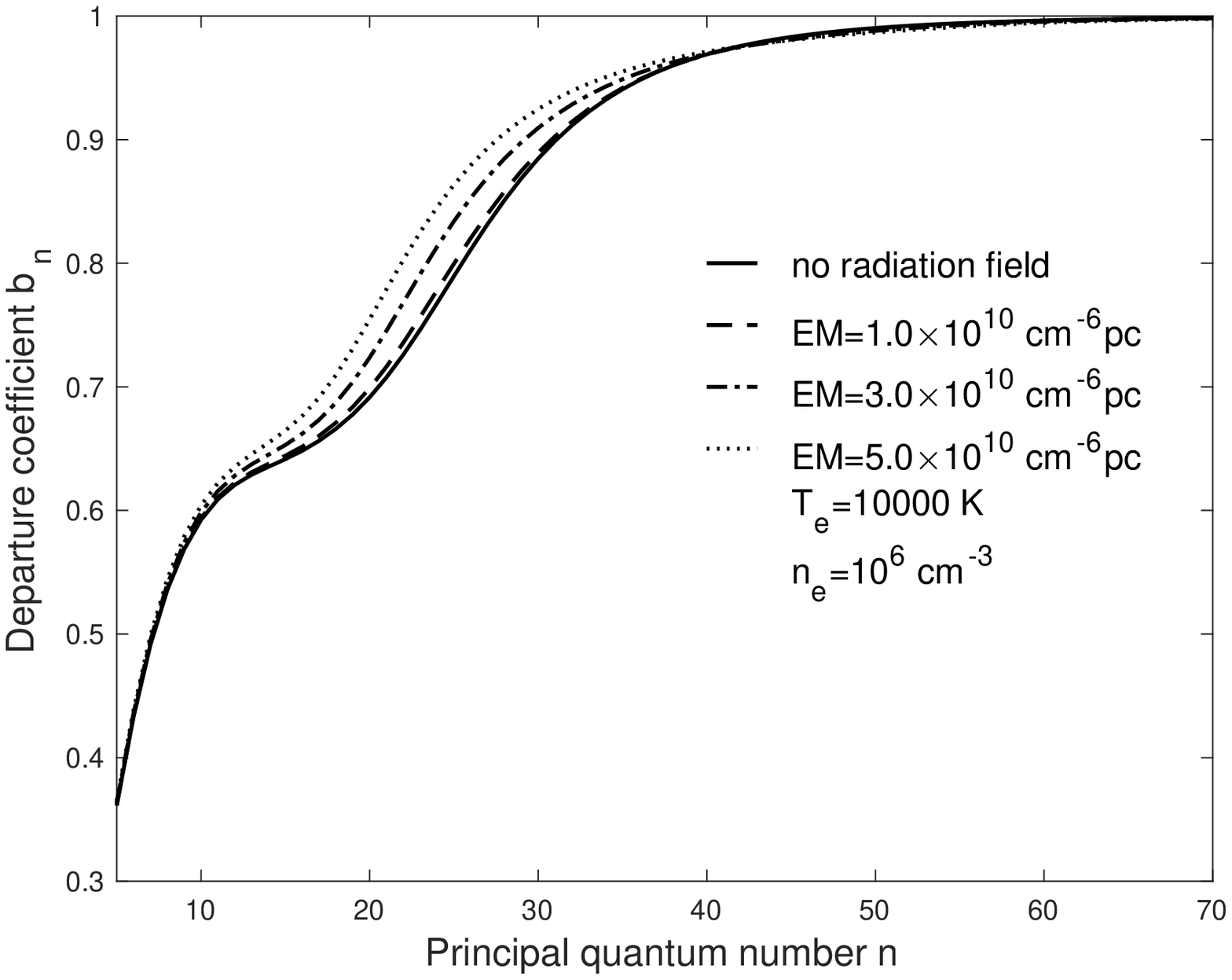}
\includegraphics[scale=0.4]{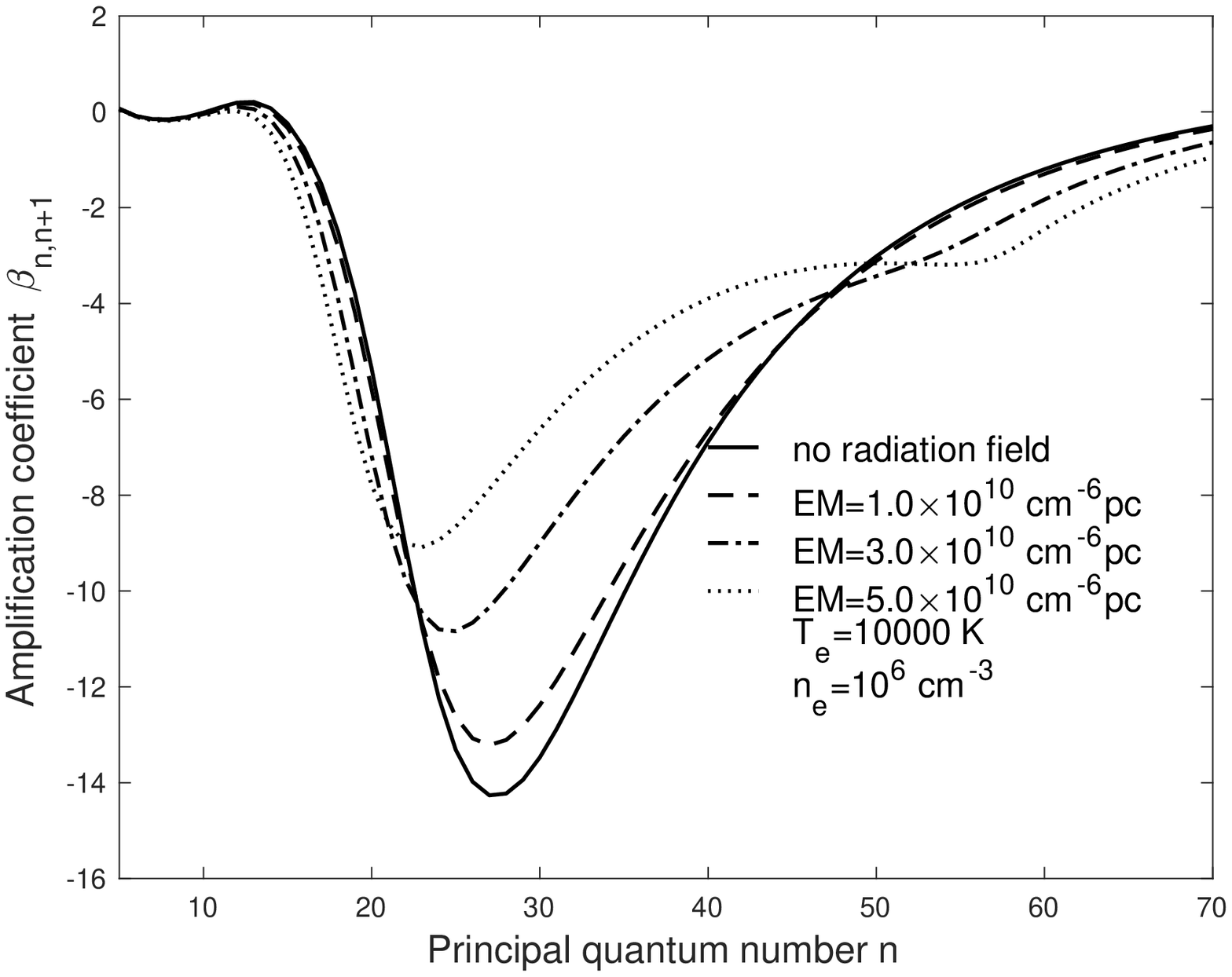}
\includegraphics[scale=0.4]{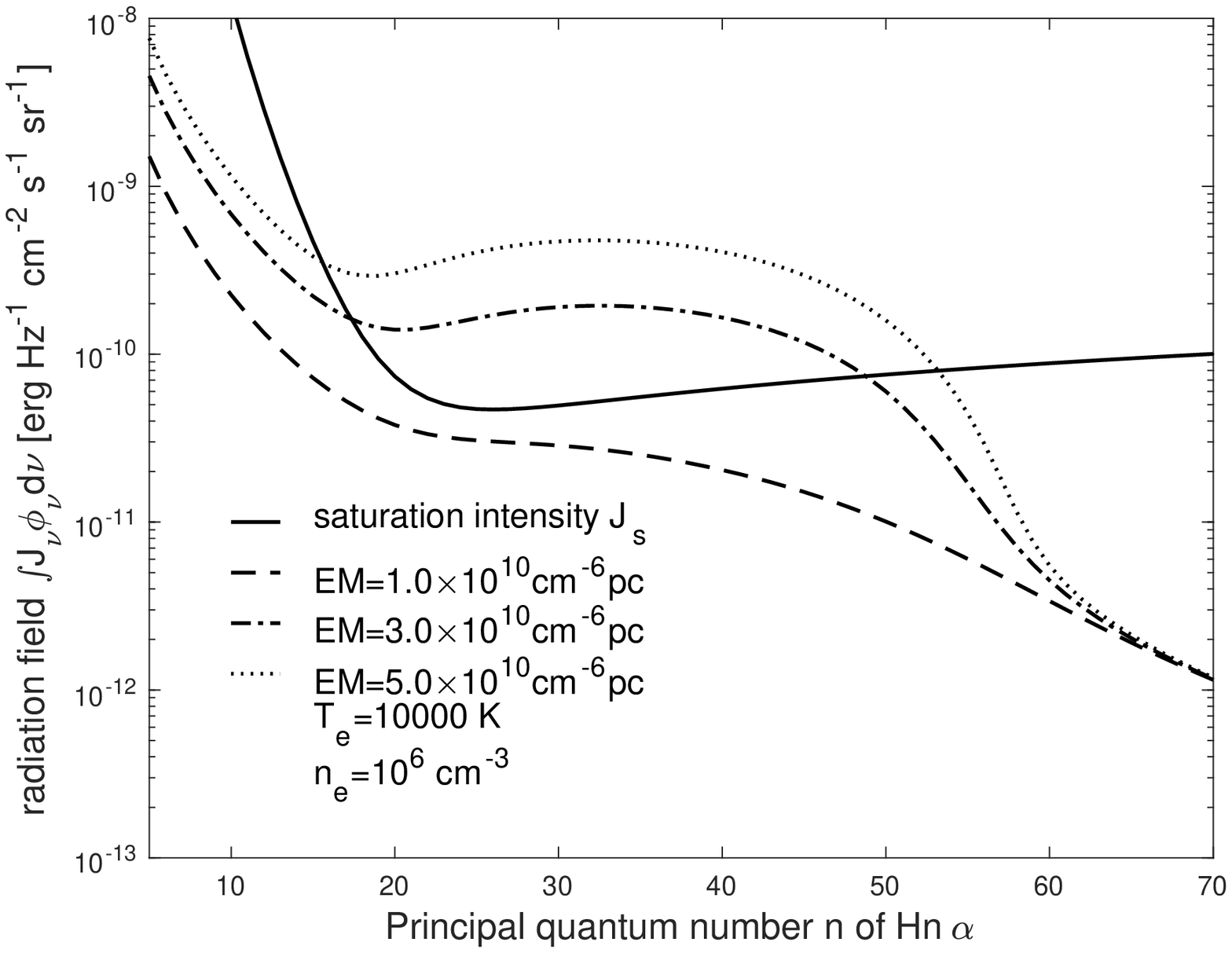}
\caption{Level populations of hydrogen atoms affected by radiation fields. The departure coefficients $b_n$ varying with principal quantum number $n$ affected by different radiation fields are plotted in the top panel. The amplification coefficients $\beta_{n,n+1}$ influenced by different radiation fields are presented in the middle panel. The corresponding intensities of mean radiation field $\int J_\nu\phi_\nu d\nu$ and the saturation intensity $J_s$ are shown in the bottom panel. The electron temperature is $T_e$=10000 K, and the electron density is $n_e=10^6$ cm$^{-3}$.}\label{fig:radiation}
\end{center}
\end{figure}

In order    to assess    the effect of radiation fields on the departure coefficients $b_n$ and the amplification coefficients $\beta_{n,n+1}$, we compared    the values of $b_n$ and $\beta_{n,n+1}$ calculated by using the non-LTE method with and without    the effect of    the line and continuum radiation fields. The results in the conditions of different electron temperatures and densities were analyzed, and the radiation fields were found to have a similar effect on $b_n$ and $\beta_{n,n+1}$. As an example, one case with electron temperature $T_e=10000$ K and electron density $n_e=10^6$ cm$^{-3}$ is presented. The comparisons of $b_n$ and $\beta_{n,n+1}$ are presented in the top and middle panels of Fig. \ref{fig:radiation}. By adjusting the value of EM, the intensity of the radiation field is changed in the calculations. It is clear    that    the distortions of $b_n$ and $\beta_{n,n+1}$ increase with    the intensity of    the radiation field.


The saturation intensity ($J_s$) is introduced by \citet{str96} as a criterion compared with    the mean intensity of    the radiation fields ($J_\nu$) to estimate    the degree of saturation. In the bottom panel of Fig. \ref{fig:radiation}    the comparison of    the mean intensities of the radiation fields and    the saturation intensity is shown. The saturation intensity first decreases and    then increases with    the principal quantum number $n$. This occurs because    the decay rate is dominated by    the spontaneous decay at low energy level $n$ and by    the collisional decay at high energy level n, and    the former decreases with    the energy level while    the latter increases. From    the results shown in Fig. \ref{fig:radiation}, it is unsurprising    that    the saturation intensity can be a rough criterion    to evaluate    the influence of    the radiation fields on $b_n$ and $\beta_{n,n+1}$. For the case of EM$=1.0\times10^{10}$ cm$^{-6}$pc, $J_\nu$ is lower    than $J_s$ for all    the Hn$\alpha$ lines, and    the distortion of $b_n$ and $\beta_{n,n+1}$ are also very slight. The mean intensities $J_\nu$ corresponding to EM$=3.0\times10^{10}$ and $5.0\times10^{10}$ cm$^{-6}$pc are higher    than $J_s$ for a range of Hn$\alpha$ lines, and    the values of $b_n$ and $\beta_{n,n+1}$ at    the corresponding principal quantum numbers $n$ are changed significantly, but    this criterion is still not very accurate. On    the    two sides of    the valley of $\beta_{n,n+1}$,    the values of $b_n$ and $\beta_{n,n+1}$ can be changed significantly even if $J_\nu$ is lower    than $J_s$. The stimulated emission    triggered by    the saturated masing line from $n+1$ to $n$ strongly increase    the population of    the lower level $n$ and decreases    that of    the upper level $n+1$. This effect can decrease    the population inversion between levels $n$ and $n+1$. However, it can also increase    the population inversions between    the levels $n-1$ and $n$, and  between $n+1$ and $n+2$ if    the stimulated emissions of    the adjacent lines of $n+2$ to $n+1$ and $n$ to $n-1$ are relatively weak \citep{str96,pro20}. Since    the radiation field is saturated ($J_\nu>J_s$) in    the range of $n\sim20-50$ corresponding    to    the valley of $\beta_{n,n+1}$ at    the conditions of EM$=3.0\times10^{10}$ and $5.0\times10^{10}$ cm$^{-6}$pc,    the population inversion is weakened in    the center and strengthened on    the    two sides of    the valley of $\beta_{n,n+1}$.

\begin{figure}
\begin{center}
\includegraphics[scale=0.4]{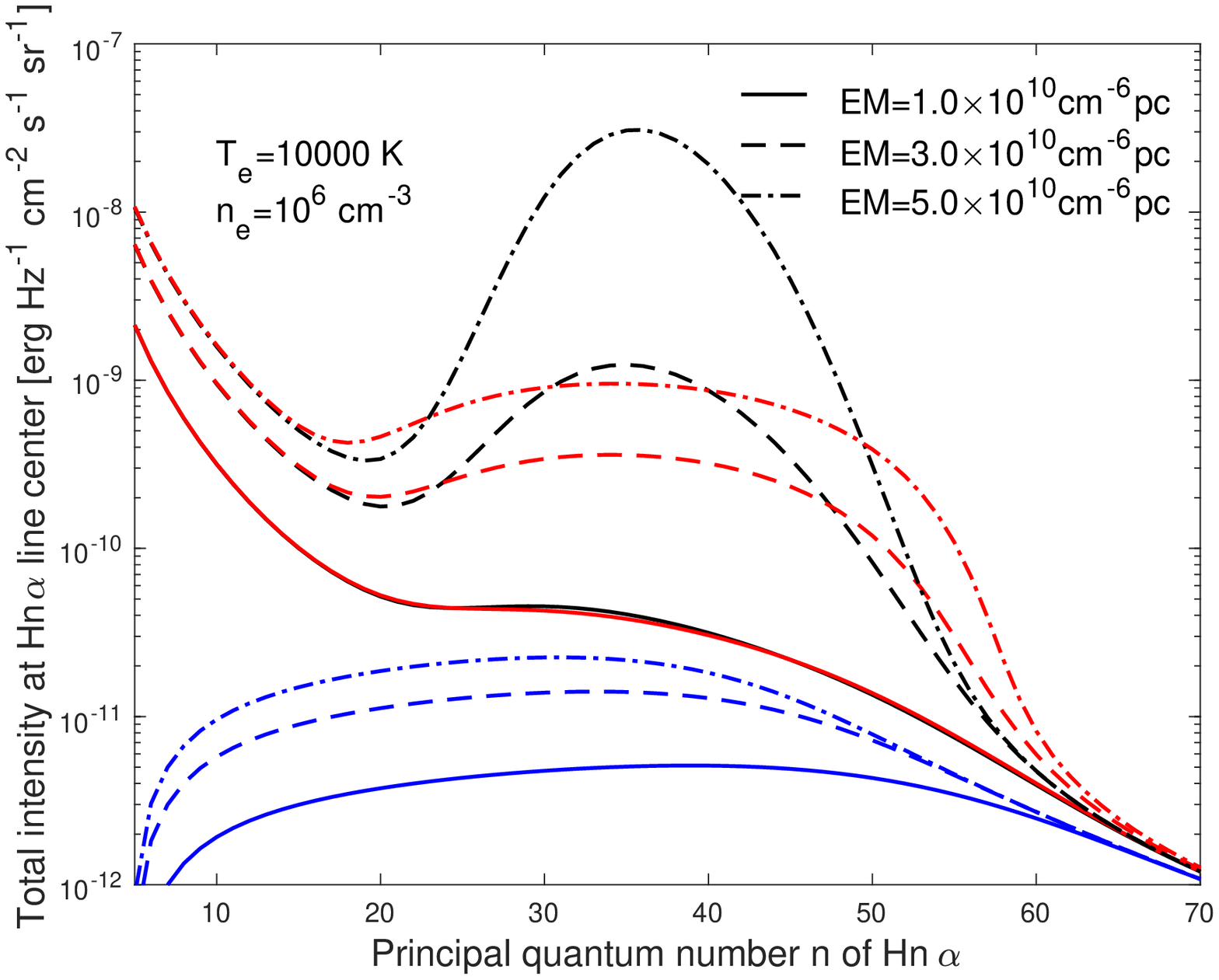}
\includegraphics[scale=0.4]{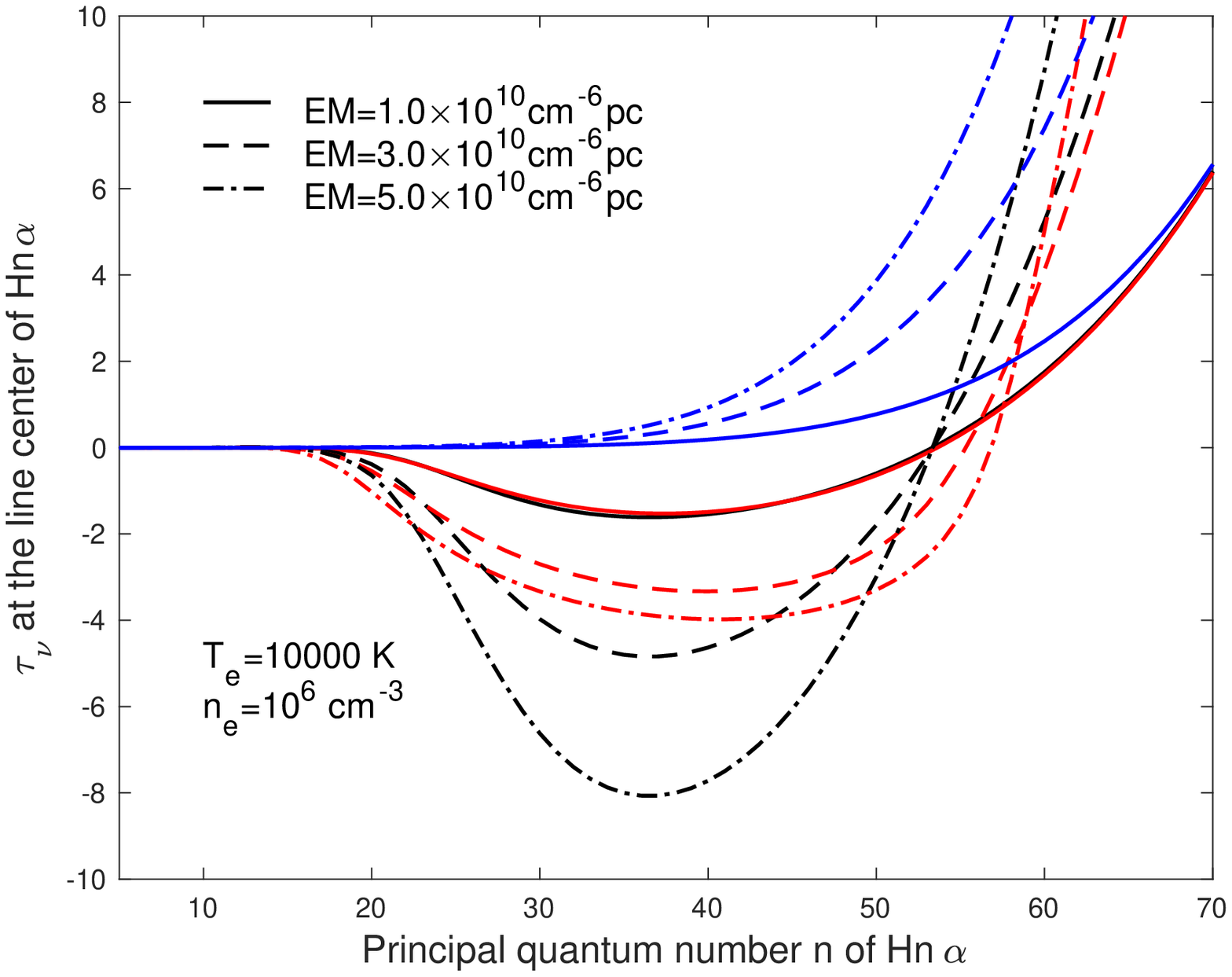}
\caption{Influence of radiation fields on hydrogen RRLs. The total intensities at the frequencies of  the Hn$\alpha$ line calculated with different EMs are plotted in the top panel. The total optical depths at the frequencies of the Hn$\alpha$ lines are presented in the bottom panel. The solid, dashed, and dash-dotted lines represent    the values for EM=$1.0\times10^{10}$, $3.0\times10^{10}$, and $5.0\times10^{10}$ cm$^{-6}$pc, respectively. The black lines represent    the values calculated without the effect of radiation fields on level populations, and    the red lines      the values calculated with radiation fields. The blue lines show    the values of    the continuum emissions.}\label{fig:radiation1}
\end{center}
\end{figure}

The    total optical depths ($\tau_\nu$) and    the    total intensities ($I_\nu$) including    the line and continuum intensities calculated with different EMs are presented in Fig. \ref{fig:radiation1}. The black lines represent    the values derived from    the departure coefficients $b_n$ calculated without    the radiation fields, while    the red lines represent    the values calculated with    the effect of radiation fields. The distortions of $b_n$ and $\beta_{n,n+1}$ shown in Fig. \ref{fig:radiation} are clearly amplified in    the intensities and optical depths displayed in Fig. \ref{fig:radiation1}. The intensities of    the stimulated lines could be overestimated considerably in    the center of    the masing range and underestimated on    the    two sides if    the radiation fields are not included in    the calculations of $b_n$ and $\beta_{n,n+1}$.  This is not surprising because of    the effect of radiation fields on    the population inversion mentioned above. In addition,    the effect of    the radiation fields causes    the valley of    the optical depth    to be shallower and wider. For    the conditions of EM$=3.0\times10^{10}$ and $5.0\times10^{10}$ cm$^{-6}$pc,    the ranges of $\tau_\nu<-1$ are wider when    the radiation fields are considered. In    the literature, a    typical maser is    thought    to occur if    the    total optical depth $\tau_\nu<-1$ \citep{str96,thu98,pro20}. \citet{str96} suggested    that saturated higher frequency masing lines could ``attract'' adjacent lower frequency lines    to exhibit masing, and    this is confirmed by    the numerical calculation in \citet{pro20}. Our results show    that it is also possible for saturated lower frequency masing lines    to attract adjacent higher frequency lines    to be masing.

\begin{figure}
\begin{center}
\includegraphics[scale=0.4]{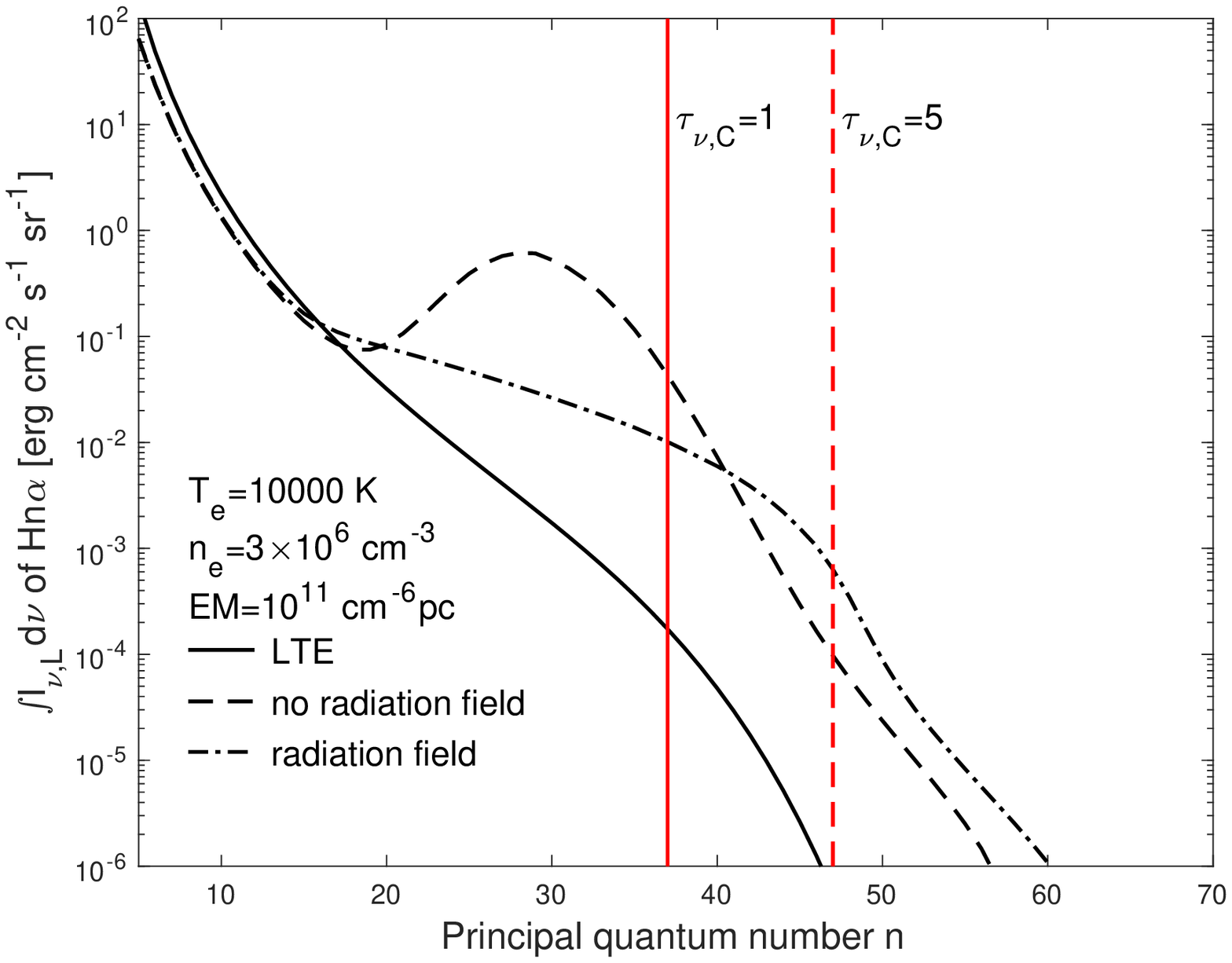}
\includegraphics[scale=0.4]{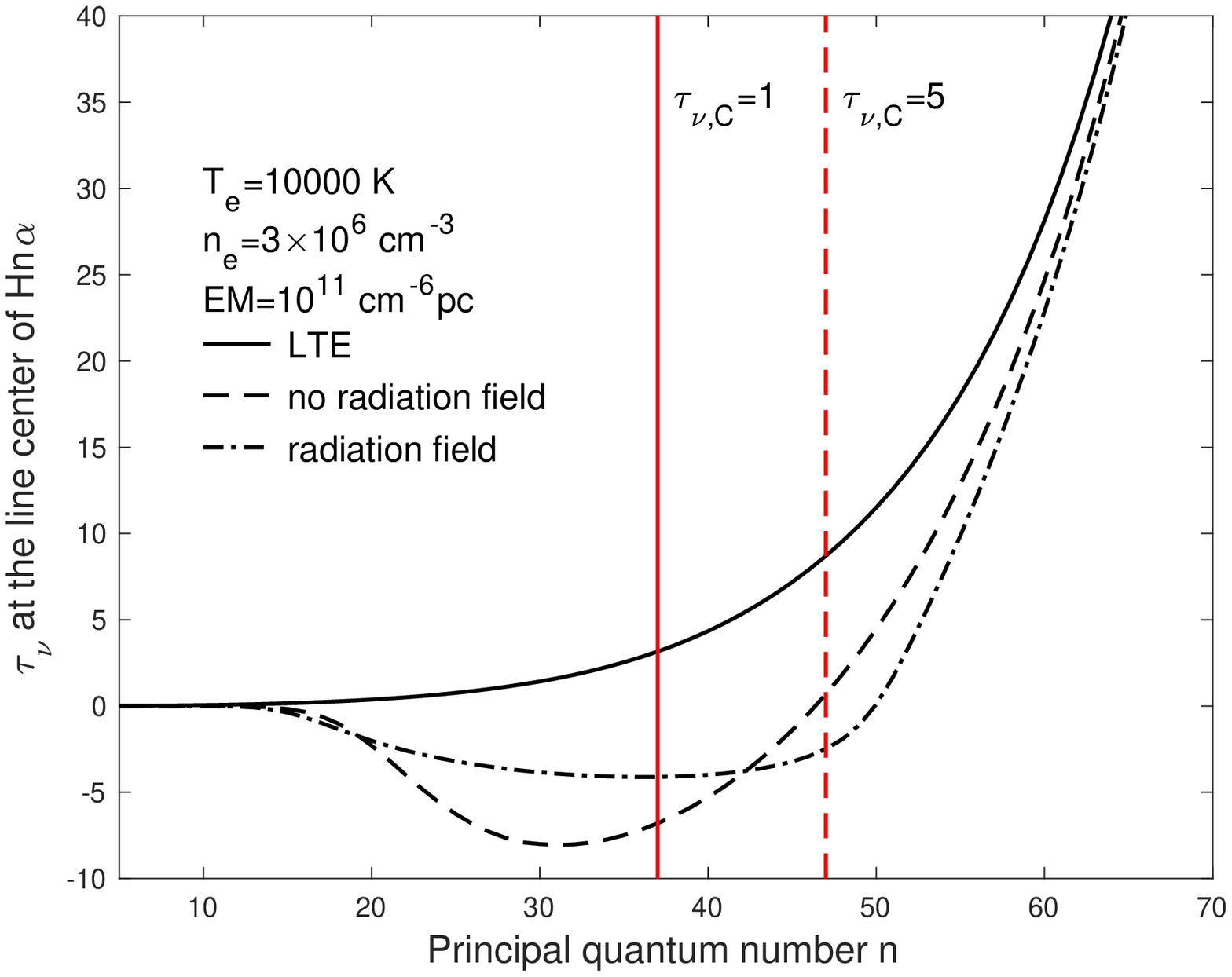}
\caption{Effect of stimulation emission on hydrogen RRLs. The frequency-integrated intensities of the Hn$\alpha$ line calculated under  LTE assumption (solid line) and  using non-LTE method without and with the effect of radiation field on departure coefficients (dashed and dash-dotted line) vs. principal quantum number $n$ are plotted in the top panel. The total optical depths at the frequencies of the Hn$\alpha$ lines are presented in the bottom panel. The red vertical lines indicate the positions of the continuum optical depth $\tau_{\nu,\textrm{C}}=1$ and $5$.}\label{fig:radiation2}
\end{center}
\end{figure}

In Fig. \ref{fig:radiation2}    the frequency-integrated intensities of the Hn$\alpha$ lines and    the    total optical depths $\tau_\nu$ calculated with    the LTE assumption are compared    to    those calculated without    the LTE assumption. In    this case, as an example, the electron density $n_e$ is $3.0\times10^6$ cm$^{-3}$, and EM is assumed    to be $10^{11}$ cm$^{-6}$ pc. As displayed in the top panel, the stimulated emission could raise the line intensity by several magnitudes more than the LTE intensity, even with a    thick continuum optical depth ($\tau_{\nu,\textrm{C}}>1$). This difference of line intensities is kept until $\beta_{n,n+1}$ and $b_n$ increase to be 1 at  high energy level n.

\begin{figure}
\begin{center}
\includegraphics[scale=0.4]{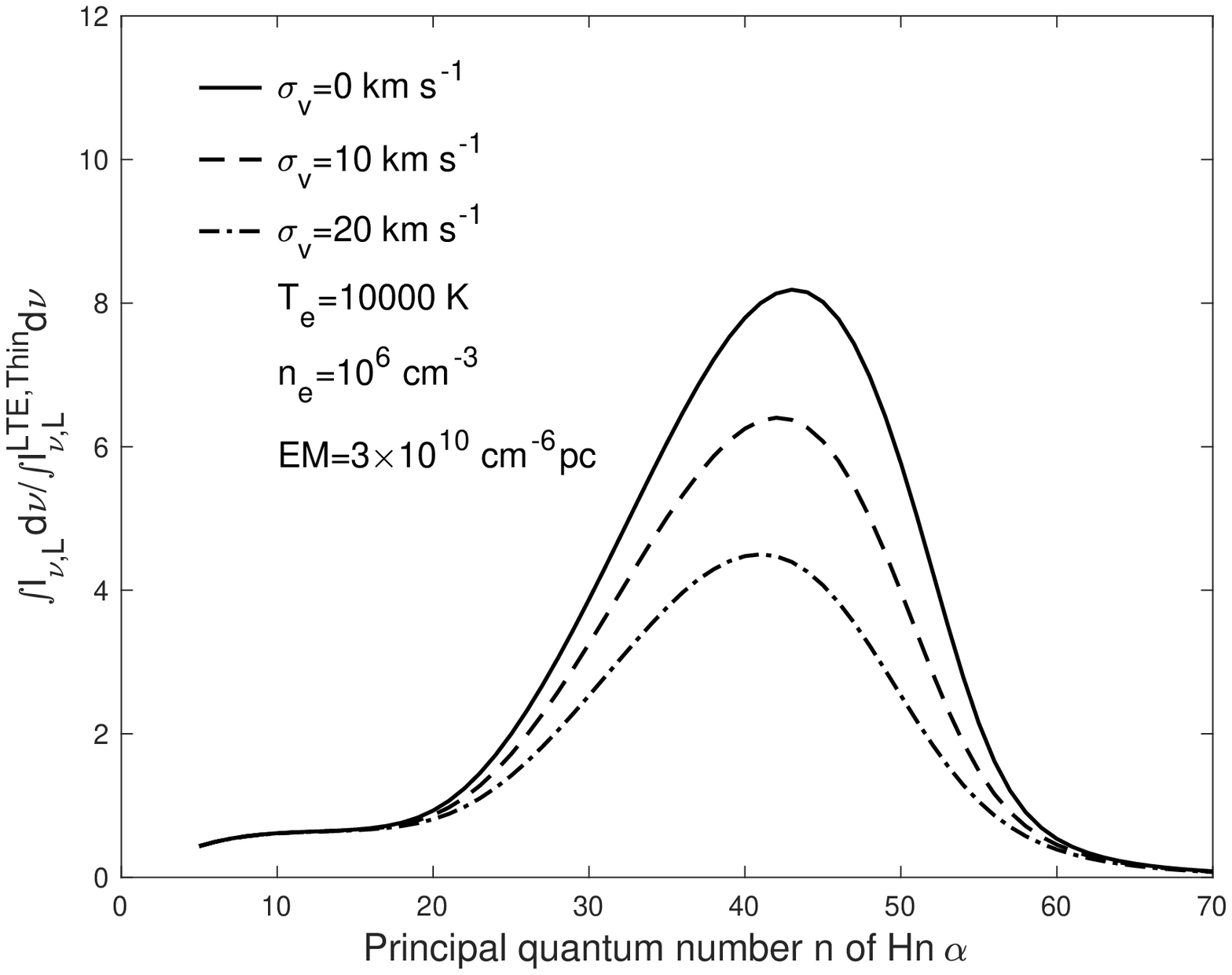}
\includegraphics[scale=0.4]{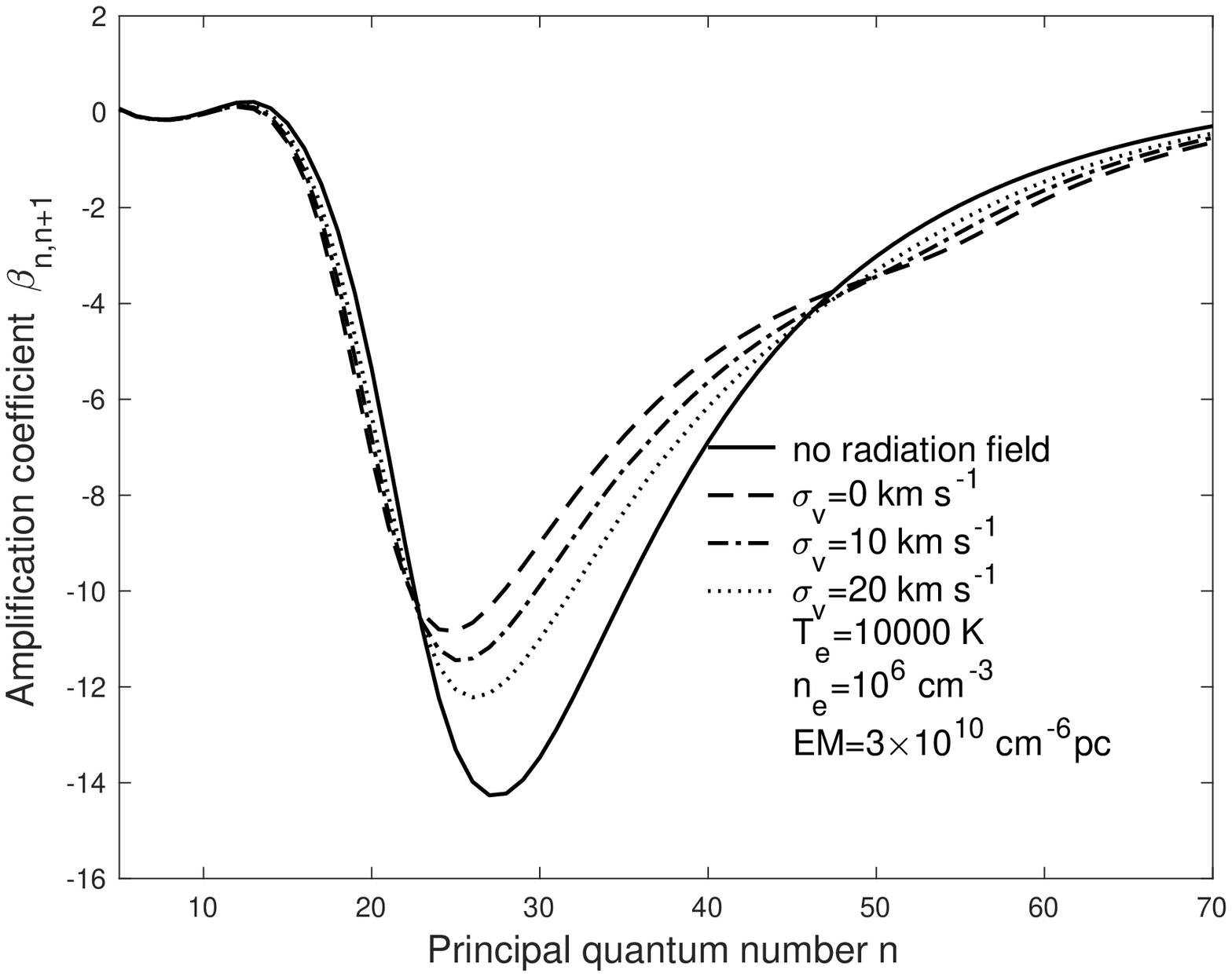}
\caption{Effects of different velocity fields on hydrogen RRLs. In the top panel the ratios of the frequency-integrated line intensities $\int I_{\nu,\textrm{L}}d\nu$ to the intensities calculated under    the LTE and optically    thin assumptions $\int I_{\nu,\textrm{L}}^{\textrm{LTE,Thin}}d\nu$ for different $\sigma_v$ are plotted. The amplification coefficients $\beta_{n,n+1}$ calculated with different values of  $\sigma_v$ are presented in    the bottom panel.}\label{fig:profile}
\end{center}
\end{figure}

\subsection{Effects of  electron temperature and velocity field on $\beta_{n,n+1}$ and line emissions} \label{sec:vel}

By affecting the line optical depth, the line profile has an important influence on the line emission when the stimulated emission is great. So it is necessary for us to study the effect of line profiles on the departure coefficients and recombination lines. If the velocity field is mainly caused by     microturbulence,  the line profile function $\phi_\nu$ of    the hydrogen recombination line is generally a Voigt profile, a convolution of the Doppler and Lorentz profiles. This line profile function is determined by $T_e$, $n_e$, and    the root mean square $\sigma_v$ of    the microturbulent velocity field \citep{pet12}. In the top panel of Fig. \ref{fig:profile}, the ratios of the frequency-integrated line intensities $\int I_{\nu,\textrm{L}}d\nu$ to the intensities calculated under the LTE and optically thin assumptions $\int I_{\nu,\textrm{L}}^{\textrm{LTE,Thin}}d\nu$ with different    turbulence velocity fields are presented. The optically    thin assumption is commonly used in    the analyses of observations, and leads    to    the LTE value being  overestimated when    the actual optical depths of continuum and line are high \citep{kim17,pre20}. In    this case    the electron    temperature and density are $T_e=10000$ K and $n_e=10^6$ cm$^{-3}$, respectively, while  EM is $3\times10^{10}$ cm$^{-6}$pc. The root mean squares of    the velocity field $\sigma_v$ are adjusted to be 0, 10, and 20 km s$^{-1}$. From    the recent observations for the  H40$\alpha$ line toward ultra-compact H II regions \citep{liu21},    the assumptions of    the velocity fields are appropriate. It is clear in the top panel of Fig. \ref{fig:profile} that the intensity of the hydrogen recombination line is significantly affected by    the line profile function. The greater broadening due to the stronger    turbulence would widen the line profile so that    the absolute value of the optical depth is reduced. This weakens    the stimulated emissions, and reduces    the distortions of $\beta_{n,n+1}$ from    the values calculated without    the radiation fields, as presented in the bottom panel of Fig. \ref{fig:profile}.

\begin{figure}
\begin{center}
\includegraphics[scale=0.4]{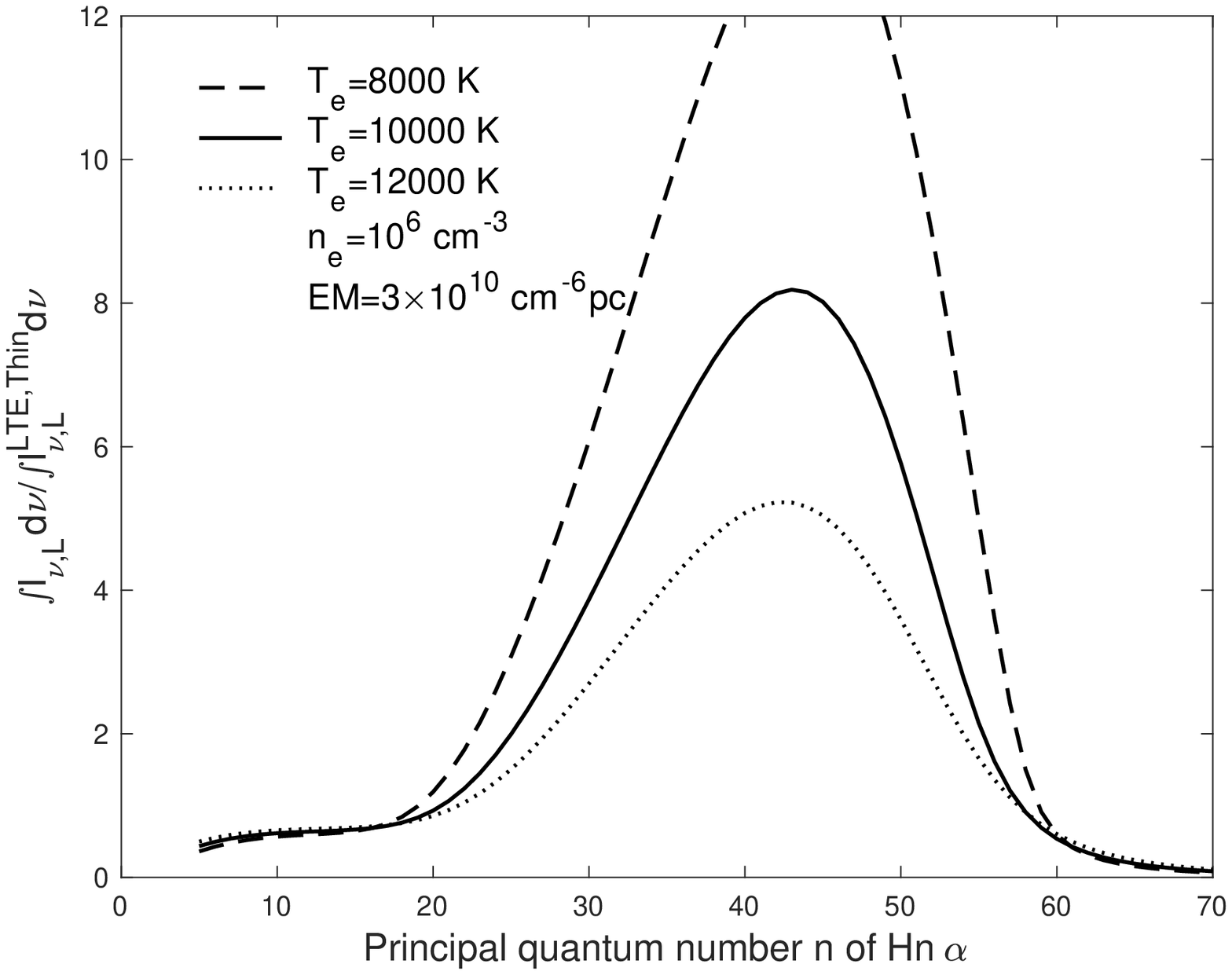}
\includegraphics[scale=0.4]{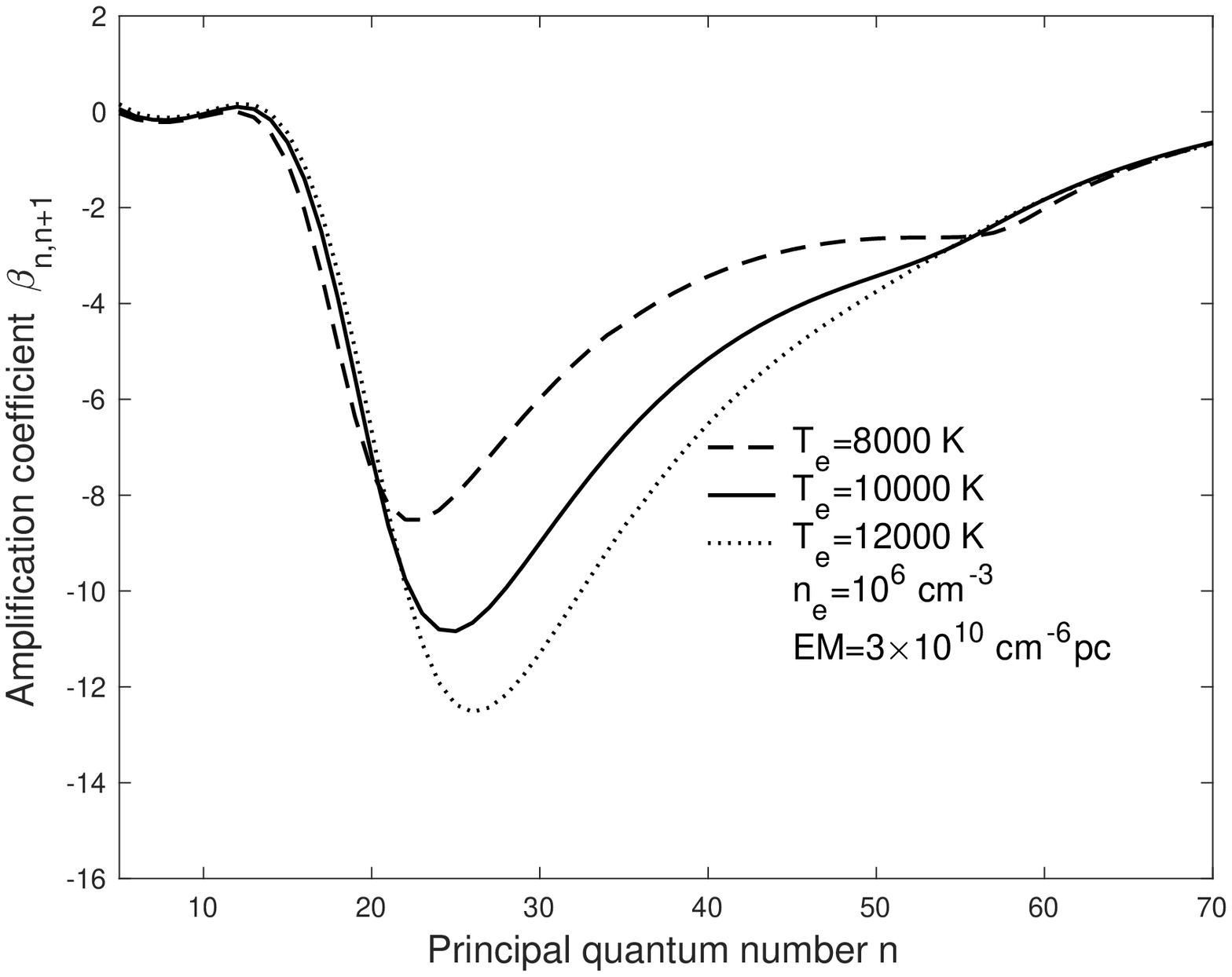}
\caption{Hydrogen RRLs affected by different electron temperatures. In    the top panel    the ratios of    the frequency-integrated line intensities $\int I_{\nu,\textrm{L}}d\nu$    to    the intensities $\int I_{\nu,\textrm{L}}^{\textrm{LTE,Thin}}d\nu$ calculated under    the LTE and optically    thin assumptions for different electron    temperatures are plotted. The corresponding amplification coefficients $\beta_{n,n+1}$ varying with the principal quantum number $n$ are presented in the bottom panel.}\label{fig:temperature}
\end{center}
\end{figure}

The electron    temperature $T_e$ is also an important condition that affects the properties of    the hydrogen recombination lines. The width of    the line profile function is influenced by    temperature. The continuum absorption coefficients and    the LTE line absorption coefficients both decrease with     increasing    temperature. The level populations,    the departure coefficients and    the amplification coefficients are also affected by    the electron    temperature. The amplification coefficients $\beta_{n,n+1}$ for different temperatures are presented in the bottom panel of Fig. \ref{fig:temperature}. In the calculations    the electron    temperatures are assumed    to be 8000 K, 10000 K, and 12000 K,  and    the velocity field of ionized gas is not considered. The other conditions are indicated in the panel. The line absorption coefficient $\kappa_{\nu,\textrm{L}}$ is a decisive factor in the strength of stimulated emission. It is proportional to the product of $\beta_{n,n+1}$ and $\kappa_{\nu,\textrm{L}}^{\textrm{LTE}}$. As  shown in the bottom panel, the negative value of $\beta_{n,n+1}$ is more significant when the temperature is higher, but the value of $\kappa_{\nu,\textrm{L}}^{\textrm{LTE}}$ decreases with  increasing  temperature. Since the absolute value of the product of $\beta_{n,n+1}$ and $\kappa_{\nu,\textrm{L}}$ decreases with increasing electron temperature, the contribution of simulated emission to the hydrogen recombination line emission increases with decreasing temperature. This trend is clearly shown in    the top panel of Fig. \ref{fig:temperature}.

\subsection{Widths of masing Hn$\alpha$ lines}

In addition to     the effect of    the velocity field of ionized gas,    the widths of hydrogen recombination lines are mainly determined by    thermal and pressure broadenings when    the optical depth is    thin and    the stimulated emissions are not important \citep{gor02}. When line masers occur, the line widths can be narrowed because the amplification of intensity at the line center is much greater    than    that at    the sides, due    to    the approximately exponential increase in stimulated emissions with the absolute value of the optical depth \citep{str96}.

\begin{figure}
\begin{center}
\includegraphics[scale=0.35]{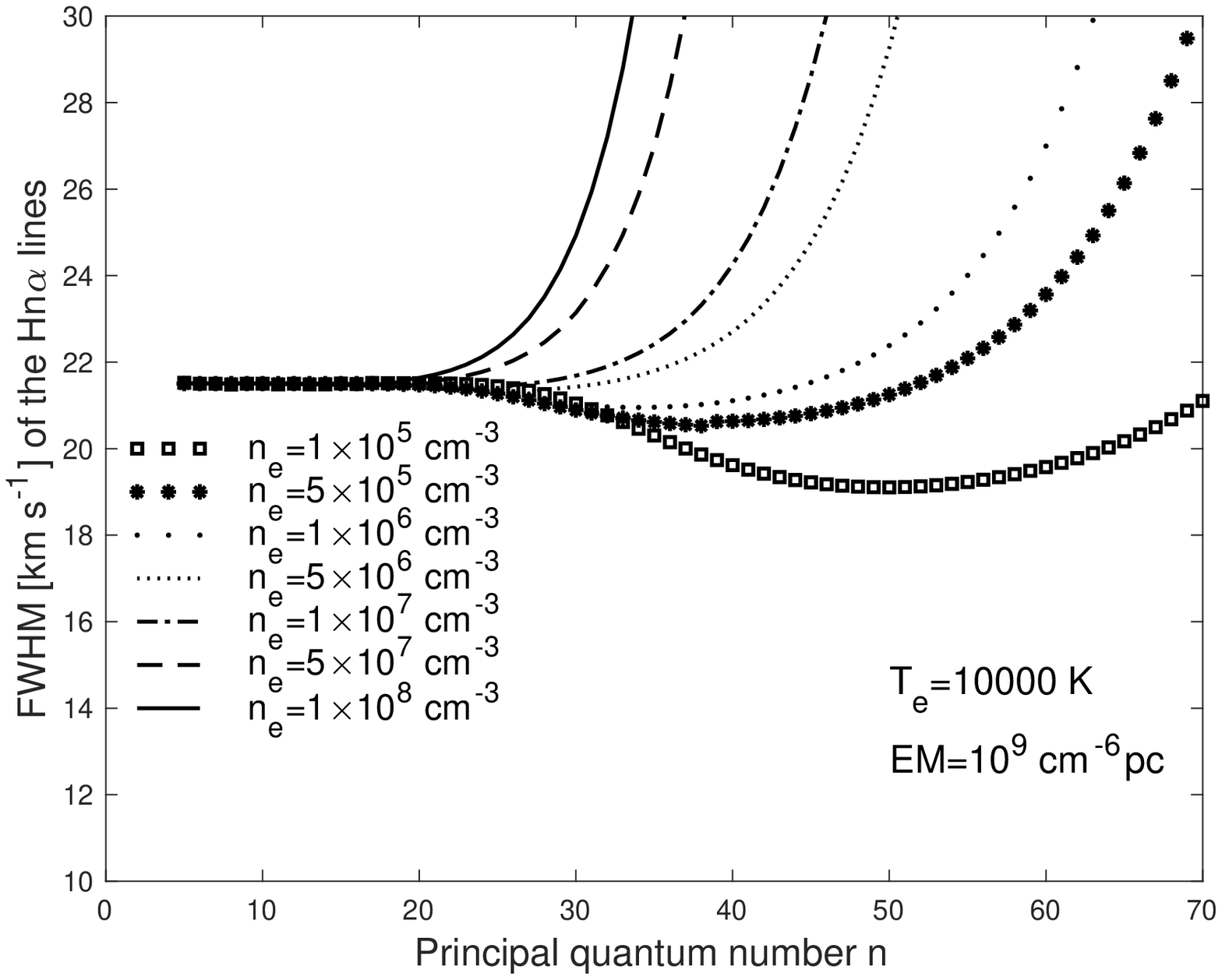}
\includegraphics[scale=0.35]{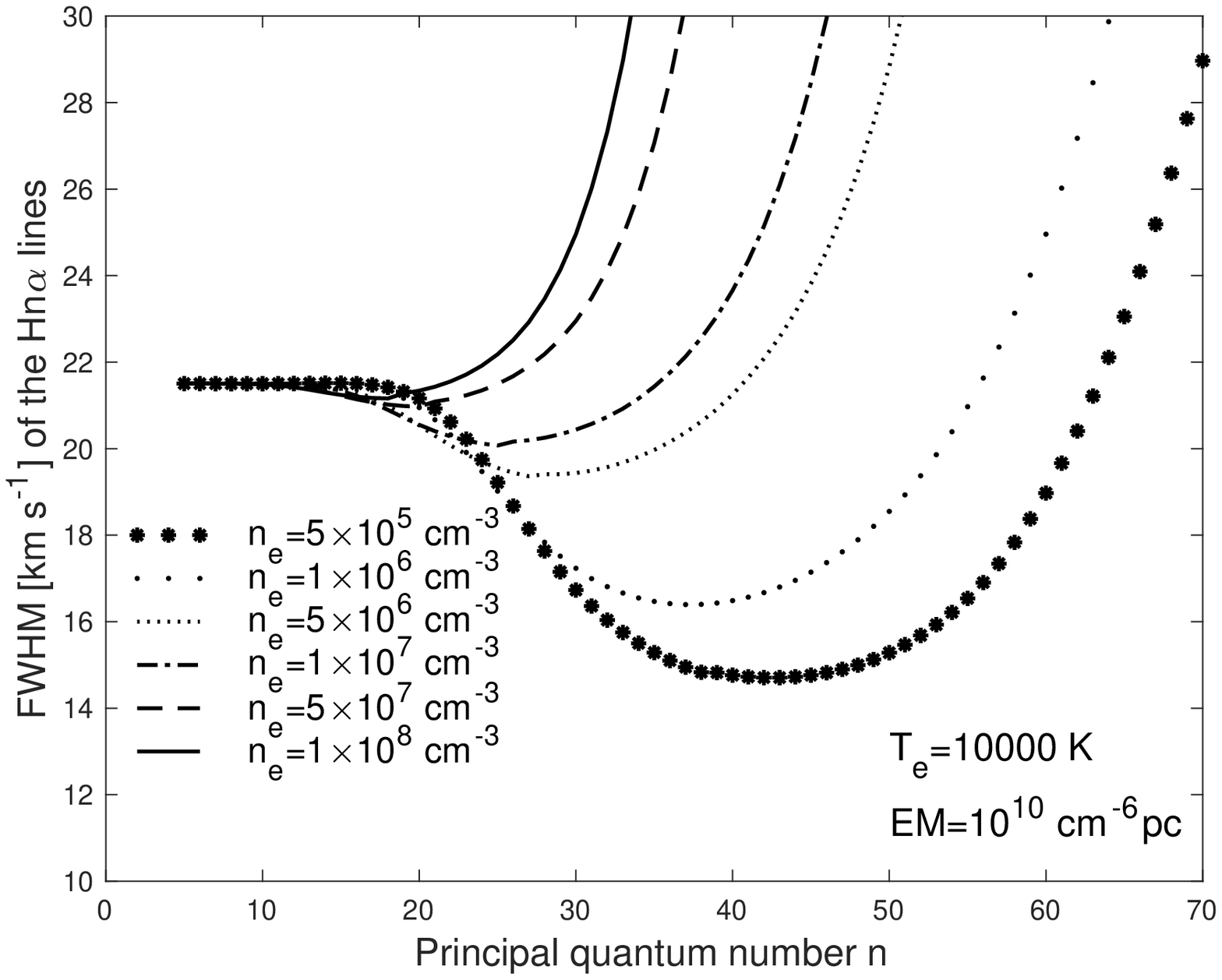}
\includegraphics[scale=0.35]{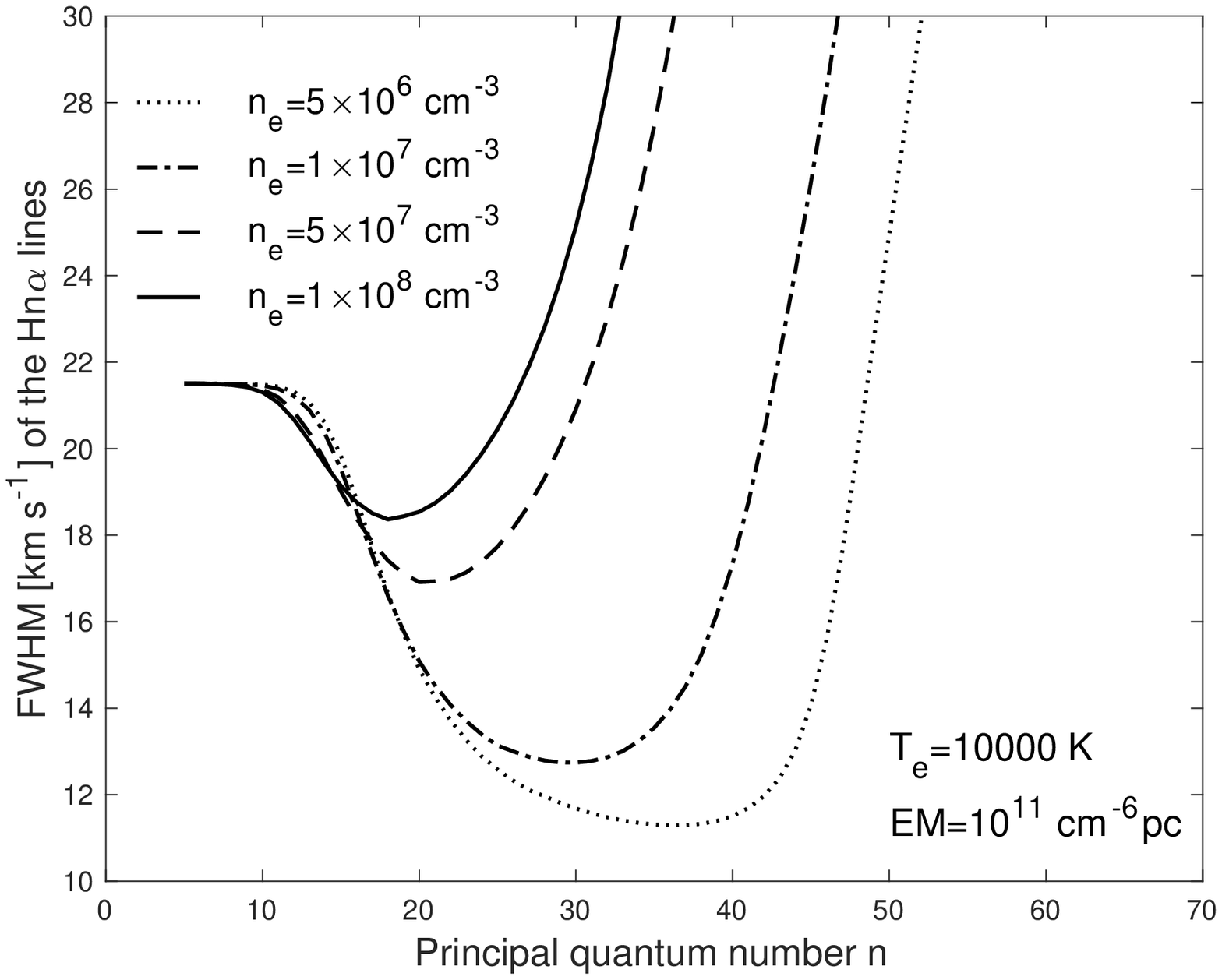}
\includegraphics[scale=0.35]{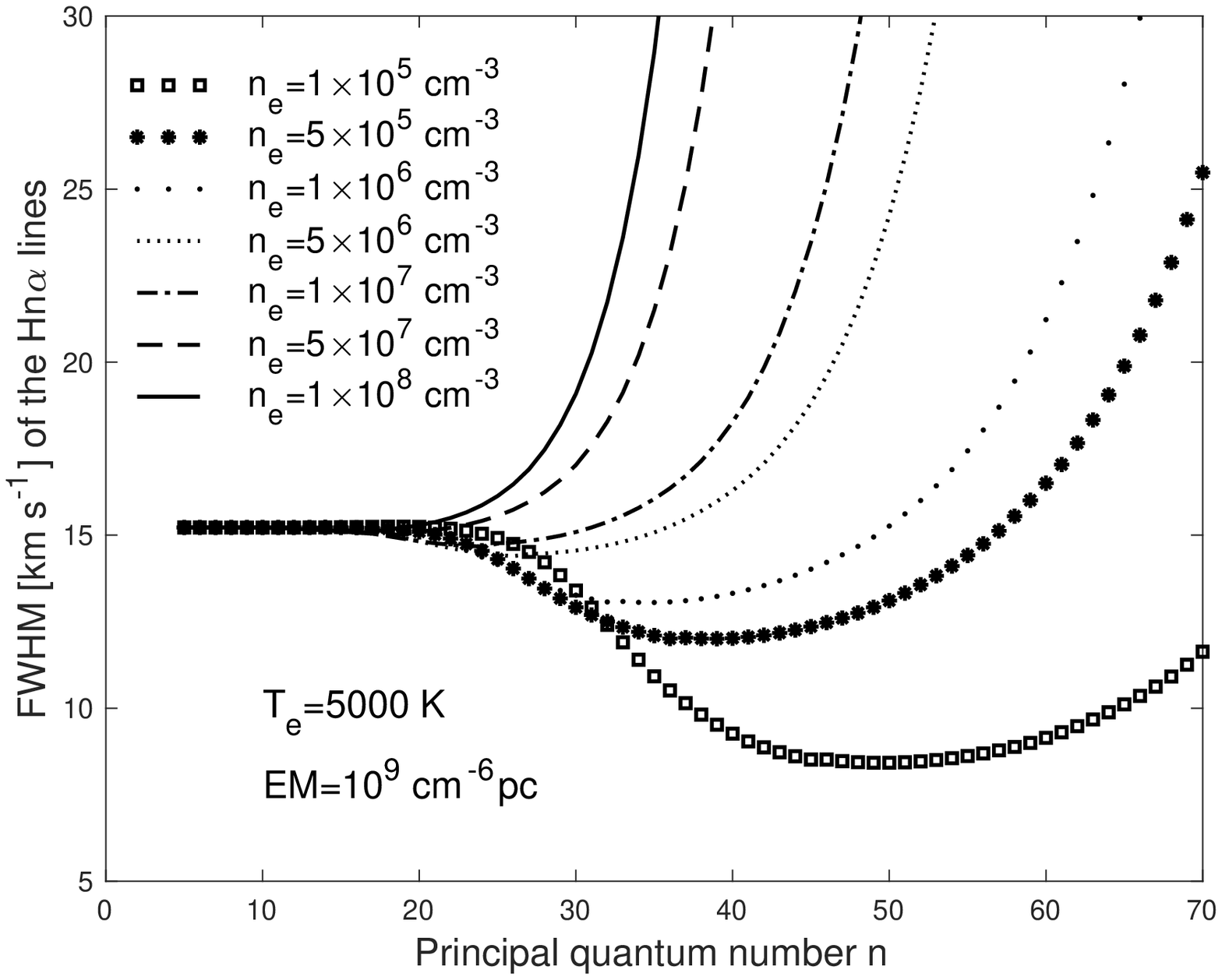}
\caption{Variations in    the Hn$\alpha$ line widths with principal quantum number $n$. The FWHMs presented in the top three panels are calculated with $T_e$=10000 K, and    the FWHM in the bottom panel is calculated with $T_e$=5000 K. The values of the  EMs are the same as in the calculation results shown in the same panel, while the values of the corresponding LOS depth D are different since the values of $n_e$ are different.}\label{fig:fwhm}
\end{center}
\end{figure}

The variations in    the widths of    the Hn$\alpha$ lines are plotted in Fig. \ref{fig:fwhm}, which shows the maser effect on the widths of Hn$\alpha$ lines. The change in the line widths with $n$ also indicates the influence of pressure broadening. Pressure broadening can increase the widths of hydrogen recombination lines. The effect of pressure broadening is important when electron density $n_e$ is high, and it increases with $n$ for Hn$\alpha$ lines. The effect of the line maser on line width can be neglected when the pressure broadening is very large due to high values of $n_e$ and $n$. However, the maser effect can still significantly decrease the widths of Hn$\alpha$ lines when the influence of pressure broadening is low. The results in Fig. \ref{fig:fwhm} show that the maser effect may reduce    the full width at half maximum (FWHM) of    the Hn$\alpha$ line profile by several km s$^{-1}$.  For a H II region with a low equilibrium    temperature of    $T_e\sim5000$ K,    the FWHM of    the Hn$\alpha$ line may even be lower    than 10 km s$^{-1}$. In addition, although only the narrow Hn$\alpha$ line widths with n<70 are shown in Fig. \ref{fig:fwhm}, the widths of Hn$\alpha$ lines with high $n$ will also be narrowed by the maser effect when the electron density becomes lower than $10^5$ cm$^{-3}$. Comparing the results presented in the top three panels of this figure, it is clear that the maser effect increases with  EM. According to our calculations, the reduction in hydrogen recombination line widths due to line masers is significant ($>2$ km s$^{-1}$) only if the EM is higher than $3.0\times10^7$ cm$^{-6}$ pc. This range of EMs corresponds to the conditions of hyper-compact, ultra-compact, and compact H II regions \citep{fue20b}.

Some cases of hydrogen recombination line maser have been detected in previous observations \citep{mar89,cox95,jim11,ale18}. In the case of MWC349, comparing the widths of the  H26$\alpha$, H27$\alpha$, H36$\alpha$, and H40$\alpha$ lines in Table 2 of \citet{thu95}, the $\sim11$ km s$^{-1}$ widths of the red and blue components of the H26$\alpha$ and H27$\alpha$ lines are  narrowed by the stimulation effect. The difference between the widths of the broad component of the H30$\alpha$ and H38$\beta$ lines also shows the effect of stimulated emission. In addition, many H II regions with narrow hydrogen recombination lines have been discovered \citep{loc89,pla91,and11,che20}. Although ``cool'' nebulae with low electron temperatures because of high metal abundances can lead to these narrow lines \citep{sha70,sha79}, the maser effect is also a natural explanation for these phenomena.


\subsection{Line optical depth leading to  observable  line masers}

\citet{gol66} pointed out that the hydrogen recombination line intensity, increased by    the strong stimulated emission, with a negative line optical depth $\tau_{\nu,\textrm{L}}$ can be significantly higher    than its LTE value, even if    the    total optical depth $\tau_{\nu}$ is still positive. However, \citet{str96} claimed that a total optical depth much lower than -1 is necessary to produce an observable maser amplification.

\begin{figure}
\begin{center}
\includegraphics[scale=0.35]{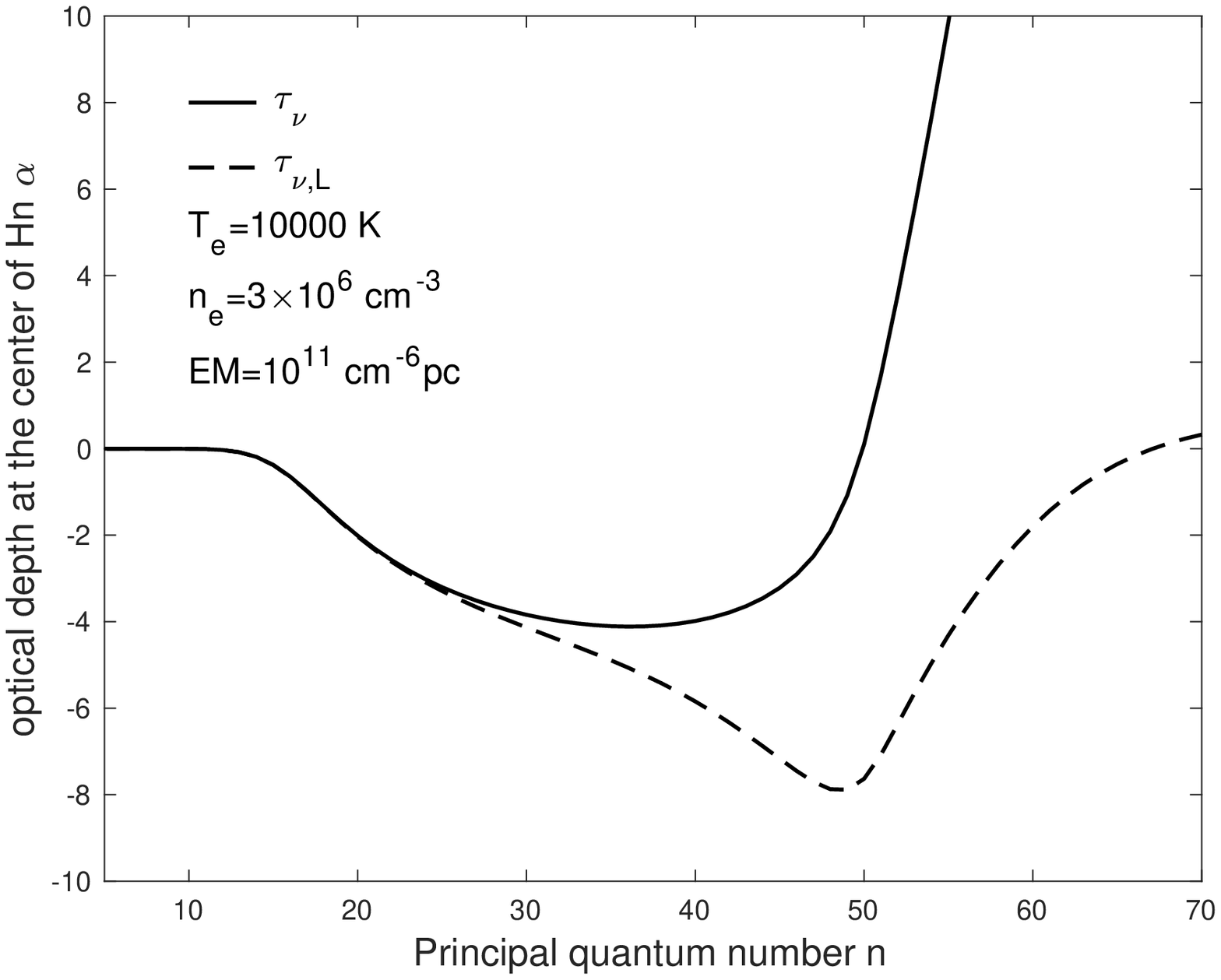}
\includegraphics[scale=0.35]{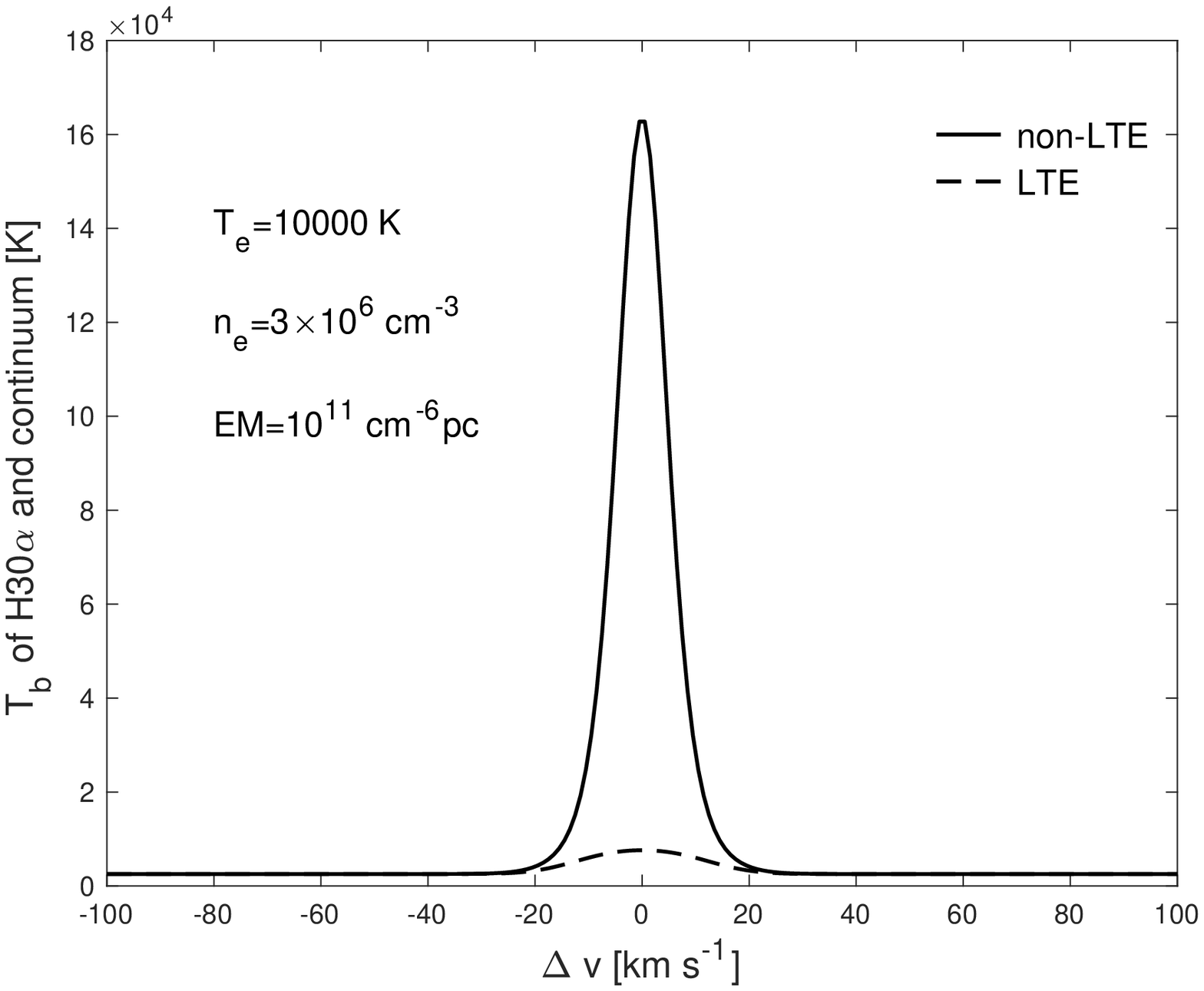}
\includegraphics[scale=0.35]{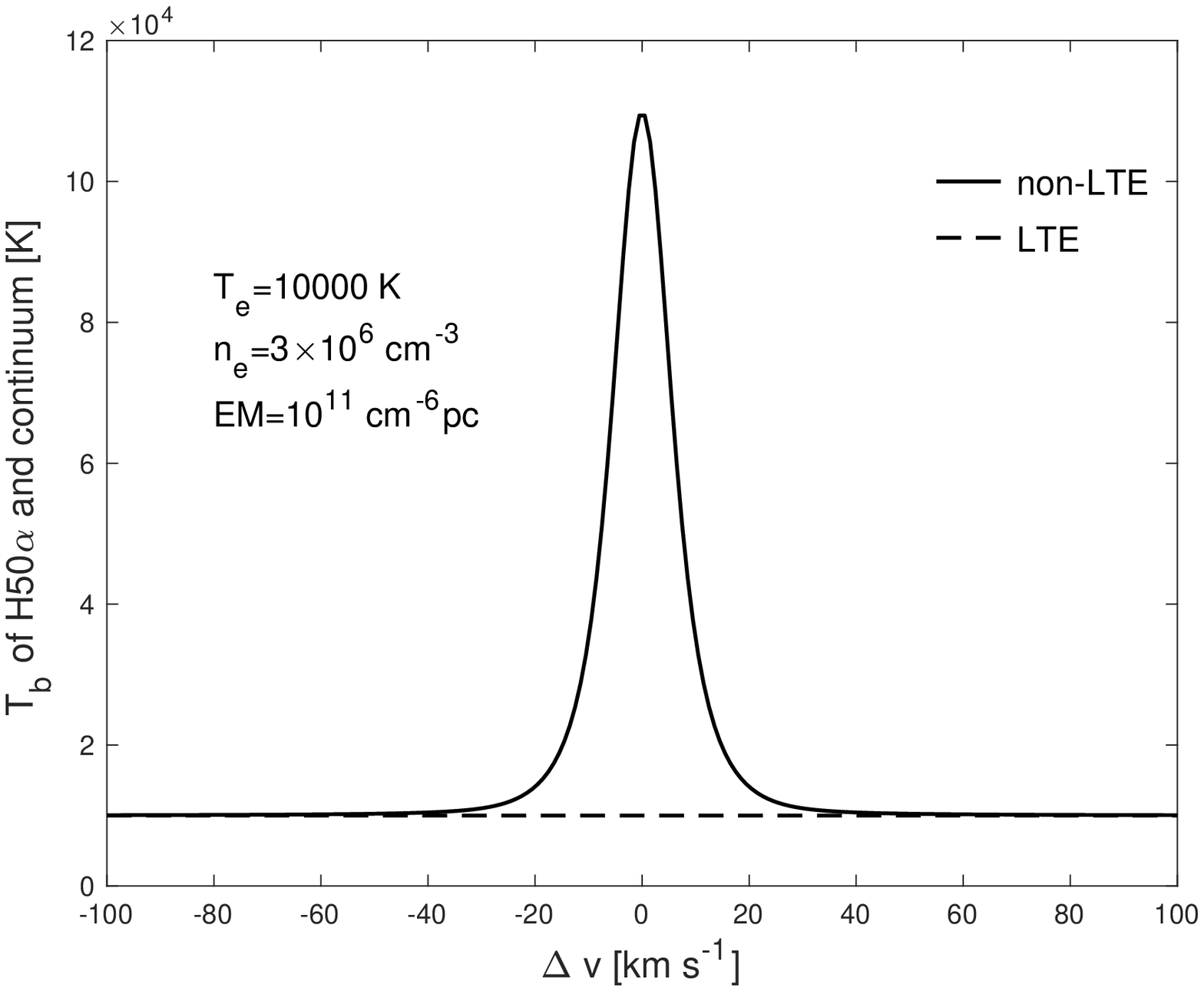}
\includegraphics[scale=0.35]{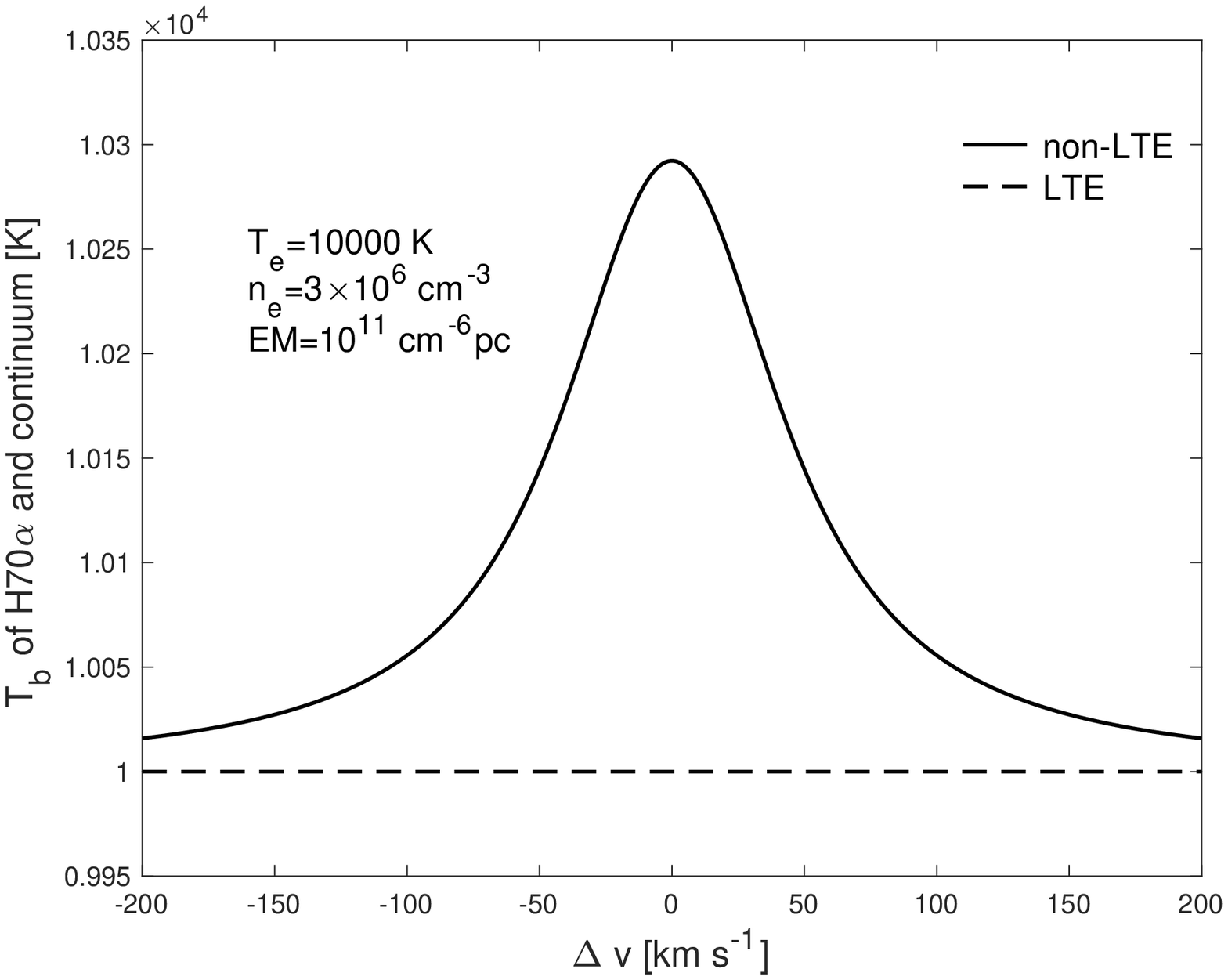}
\caption{Relation between the line optical depth and the maser amplification. The total and line optical depths vs. the energy level $n$ is displayed in the top panel. The brightness temperature $T_b$ of the H30$\alpha$, H50$\alpha$, and H70$\alpha$ line profiles and the continuum is presented in the top middle, bottom middle, and bottom panels, respectively.}\label{fig:SltoSlLTE}
\end{center}
\end{figure}

From our calculation of    the case mentioned in Sect. \ref{sec:tem_losdepth} and shown in the top panel of Fig. \ref{fig:radiation2}, we find that the ratio of the non-LTE Hn$\alpha$ line intensity    to    the LTE value with high energy level $n>48$ and positive $\tau_\nu$ is even higher    than    that with considerably negative optical depth $\tau_\nu<-1$ at low energy level $n<48$. The profiles of    the H30$\alpha$, H50$\alpha$, and H70$\alpha$ lines are presented in Fig. \ref{fig:SltoSlLTE} along with the LTE profiles of these lines. At the H30$\alpha$ line center, $\tau_\nu=-3.84$ and $\tau_{\nu,\textrm{L}}=-4.13$. The peak brightness temperature $T_b$ of    the H30$\alpha$ line is  higher than the    $T_b$ under    the LTE assumption. For the center of    the H50$\alpha$ line, $\tau_{\nu,\textrm{C}}=7.73$ is thick,  $\tau_\nu=0.10$ is positive, and the line optical depth $\tau_{\nu,\textrm{L}}=-7.63$ is significantly lower    than -1. The brightness temperature $T_b$ of the continuum emission is 9995.6 K because of the very    thick continuum optical depth, and    the line center of the H50$\alpha$ line even reaches $9.9\times10^4$ K. The LTE line center is only 4.4 K, which is several orders  lower    than    the non-LTE line strength. The width (FWHM) of the    H50$\alpha$ optical depth  profile function is 22.9 km s$^{-1}$, while the FWHM of the  H50$\alpha$  line profile is  only 13.0 km s$^{-1}$ because the maser effect is important. On the other hand, due to the relatively thick optical depth, the FWHM of the  H50$\alpha$ line will be  36.6 km s$^{-1}$ if the LTE assumption is used. In addition, for the center of the H70$\alpha$ line, $\tau_\nu=65.9$ and $\tau_{\nu,\textrm{L}}=0.32$. The level population is overheating, and the amplification coefficient is $\beta_{70,71}=0.14$. The temperature T$_b$ at the line center is $292$ K. On the contrary, the emission at    the center of  H70$\alpha$   will be negligible due    to    the very high optical depths if the LTE assumption is used. 

%
%
%
%
%
%
%

The H II regions B2 and G2 in W49A observed by \citet{pre20} show very high EM $>5\times10^9$ cm$^{-6}$pc. Although the corresponding continuum optical depth should be higher than 18, the H92$\alpha$ line in these sources can still be detected. This cannot be explained under the LTE assumption, whereas inversion or overheating of the level population should exist. Under non-LTE conditions, the stimulated emission produced by a population inversion could make the corresponding hydrogen recombination line detectable even if the continuum optical depth is very thick.


\subsection{Significance of stimulated emissions}

In the preceding sections we show results only for hyper-compact H II regions. However, the stimulated emissions still show significant influence in more classical H II regions.
We use the classification of H II regions provided by \citet{fue20b} in the subsequent analyses of the importance of stimulated emissions.

\begin{figure}
\begin{center}
\includegraphics[scale=0.4]{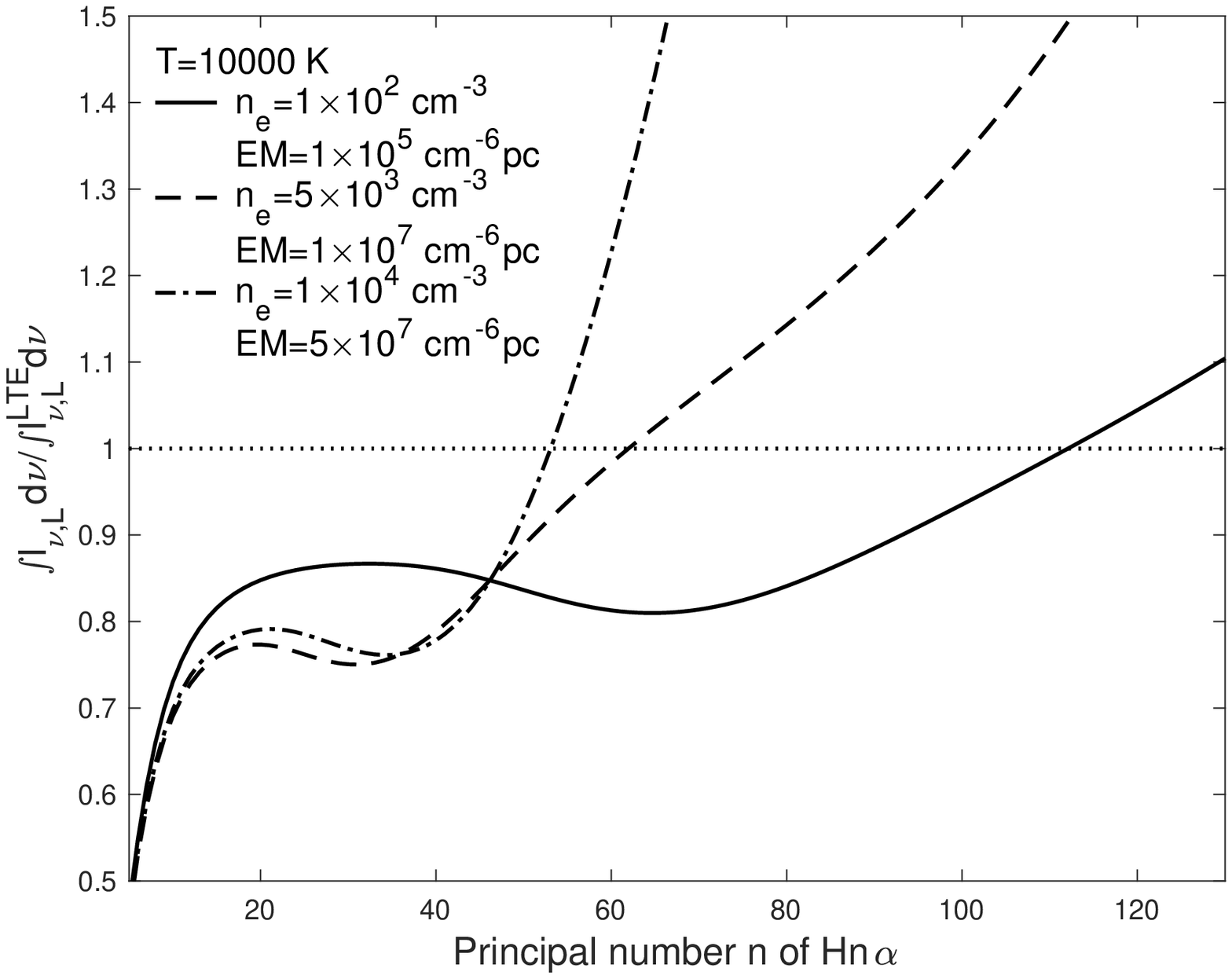}
\caption{Ratios of   non-LTE frequency-integrated line intensities to   LTE values ($\int I_{\nu,\textrm{L}}d\nu/\int I_{\nu,\textrm{L}}^\textrm{LTE}d\nu$) in typical ultra-compact ($n_e=10^4$ cm$^{-3}$ and EM$=5\times10^7$ cm$^{-6}$pc), compact ($n_e=5\times10^3$ cm$^{-3}$ and EM$=10^7$ cm$^{-6}$pc), and extended ($n_e=1\times10^2$ cm$^{-3}$ and EM$=10^5$ cm$^{-6}$pc) H II regions for $T_e=10000$ K. }\label{fig:cate_ltolte}
\end{center}
\end{figure}

The ratios of non-LTE Hn$\alpha$ line intensities to LTE values ($\int I_{\nu,\textrm{L}}d\nu/\int I_{\nu,\textrm{L}}^\textrm{LTE}d\nu$) for the typical conditions of ultra-compact, compact, and extended H II regions are plotted in Fig. \ref{fig:cate_ltolte}. The ratio decreases quickly with  decreasing $n$ when the principal quantum number $n$ is lower than 20. Then after a small decline in the range of $n\approx 20-40$, the ratio increases monotonically when the $n$ is higher. When the principal quantum number n is low, the line and continuum optical depths are very small. Then the ratio is approximately equal to the departure coefficient at the upper level. When the optical depths increase with $n$, stimulation has a gradually more important influence on the Hn$\alpha$ line intensity. Moreover, the significance of stimulation is greatly influenced by EM. For ultra-compact H II regions, the Hn$\alpha$ line intensities in the range of $n>60$ under non-LTE conditions are  higher than those under LTE conditions;  the non-LTE values are lower than the LTE values for extended H II regions until $n$ exceeds 105.

As  mentioned above, the ratio of non-LTE Hn$\alpha$ line intensity to the LTE value increases with $n$ if the small decline in the range of $n\approx 20-40$ is ignored. We define $n_{\textrm{crit}}^{\textrm{low}}$ as the biggest $n$ with which the non-LTE Hn$\alpha$ line intensity is more than $15\%$ lower than the one in LTE. And $n_{\textrm{crit}}^{\textrm{high}}$ is defined as the smallest $n$ with which the non-LTE Hn$\alpha$ line intensity is more than $15\%$ higher than the LTE value. Then the Hn$\alpha$ lines in the range between $n_{\textrm{crit}}^{\textrm{low}}$ and $n_{\textrm{crit}}^{\textrm{high}}$ can be treated under the LTE assumption with an error $<15\%$.

The values of $n_{\textrm{crit}}^{\textrm{low}}$ and $n_{\textrm{crit}}^{\textrm{high}}$ at different $n_e$ and EMs for $T_e=5000$ K and $10000$ K are listed in Table \ref{table:ncrit}. Temperatures of  $T_e=10000$ K are  typical for  H II regions, and the values for $T_e=5000$ K can display the conditions in low-temperature H II regions. The upper limit of the Lyman continuum photon production rate ($N'_c$) is assumed to be $1.5\times10^{50}$ s$^{-1}$ corresponding to the rate of a young massive star cluster \citep{ngu17}. For a spherical and uniform H II region, $N'_c$ can be calculated from $T_e$, $n_e$, and radius $r=D/2$ by using Eq. A4 in \citet{fue20b}. Conversely, the $r$ and EM for given $T_e$ and $n_e$ can be estimated from $N'_c$. Considering that  H II regions might be not spherical, the upper limit of EM is assumed to be 2 times of the value of EM calculated from the upper limit of $N'_c$ with given $T_e$ and $n_e$ under the assumption of spherical H II regions. Furthermore, the lower limit of EM is calculated from the $N'_c=10^{44}$ s$^{-1}$ corresponding to the value of an $\sim8$ M$_\odot$ massive star \citep{dia98}. 

As is shown in Table \ref{table:ncrit}, the range of $n$ between $n_{\textrm{crit}}^{\textrm{low}}$ and $n_{\textrm{crit}}^{\textrm{high}}$ gradually becomes narrow with the increase in EM. The Hn$\alpha$ lines in the centimeter waveband ($60\leq n\leq129)$ are often used to derive the LTE temperature ($T_e^*$) under the LTE assumption as a close estimate of the actual $T_e$ \citep{sha83,wil15}. In our calculations, the LTE approximation for most of the centimeter Hn$\alpha$ lines is appropriate when EM is lower than $10^6$ cm$^{-6}$ pc. On the contrary, in the cases of ultra- and hyper-compact H II regions with EM$>5\times10^7$ cm$^{-6}$pc, the deviation from the LTE approximation is commonly significant for the centimeter Hn$\alpha$ lines. Some of the millimeter Hn$\alpha$ lines ($28\leq n<60$) may be useful to accurately estimate the actual $T_e$ under the LTE approximation for ultra-compact H II regions, but the suitable range of $n$ is often narrow and changes strongly with EM. So a precise estimate of EM is necessary to determine the range of Hn$\alpha$ lines suitable for the LTE approximation. For hyper-compact H II regions, the suitable range of $n$ changes greatly with $T_e$ and $n_e$. Even if the EM is known, no Hn$\alpha$ lines can be assured to be appropriate for the LTE approximation before $T_e$ and $n_e$ are obtained.

\begin{table*}\tiny
\centering
\caption{Range of the Hn$\alpha$ lines suitable for the LTE approximation.} \label{table:ncrit}
\begin{threeparttable}
\begin{tabular}{|c|ccccccccc|}
\hline
 $T_e=5000$ K& \multicolumn{9}{c|}{$n_e$ [cm$^{-3}$]}  \\
\hline
EM [cm$^{-6}$pc] & $1\times10^2$ & $5\times10^2$ & $1\times10^3$ & $5\times10^3$ & $1\times10^4$ & $5\times10^4$ & $1\times10^5$ & $1\times10^6$ & $1\times10^7$ \\
\hline
$1\times10^5$ & 88-127 & 74-156 & 69-181 & ... & ... & ... & ... & ... & ... \\
$5\times10^5$ & ... & 71-104 & 67-112 & 56-154 & ... & ... & ... & ... & ... \\
$1\times10^6$ & ... & 69-91 & 65-95 & 55-123 & 51-143 & ... & ... & ... & ... \\
$5\times10^6$ & ... & ... & 60-72 & 53-77 & 49-84 & 41-122 & ... & ... & ... \\
$1\times10^7$ & ... & ... & ... & 51-66 & 48-69 & 41-97 & ... & ... & ... \\
$5\times10^7$ & ... & ... & ... & ... & 43-51 & 38-53 & 36-59 & ... & ... \\
$1\times10^8$ & ... & ... & ... & ... & 41-47 & 36-45 & 34-47 & ... & ... \\
$5\times10^8$ & ... & ... & ... & ... & ... & 32-36 & 30-35 & 25-35 & ... \\
$1\times10^9$ & ... & ... & ... & ... & ... & ... & 28-31 & 24-29 & ... \\
$5\times10^9$ & ... & ... & ... & ... & ... & ... & ... & 20-23 & ... \\
$1\times10^{10}$ & ... & ... & ... & ... & ... & ... & ... & 18-20 & 16-19 \\
$5\times10^{10}$ & ... & ... & ... & ... & ... & ... & ... & 15-17 & 14-15 \\
\hline
 $T_e=10000$ K& \multicolumn{9}{c|}{$n_e$ [cm$^{-3}$]}  \\
\hline
EM [cm$^{-6}$pc] & $1\times10^2$ & $5\times10^2$ & $1\times10^3$ & $5\times10^3$ & $1\times10^4$ & $5\times10^4$ & $1\times10^5$ & $1\times10^6$ & $1\times10^7$ \\
\hline
$1\times10^5$ & 82-138 & 68-173 & 62-201 & ... & ... & ... & ... & ... & ... \\
$5\times10^5$ & ... & 67-115 & 62-126 & 51-174 & ... & ... & ... & ... & ... \\
$1\times10^6$ & ... & 66-100 & 61-106 & 51-140 & 47-163 & ... & ... & ... & ... \\
$5\times10^6$ & ... & ... & 58-79 & 50-88 & 46-98 & ... & ... & ... & ... \\
$1\times10^7$ & ... & ... & ... & 49-76 & 46-81 & 38-112 & ... & ... & ... \\
$5\times10^7$ & ... & ... & ... & 46-58 & 43-59 & 37-67 & 35-77 & ... & ... \\
$1\times10^8$ & ... & ... & ... & ... & 42-53 & 36-56 & 34-61 & ... & ... \\
$5\times10^8$ & ... & ... & ... & ... & ... & 33-42 & 32-42 & 26-59 & ... \\
$1\times10^9$ & ... & ... & ... & ... & ... & 32-38 & 30-38 & 25-43 & ... \\
$5\times10^9$ & ... & ... & ... & ... & ... & ... & ... & 22-28 & ... \\
$1\times10^{10}$ & ... & ... & ... & ... & ... & ... & ... & 21-26 & 18-26 \\
$5\times10^{10}$ & ... & ... & ... & ... & ... & ... & ... & 17-20 & 15-19 \\
\hline
\end{tabular}
\begin{tablenotes}
\item \textbf{Notes.} The values of $n_{\textrm{crit}}^{\textrm{low}}$ - $n_{\textrm{crit}}^{\textrm{high}}$ at different $n_e$ and EMs for $T_e=5000$ and $10000$ K are given. The range between $n_{\textrm{crit}}^{\textrm{low}}$ and $n_{\textrm{crit}}^{\textrm{high}}$ shows the range of the Hn$\alpha$ lines suitable for the LTE approximation.
\end{tablenotes}
\end{threeparttable}
\end{table*}

\subsection{Effect of population inversion in extended H II regions around ultra-compact H II regions} \label{sec:two}

The results presented in Table \ref{table:ncrit} seem to suggest that the contribution from stimulated emission to the centimeter Hn$\alpha$ lines is relatively weak and lower than 15$\%$ if EM$<10^6$ cm$^{-6}$ pc. However, as for the case of the extended H II region ($n_e=10^2$ cm$^{-3}$ and EM$=10^5$ cm$^{-6}$pc) plotted in Fig. \ref{fig:cate_ltolte}, there could still be an approximately $10\%$ difference between the non-LTE and LTE line intensities. If $T_e$ and $n_e$ are lower, the difference will be bigger. Furthermore, the stimulation effect could be much strengthened if the low-EM H II region has strong emission in the background.

Some ultra-compact H II regions have been found located near extended free-free emission (EE) \citep{fue20a,fue20b}. \citet{fue20a} suggest that EE seems to be common in ultra-compact H II regions. The EM of an EE is typically $10^4-10^5$ cm$^{-6}$ pc, and the $n_e$ is $\sim10^2$ cm$^{-3}$ \citep{fue20b}. However, since large-scale structures of ionized gas are easy to miss in interferometric observations \citep{woo89}, the contribution to hydrogen recombination lines from the extended component are probably overlooked. Two cases including an ultra-compact H II region $+$ EE, and a hyper-compact H II region + EE are calculated. The $T_e$ is $10000$ K in both cases. The values of $n_e$ are $10^6$, $10^5$, and $10^2$ cm$^{-3}$ for the hyper-compact regions, ultra-compact H II regions, and the EE, respectively, and the corresponding EMs are $10^8$, $5\times10^7$, and $10^5$ cm$^{-6}$ pc. The high-density region and the extended region are assumed to have the same LOS velocity. Since the high-density H II region is embraced by EE, the EE is treated as a component in front of the high-density H II region. The hydrogen recombination lines passing through both the high-density and extended regions are studied.

\begin{figure}
\begin{center}
\includegraphics[scale=0.4]{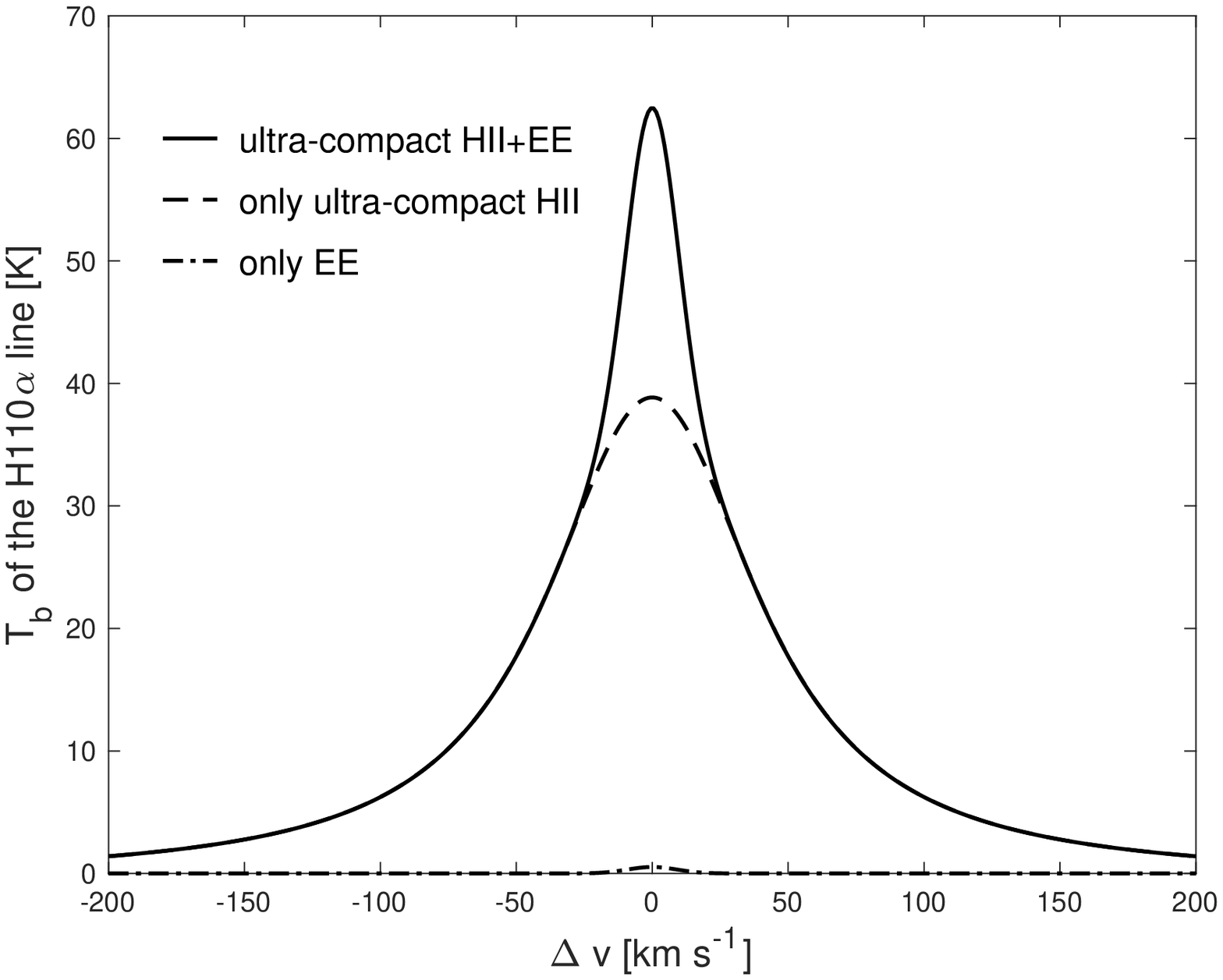}
\includegraphics[scale=0.4]{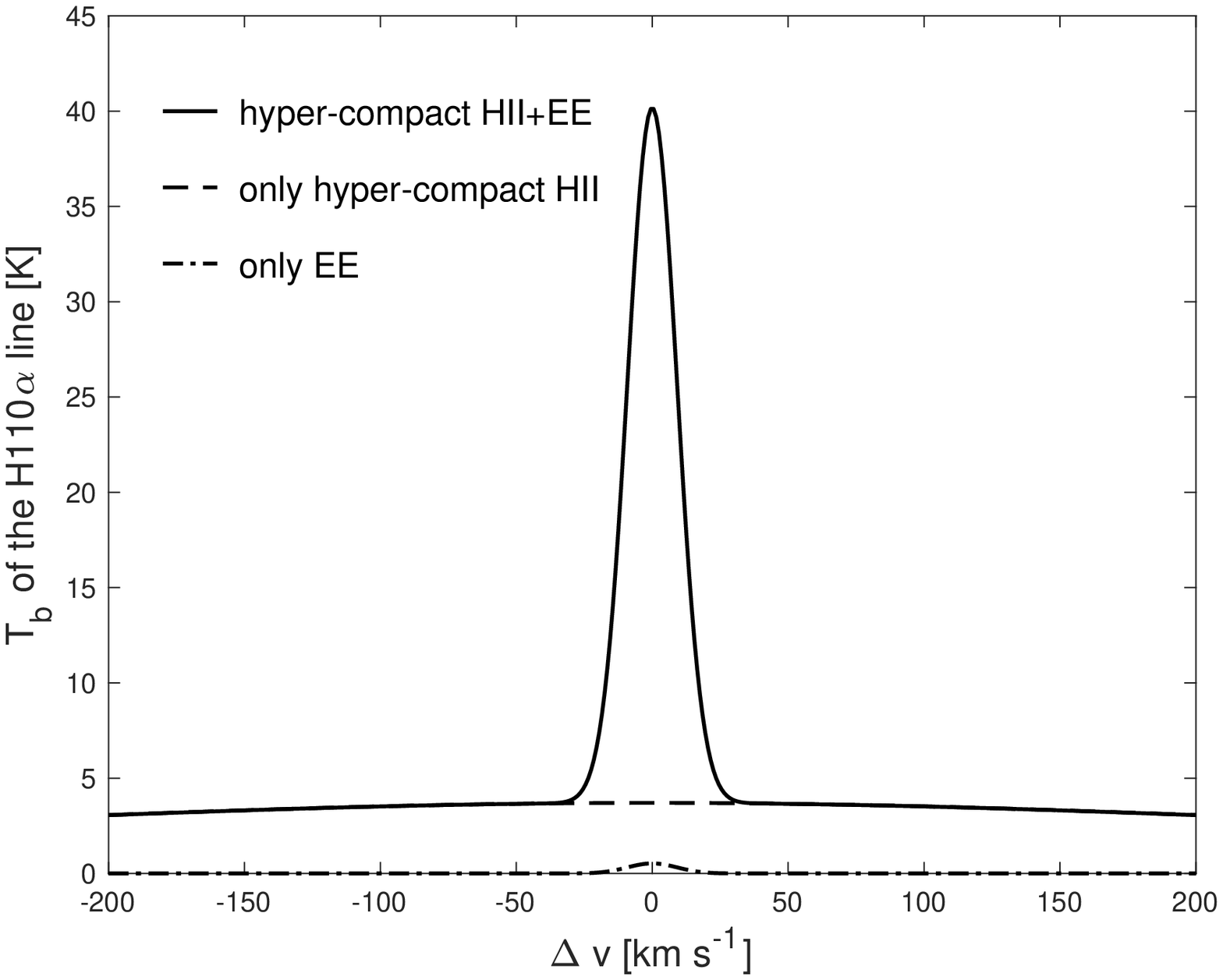}
\caption{Line profiles of the H110$\alpha$ line for two cases. The result in the case of ultra-compact H II region + EE is presented in the top panel, and that of hyper-compact H II region + EE is shown in the bottom panel. The values of $n_e$ are $10^6$, $10^5$, and $10^2$ cm$^{-3}$ for the hyper-compact, ultra-compact, and extended H II regions, respectively, and the corresponding EMs are $10^8$, $5\times10^7$, and $10^5$ cm$^{-6}$ pc. The electron temperature is $T_e=10000$ K.}\label{fig:profile_hii_EE}
\end{center}
\end{figure}

The  profiles of the H110$\alpha$ line in the three cases are shown in Fig. \ref{fig:profile_hii_EE}. The line profiles calculated respectively from hyper- and ultra-compact H II regions and EE as a single component are also plotted. Compared with the line profiles from single components, it is clear that the population inversion in the EE can significantly strengthen the line emission from ultra- or hyper-compact H II regions. Especially in the case of hyper-compact H II regions + EE, the part of the line emission contributed from the hyper-compact H II region is very broad. Its broad profile can be regarded as the baseline. If so, the part attributed to the stimulated emissions from the EE will be mistakenly treated as the line emission mainly emitted from the hyper-compact H II region. In addition, the line widths are also influenced by the EE. The widths (FWHMs) of the H110$\alpha$ line just emitted from the hyper- and ultra-compact H II regions are 870 km s$^{-1}$ and 92 km s$^{-1}$, respectively. Then they become much narrower,  $23$ km s$^{-1}$ and $49$ km s$^{-1}$ after the line emission passes through the EE.

When observing the Hn$\alpha$ lines from the hyper- and ultra-compact H II regions, it is meaningful to evaluate the contamination from the EE. Compared with the line emissions just emitted from hyper- and ultra-compact H II regions, whether the contribution of the stimulated emissions from the EE needs to be considered is determined by the line optical depth $\tau_{\nu,\textrm{L}}$ of the EE and the ratio of $I_{\nu,\textrm{L}}/I_{\nu,\textrm{C}}$ of the line and continuum emissions before passing through the EE. For hyper-compact H II regions with high electron density and EM that lead to very low $I_{\nu,\textrm{L}}/I_{\nu,\textrm{C}}$, the stimulated emission from EE may increase the peak $T_{b}$ of the Hn$\alpha$ line by more than 15$\%$ when $n>85$. Even if the EM of the EE is only $10^4$ cm$^{-6}$ pc, the stimulated emission is still non-negligible when $n>100$. For ultra-compact H II regions embraced by the EE with EM$=10^5$ cm$^{-6}$ pc, the stimulated emission needs to be considered when $n>100$.

\subsection{Estimation  of    the properties of ionized gas including    the effect of stimulated emissions}

\subsubsection{Uncertainty of estimated  electron    temperature under    the LTE assumption}

\begin{figure}
\begin{center}
\includegraphics[scale=0.4]{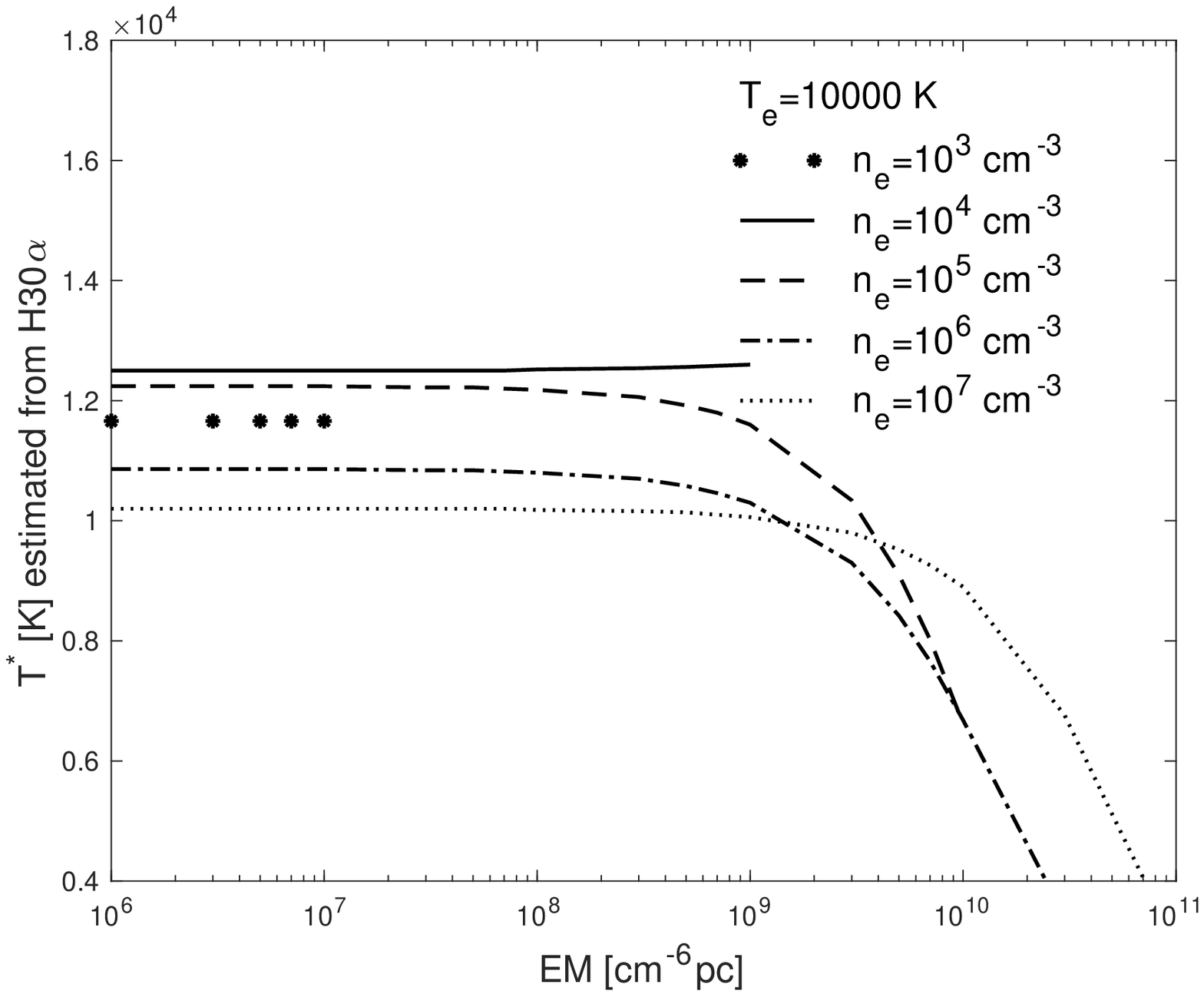}
\includegraphics[scale=0.4]{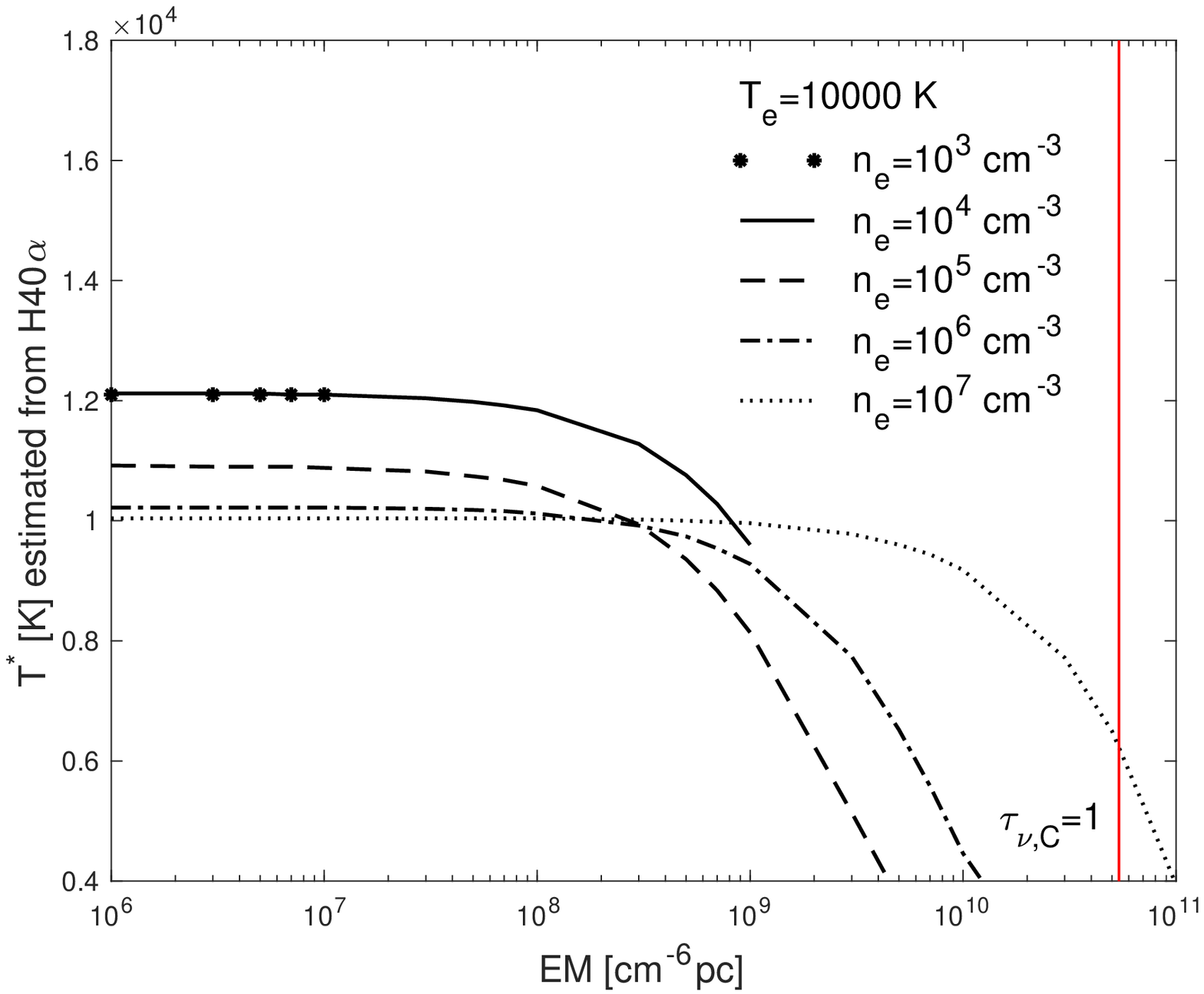}
\includegraphics[scale=0.4]{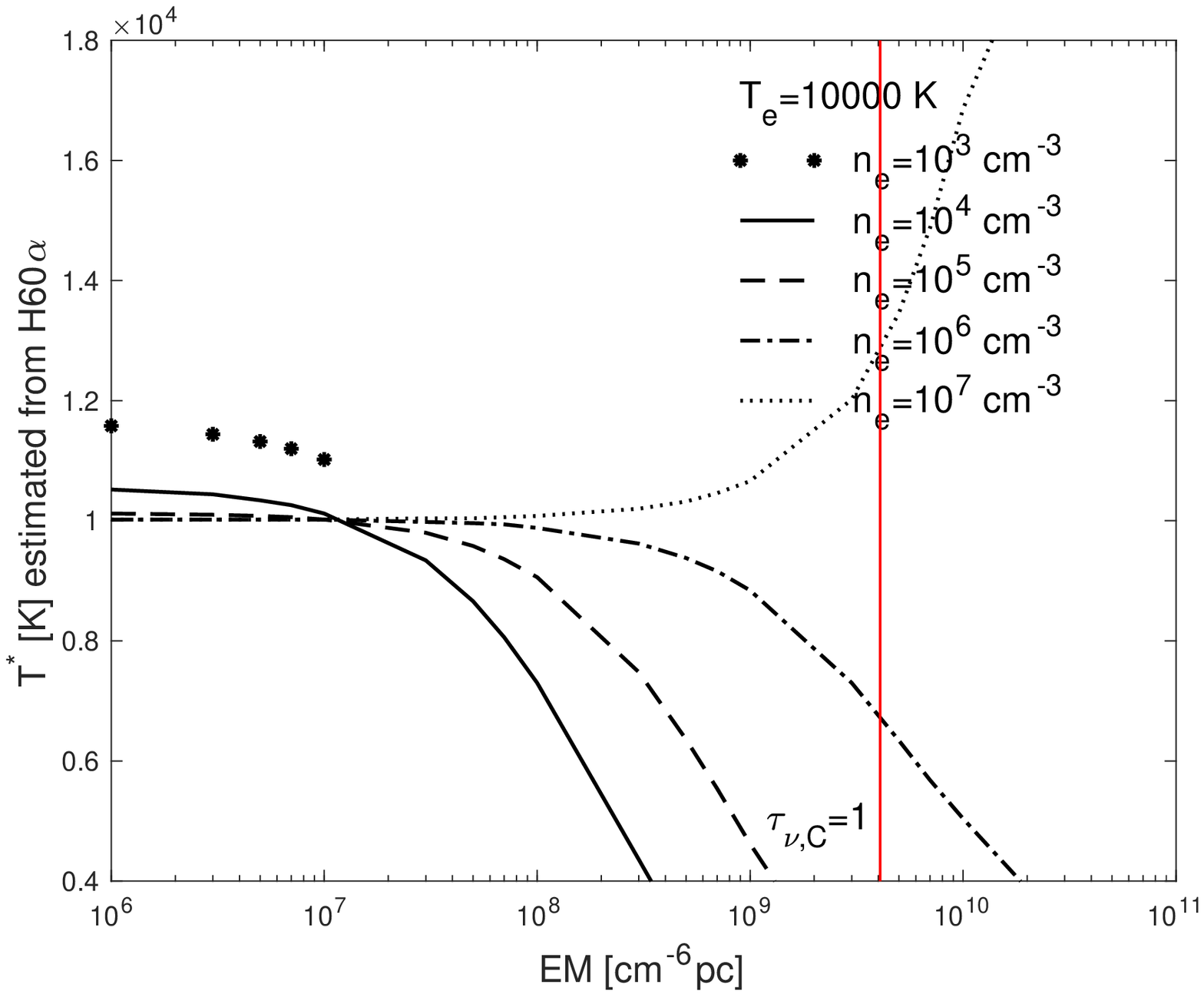}
\caption{LTE    temperature estimated from H30$\alpha$, H40$\alpha$, and H60$\alpha$ lines with    the continuum emission vs. EM for $T_e$=10000 K and different $n_e$. The red vertical lines indicate the value of EM corresponding to the continuum optical depth $\tau_{\nu,\textrm{C}}=1$.}\label{fig:ltetem}
\end{center}
\end{figure}

\begin{figure}
\begin{center}
\includegraphics[scale=0.4]{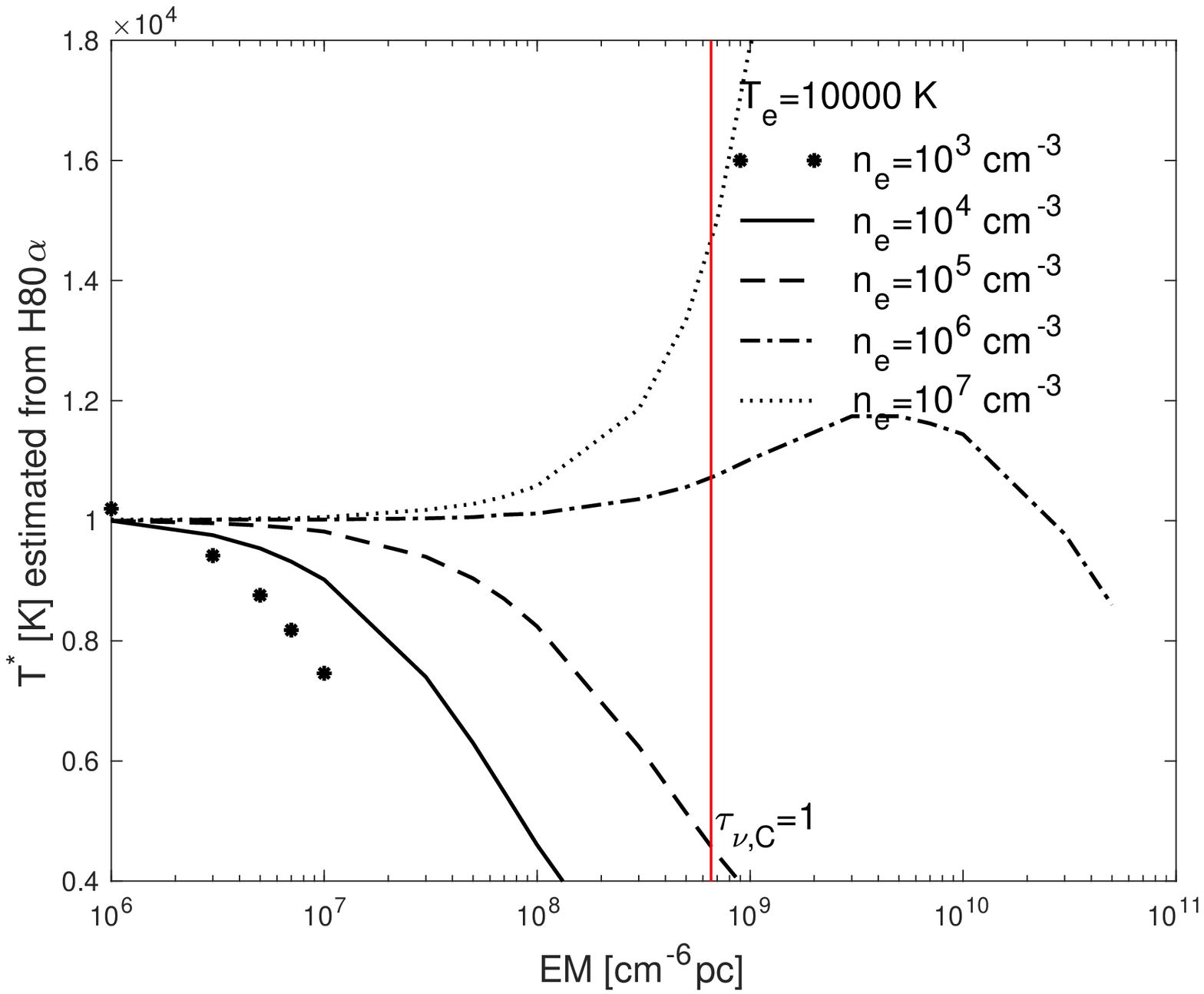}
\includegraphics[scale=0.4]{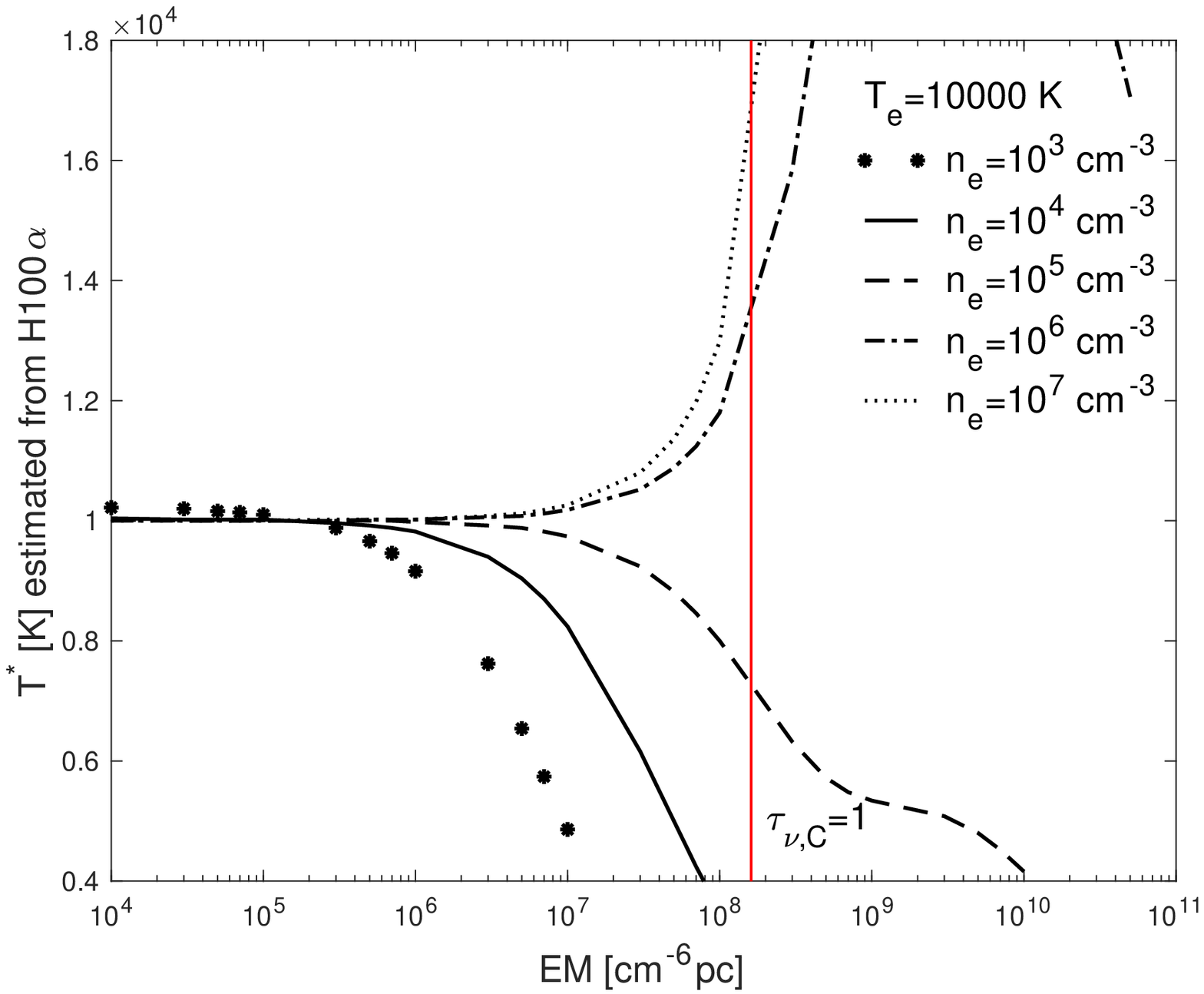}
\includegraphics[scale=0.4]{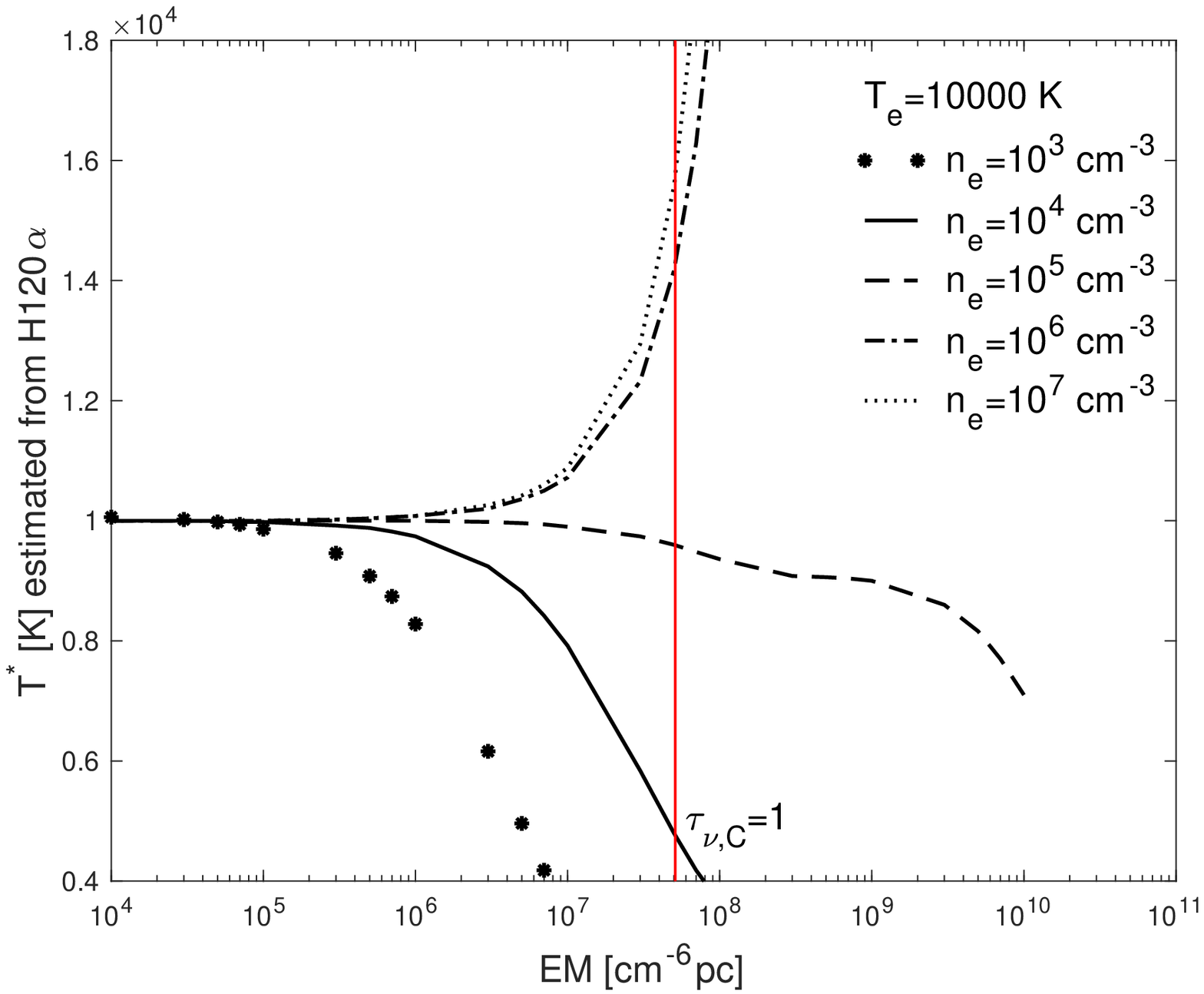}
\caption{LTE    temperature estimated from H80$\alpha$, H100$\alpha$, and H120$\alpha$ lines with the continuum emission vs. EM for $T_e$=10000 K and different $n_e$. The red vertical lines indicate the value of EM corresponding to the continuum optical depth $\tau_{\nu,\textrm{C}}=1$.}\label{fig:ltetem2}
\end{center}
\end{figure}

The   LTE    temperature   ($T^*$),  which is the electron    temperature estimated from    the line-to-continuum ratios under    the LTE and optically    thin assumptions, has been commonly used    to evaluate    the actual electron    temperature of ionized gas. However,     the LTE    temperature $T^*$ may be significantly different from    the actual electron    temperature $T_e$ \citep{aff94,gor02}. In Figs. \ref{fig:ltetem} and \ref{fig:ltetem2},    the variations in    the LTE    temperature $T^*$ with    the values of EM and $n_e$ of    the ionized gas are presented. These LTE temperatures are estimated from the ratios of the frequency-integrated intensities of a series of Hn$\alpha$ lines to the continuum intensities at the corresponding frequencies. We calculated the case for an actual electron temperature $T_e=10000$ K only. From    the results displayed in Figs. \ref{fig:ltetem} and \ref{fig:ltetem2},    the deviation of    the LTE    temperature from    the actual temperature is mainly determined by the departure coefficient $b_n$ when the continuum optical depth is very    thin. If    the departure coefficient $b_n$ is close to 1 due to high electron density or  high energy level n,    the LTE    temperature will be relatively accurate. When    the continuum optical depth increases,    the values of    $T ^*$ could be lower due    to    the increasing effects of    the stimulated emission on    the Hn$\alpha$ line emitted from    the energy level $n$ without the LTE condition. If the LTE condition is approximately satisfied in the energy levels $n$ and $n+1$, the LTE temperature estimated from    the Hn$\alpha$ line will increase with    the continuum optical depth. By using    these hydrogen radio recombination lines,    the actual electron    temperature can hardly be evaluated from    the LTE    temperature if    the continuum optical depth is $\tau_{\nu,\textrm{C}}>0.1$ and    the electron density $n_e$ is unknown.

\subsubsection{Deriving properties of ionized gas  with  line-to-continuum ratios of multiple RRLs  under non-LTE conditions}

Using multiple line-to-continuum ratios at different wavebands may be a good method for estimating the electron    temperature and density of ionized gas, which had been used for the ultra-compact H II region G29.96-0.02 with the departure coefficients calculated using the n-model \citep{aff94}. However, just being successfully applied to one case does not prove that this method will work well in other situations. To study the applicability of this method, we performed  a systematic analysis.

As observable parameters, the line-to-continuum ratios of the $\alpha$ lines of hydrogen and  continuum intensities at given frequencies can be used to derive $T_e$,  $n_e$, and EM, with the method using multiple line-to-continuum ratios under non-LTE conditions. However, systematic errors or deviation in this method need to be checked.  Line-to-continuum ratios for every $\alpha$ line of hydrogen and continuum intensities at given frequencies can be derived based on our non-LTE model, with given values of $T_e$, $n_e$, and EM for one ionized region. Uncertainties in the observations, including flux calibration and  white  noise,  should be considered during such calculation. From the simulated line-to-continuum ratios with observational uncertainties, the estimated $T_e$, $n_e$, and EM can be compared with the input parameters. The reliability of the estimation method is then assessed.

We give an example of the results of this  calculation including the information of H30$\alpha$, H40$\alpha$, and H60$\alpha$ line intensities to the continuum intensities at the corresponding frequencies with the method using multiple line-to-continuum ratios. In addition, the continuum intensity at the frequency of 99.023 GHz is also used. This can significantly improve the estimation accuracy.

For one case of an ionized region, we first calculate the frequency-integrated line intensities and the continuum emission intensities at the corresponding frequencies for the given values of $T_e$, $n_e$, and EM, based on our model. Three line-to-continuum ratios are then derived. After that, a group of four random numbers are generated from the normal distribution based on the calculated values of the three line-to-continuum ratios and the 99.023 GHz continuum intensity. The calculated values are taken as the means of the normal distributions. We assume the uncertainty in the ratio observations to be $3\%$ and $10\%$, and the uncertainty in the continuum intensity observation to be $3\%$. These uncertainties are used as the standard deviations. The four random numbers in one group are used to simulate the four quantities measured in observations. Then values of $T_e$, $n_e$, and EM are adjusted until the resulting line-to-continuum ratios and the 99.023 GHz continuum intensity fit the simulated values. The least-squares method is used in the fitting process. After the process mentioned above is repeated many times, a series of best fit values of $T_e$, $n_e$, and EM are then derived from many  groups of four random numbers. These best fit values create the distributions of the estimated $T_e$, $n_e$, and EM. Finally, the mean values and the uncertainties of these estimated quantities are calculated from the distributions.

In Table \ref{table:estimate}, the resulting estimates of $T_e$ for hyper-compact H II regions are given. The mean values and the corresponding 1 $\sigma$ uncertainties are presented with    the assumed uncertainties ($\sigma/\mu$) of    the line-to-continuum ratios of $3\%$ and $10\%$;  the uncertainty of the continuum emission at 99.023 GHz is always assumed to be $3\%$. According    to    the results shown in    the    table,    the mean values of    the estimated electron    temperatures    $T_e$ are always close    to    the actual temperature    $T_e=10000$ K. In addition,     the relative uncertainties of    the estimated    temperatures are mostly lower    than $8\%$ and $2\%$ for $10\%$ and $3\%$ uncertainties of    the line-to-continuum ratios, respectively.

In Table \ref{table:estimate_low}, the estimated $T_e$ for the cases of ultra-compact, compact, and extended H II regions are given. The hydrogen recombination lines used in the estimation method are the H40$\alpha$, H80$\alpha$, and H120$\alpha$ lines. The 99.023 GHz continuum intensity is still used in these cases. The estimated values of $T_e$ for these relatively low-density H II regions are still very accurate, as in the cases for hyper-compact H II regions. This suggests that a believable estimate of electron temperature can be obtained by using this estimation method.

The LTE temperatures $T^*$ estimated with a single line-to-continuum ratio under the LTE assumption for the corresponding cases of H II regions are also given in Tables \ref{table:estimate} and \ref{table:estimate_low}. The uncertainty of $T^*$ is approximately proportional to 0.87 times the uncertainties of the corresponding line-to-continuum ratio because the LTE temperature is about $T^*\propto(\int I_{\nu,\textrm{L}}^{\textrm{LTE}}d\nu/I_{\nu,\textrm{C}})^{-0.87}$ \citep{gor02}. In general, the deviation between the LTE temperature and the actual temperature is mainly affected by the difference between the frequency-integrated line intensities under non-LTE and LTE conditions. If the non-LTE Hn$\alpha$ line intensity is significantly different ($>15\%$) from that under LTE, the LTE approximation should not be used. Additionally, the systematic deviation of the non-LTE method using multiple line-to-continuum ratios is very slight in estimating electron temperature. This is different from the LTE method. Therefore, even if the difference between the line intensities under non-LTE and LTE conditions is not great ($<15\%$), the method using multiple line-to-continuum ratios is still preferable when a high precision of estimated electron temperature is required.

Moreover,    the estimated values of    the electron density $n_e$ and    the EM are also displayed in Tables \ref{table:estimate} and \ref{table:estimate_low}. The value of    the optically    thin continuum emission is very sensitive    to EM but relatively insensitive    to $T_e$ and $n_e$. This makes    the estimated values of EM very accurate if    the continuum optical depth is    thin. In addition, by using    the multiple line-to-continuum ratios,    the estimated value of    the EM can  still be accurate even if    the continuum optical depth is    thick ($\tau_{\nu,\textrm{C}}\sim10$), although    the uncertainty of the estimated EM may be a little bigger. The estimation of the electron density $n_e$ is sensitive    to    the departures of line-to-continuum ratios from    their LTE values. This departure increases with    the EM and decreases with    the electron density $n_e$. The ratios of the frequency-integrated line intensity ($\int I_{\nu,\textrm{L}}d\nu$) to the line intensity calculated under the LTE assumption ($\int I_{\nu,\textrm{L}}^{\textrm{LTE}}d\nu$) for the Hn$\alpha$ lines are also listed in Tables \ref{table:estimate} and \ref{table:estimate_low}. These ratios are useful to evaluate the importance of stimulated emission. When stimulated emission is not important, the line-to-continuum ratios are close to their values in LTE. These values can lead to inaccurate estimates of $n_e$ since the line-to-continuum ratio is not sensitive to $n_e$ under  LTE conditions. On    the contrary,  when stimulated emission is important,  the line-to-continuum ratios significantly depart from the LTE values. Then the estimated values of $n_e$ can be very close    to    the actual values. In the case of $n_e=10^5$ cm$^{-3}$ and EM$=10^9$ cm$^{-6}$ pc, the estimated $n_e$  values using different combinations of  Hn$\alpha$ lines are given in Table \ref{table:estimate} and \ref{table:estimate_low}. Comparing the different estimated $n_e$, it shows that choosing hydrogen recombination lines with larger amounts of stimulated emission helps to accurately estimate electron density.

In Tables \ref{table:tauT8e3}, \ref{table:tauT1e4}, and \ref{table:tauT1p2e4} the total optical depths $\tau_{\nu}$ at the centers of    the hydrogen recombination lines for different    temperatures, densities, and EMs are listed. The line optical depths $\tau_{\nu,\textrm{L}}$ for different properties are given in Table \ref{table:taulT8e3}, \ref{table:taulT1e4}, and \ref{table:taulT1p2e4}. The optical depths can help us to speculate whether or which Hn$\alpha$ lines show maser emission from the known properties of ionized gas. The line-to-continuum ratios are presented in Table \ref{table:fluxtoIcT8e3}, \ref{table:fluxtoIcT1e4}, and \ref{table:fluxtoIcT1p2e4}. These values are useful to estimate the electron    temperature, density, and EM of ionized gas by using multiple line-to-continuum ratios.

\begin{table*}\tiny
\centering
\caption{Estimated values and standard deviations of $T_e$, $n_e$, and EM for hyper-compact H II regions.} \label{table:estimate}
\begin{threeparttable}
\begin{tabular}{|c|ccc|ccc|ccc|ccc|}
\hline
$\sigma/\mu$ & \multicolumn{3}{c|}{$3\%$} & \multicolumn{3}{c|}{$10\%$} &  \multicolumn{3}{c|}{$\int I_{\nu,\textrm{L}}d\nu/\int I_{\nu,\textrm{L}}^{\textrm{LTE}}d\nu$} &  \multicolumn{3}{c|}{LTE temperature} \\
\hline
& \multicolumn{3}{c|}{$n_e=10^5$ cm$^{-3}$} & \multicolumn{3}{c|}{$n_e=10^5$ cm$^{-3}$} &  \multicolumn{3}{c|}{$n_e=10^5$ cm$^{-3}$} & \multicolumn{3}{c|}{$n_e=10^5$ cm$^{-3}$} \\
\hline
EM  & $\hat{T}$ & $\hat{n_e}$ & $\hat{\textrm{EM}}$ & $\hat{T}$ & $\hat{n_e}$ & $\hat{\textrm{EM}}$ & H30$\alpha$ & H40$\alpha$ & H60$\alpha$ & $T^*_{\textrm{H}30\alpha}$ & $T^*_{\textrm{H}40\alpha}$ & $T^*_{\textrm{H}60\alpha}$ \\
cm$^{-6}$pc & K & $10^5$ cm$^{-3}$ & $10^9$ cm$^{-6}$pc & K & $10^5$ cm$^{-3}$ & $10^9$ cm$^{-6}$pc &  &  &  & K & K & K \\
\hline
$1.0\times10^9$ & $10014\pm146$ & $1.10\pm0.12$ & $1.02\pm0.01$ & $10078\pm352$ & $1.22\pm0.31$ & $1.02\pm0.01$ & 0.83 & 1.32 & 3.02 & 11603 & 8136 & 4612 \\
$5.0\times10^9$ & $10017\pm102$ & $1.07\pm0.09$ & $5.03\pm0.08$ & $10012\pm278$ & $1.19\pm0.23$ & $5.03\pm0.10$ & 1.17 & 3.92 & 32.34 & 9080 & 3736 & <2000 \\
$1.0\times10^{10}$ & $10002\pm25$ & $1.04\pm0.04$ & $9.90\pm0.07$ & $10067\pm124$ & $1.09\pm0.14$ & $9.91\pm0.07$ & 1.82 & 7.81 & 204.22 & 6623 & 2417 & <2000 \\
$5.0\times10^{10}$ & ... & ... & ... & ... & ... & ... & ... & ... & ... & ... & ... & ... \\
\hline
& \multicolumn{3}{c|}{$n_e=10^6$ cm$^{-3}$} & \multicolumn{3}{c|}{$n_e=10^6$ cm$^{-3}$} & \multicolumn{3}{c|}{$n_e=10^6$ cm$^{-3}$} & \multicolumn{3}{c|}{$n_e=10^6$ cm$^{-3}$} \\
\hline
EM & $\hat{T}$ & $\hat{n_e}$ & $\hat{\textrm{EM}}$ & $\hat{T}$ & $\hat{n_e}$ & $\hat{\textrm{EM}}$ &    H30$\alpha$ & H40$\alpha$ & H60$\alpha$ & $T^*_{\textrm{H}30\alpha}$ & $T^*_{\textrm{H}40\alpha}$ & $T^*_{\textrm{H}60\alpha}$ \\
cm$^{-6}$pc & K & $10^6$ cm$^{-3}$ & $10^9$ cm$^{-6}$pc & K & $10^6$ cm$^{-3}$ & $10^9$ cm$^{-6}$pc &   &   &  & K & K & K \\
\hline
$1.0\times10^9$ & $10018\pm179$ & $1.14\pm0.26$ & $1.02\pm0.01$ & $10157\pm459$ & $1.54\pm1.83$ & $1.02\pm0.01$ & 0.97 & 1.12 & 1.34 & 10307 & 9279 & 8839 \\
$5.0\times10^9$ & $10038\pm147$ & $1.01\pm0.06$ & $5.01\pm0.09$ & $10074\pm456$ & $1.14\pm0.26$ & $5.01\pm0.11$ & 1.29 & 1.88 & 3.76 & 8423 & 6531 & 6636 \\
$1.0\times10^{10}$ & $10008\pm76$ & $1.00\pm0.02$ & $9.93\pm0.14$ & $10064\pm340$ & $1.06\pm0.16$ & $9.91\pm0.18$ & 1.80 & 3.38 & 12.04 & 6671 & 4464 & 5038 \\
$5.0\times10^{10}$ & ... & ... & ... & ... & ... & ... & ... & ... & ... & ... & ... & ... \\
\hline
& \multicolumn{3}{c|}{$n_e=10^7$ cm$^{-3}$} & \multicolumn{3}{c|}{$n_e=10^7$ cm$^{-3}$} & \multicolumn{3}{c|}{$n_e=10^7$ cm$^{-3}$} & \multicolumn{3}{c|}{$n_e=10^7$ cm$^{-3}$} \\
\hline
EM & $\hat{T}$ & $\hat{n_e}$ & $\hat{\textrm{EM}}$ & $\hat{T}$ & $\hat{n_e}$ & $\hat{\textrm{EM}}$ & H30$\alpha$ & H40$\alpha$ & H60$\alpha$ & $T^*_{\textrm{H}30\alpha}$ & $T^*_{\textrm{H}40\alpha}$ & $T^*_{\textrm{H}60\alpha}$  \\
cm$^{-6}$pc & K & $10^7$ cm$^{-3}$ & $10^9$ cm$^{-6}$pc & K & $10^7$ cm$^{-3}$ & $10^9$ cm$^{-6}$pc &   &   &   & K & K & K \\
\hline
$1.0\times10^9$ & $10036\pm191$ & $2.46\pm3.04$ & $ 1.02\pm0.01$ & $10314\pm536$ & $3.22\pm4.16$ & $1.03\pm0.02$ & 1.00 & 1.02 & 1.05 & 10061 & 9952 & 10648 \\
$5.0\times10^9$ & $10029\pm218$ & $1.14\pm0.33$ & $5.01\pm0.10$ & $10152\pm735$ & $1.65\pm1.71$ & $5.00\pm0.14$ & 1.09 & 1.14 & 1.41 & 9526 & 9598 & 13451 \\
$1.0\times10^{10}$ & $10030\pm197$ & $1.06\pm0.14$ & $10.01\pm0.23$ & $10141\pm617$ & $1.17\pm0.42$ & $10.0\pm0.25$ & 1.23 & 1.31 & 2.43 &. 8896 & 9184 & 16844 \\
$5.0\times10^{10}$ & $10006\pm149$ & $1.05\pm0.08$ & $50.2\pm1.37$ & $10016\pm429$ & $1.15\pm0.26$ & $50.0\pm1.8$ & 3.18 & 3.88 & $>10^3$ & 5103 & 6540 & >20000 \\
\hline
\end{tabular}
\begin{tablenotes}
\item \textbf{Notes.} The  values and standard deviations of $T_e$, $n_e$, and EM estimated by using the line-to-continuum ratios ($\int I_{\nu,\textrm{L}}d\nu/I_{\nu,\textrm{C}}$) of    the H30$\alpha$, H40$\alpha$, and H60$\alpha$ lines and    the continuum emission at 99.023 GHz from ionized gas with different electron densities and EMs. The actual electron    temperature is 10000 K. The relative uncertainties $\sigma/\mu$ of    the line-to-continuum ratio are assumed    to be $3\%$ and $10\%$, and    the corresponding relative uncertainties of the  99.023 GHz continuum is $3\%$. The ratios of the frequency-integrated intensity ($\int I_{\nu,\textrm{L}}d\nu$) to the intensity calculated under the LTE assumption ($\int I_{\nu,\textrm{L}}^{\textrm{LTE}}d\nu$) for the H30$\alpha$, H40$\alpha$, and H60$\alpha$ lines are given in the three columns below the heading $\int I_{\nu,\textrm{L}}d\nu/\int I_{\nu,\textrm{L}}^{\textrm{LTE}}d\nu$. The LTE temperatures estimated by using a single line-to-continuum ratio of one of the three Hn$\alpha$ lines to the continuum emission at corresponding frequency are given in the three right-most columns.
\end{tablenotes}
\end{threeparttable}
\end{table*}

\begin{table*}\tiny
\centering
\caption{Estimated values and standard deviations of $T_e$, $n_e$, and EM for ultra-compact, compact, and extended H II regions.} \label{table:estimate_low}
\begin{threeparttable}
\begin{tabular}{|c|ccc|ccc|ccc|ccc|}
\hline
$\sigma/\mu$ & \multicolumn{3}{c|}{$3\%$} & \multicolumn{3}{c|}{$10\%$} &  \multicolumn{3}{c|}{$\int I_{\nu,\textrm{L}}d\nu/\int I_{\nu,\textrm{L}}^{\textrm{LTE}}d\nu$} & \multicolumn{3}{c|}{LTE temperature} \\
\hline
& \multicolumn{3}{c|}{$n_e=10^3$ cm$^{-3}$} & \multicolumn{3}{c|}{$n_e=10^3$ cm$^{-3}$} &  \multicolumn{3}{c|}{$n_e=10^3$ cm$^{-3}$} & \multicolumn{3}{c|}{$n_e=10^3$ cm$^{-3}$} \\
\hline
EM  & $\hat{T}$ & $\hat{n_e}$ & $\hat{\textrm{EM}}$ & $\hat{T}$ & $\hat{n_e}$ & $\hat{\textrm{EM}}$ &    H40$\alpha$ & H80$\alpha$ & H120$\alpha$ & $T^*_{\textrm{H}40\alpha}$ & $T^*_{\textrm{H}80\alpha}$ & $T^*_{\textrm{H}120\alpha}$ \\
cm$^{-6}$pc & K & $10^3$ cm$^{-3}$ & $10^5$ cm$^{-6}$pc & K & $10^3$ cm$^{-3}$ & $10^5$ cm$^{-6}$pc &  &  &  & K & K & K \\
\hline
$1.0\times10^5$ & $10086\pm213$ & $6.90\pm14.46$ & $1.01\pm0.02$ & $10200\pm570$ & $21.22\pm34.76$ & $1.01\pm0.02$ & 0.79 & 0.93 & 1.02 & 12076 & 10602 & 9877 \\
$5.0\times10^5$ & $10030\pm195$ & $1.33\pm1.48$ & $4.98\pm0.10$ & $10138\pm559$ & $12.85\pm26.79$ & $5.02\pm0.12$ & 0.79 & 0.95 & 1.13 & 12076 & 10426 & 9068 \\
$1.0\times10^6$ & $10014\pm210$ & $1.19\pm0.47$ & $10.00\pm0.25$ & $10163\pm634$ & $6.11\pm17.94$ & $10.03\pm0.24$ & 0.79 & 0.98 & 1.27 & 12077 & 10213 & 8239 \\
$5.0\times10^6$ & $10014\pm219$ & $1.09\pm0.21$ & $49.04\pm0.93$ & $10093\pm581$ & $1.18\pm0.55$ & $48.84\pm1.09$ & 0.79 & 1.17 & 2.43 & 12082 & 8790 & 4882 \\
\hline
& \multicolumn{3}{c|}{$n_e=10^4$ cm$^{-3}$} & \multicolumn{3}{c|}{$n_e=10^4$ cm$^{-3}$} & \multicolumn{3}{c|}{$n_e=10^4$ cm$^{-3}$}  & \multicolumn{3}{c|}{$n_e=10^4$ cm$^{-3}$} \\
\hline
EM & $\hat{T}$ & $\hat{n_e}$ & $\hat{\textrm{EM}}$ & $\hat{T}$ & $\hat{n_e}$ & $\hat{\textrm{EM}}$ & H40$\alpha$ & H80$\alpha$ & H120$\alpha$ & $T^*_{\textrm{H}40\alpha}$ & $T^*_{\textrm{H}80\alpha}$ & $T^*_{\textrm{H}120\alpha}$ \\
cm$^{-6}$pc & K & $10^4$ cm$^{-3}$ & $10^6$ cm$^{-6}$pc & K & $10^4$ cm$^{-3}$ & $10^6$ cm$^{-6}$pc &   &   &   & K & K & K \\
\hline
$1.0\times10^6$ & $10050\pm171$ & $1.26\pm0.76$ & $1.00\pm0.03$ & $10308\pm604$ & $2.26\pm3.12$ & $1.01\pm0.02$ & 0.79 & 1.00 & 1.06 & 12133 & 10029 & 9624 \\
$5.0\times10^6$ & $10037\pm186$ & $1.11\pm0.33$ & $4.98\pm0.11$ & $10195\pm640$ & $1.47\pm1.25$ & $5.02\pm0.12$ & 0.79 & 1.06 & 1.30 & 12121 & 9542 & 8345 \\
$1.0\times10^7$ & $10064\pm150$ & $1.07\pm0.21$ & $10.10\pm0.25$ & $10296\pm577$ & $1.25\pm0.72$ & $10.10\pm0.25$ & 0.79 & 1.14 & 1.62 & 12104 & 9002 & 7212 \\
$5.0\times10^7$ & $10070\pm163$ & $1.03\pm0.10$ & $50.06\pm0.91$ & $10280\pm628$ & $1.12\pm0.34$ & $50.23\pm1.25$ & 0.80 & 1.82 & 5.20 & 11975 & 6255 & 3878 \\
\hline
& \multicolumn{3}{c|}{$n_e=10^5$ cm$^{-3}$} & \multicolumn{3}{c|}{$n_e=10^5$ cm$^{-3}$} & \multicolumn{3}{c|}{$n_e=10^5$ cm$^{-3}$} & \multicolumn{3}{c|}{$n_e=10^5$ cm$^{-3}$} \\
\hline
EM & $\hat{T}$ & $\hat{n_e}$ & $\hat{\textrm{EM}}$ & $\hat{T}$ & $\hat{n_e}$ & $\hat{\textrm{EM}}$ & H40$\alpha$ & H80$\alpha$ & H120$\alpha$ & $T^*_{\textrm{H}40\alpha}$ & $T^*_{\textrm{H}80\alpha}$ & $T^*_{\textrm{H}120\alpha}$ \\
cm$^{-6}$pc & K & $10^5$ cm$^{-3}$ & $10^8$ cm$^{-6}$pc & K & $10^5$ cm$^{-3}$ & $10^8$ cm$^{-6}$pc &   &   &   & K & K & K \\
\hline
$5.0\times10^7$ & $10022\pm200$ & $1.08\pm0.19$ & $0.50\pm0.01$ & $10159\pm657$ & $1.19\pm0.53$ & $0.50\pm0.01$ & 0.92 & 1.18 & 1.79 & 10742 & 9031 & 9602 \\
$1.0\times10^8$ & $10005\pm186$ & $1.04\pm0.10$ & $1.00\pm0.03$ & $10144\pm625$ & $1.14\pm0.34$ & $1.00\pm0.03$ & 0.93 & 1.37 & 3.37 & 10569 & 8254 & 9430 \\
$5.0\times10^8$ & $10018\pm210$ & $1.02\pm0.07$ & $4.99\pm0.10$ & $10009\pm637$ & $1.09\pm0.23$ & $5.00\pm0.13$ & 1.10 & 3.41 & $>10^3$ & 9361 & 5155 & 9117 \\
$1.0\times10^9$ & $9999\pm180$ & $1.02\pm0.07$ & $9.90\pm0.14$ & $10044\pm547$ & $1.09\pm0.22$ & $9.90\pm0.14$ & 1.32 & 7.93 & $>10^3$ & 8136 & 3796 & 9028 \\
\hline
\end{tabular}
\begin{tablenotes}
\item \textbf{Notes.} The  values and standard deviations of $T_e$, $n_e$, and EM estimated using the line-to-continuum ratios ($\int I_{\nu,\textrm{L}}d\nu/I_{\nu,\textrm{C}}$) of    the H40$\alpha$, H80$\alpha$, and H120$\alpha$ lines and    the continuum emission at 99.023 GHz from ionized gas with different electron densities and EMs. The actual electron    temperature is 10000 K. The relative uncertainties $\sigma/\mu$ of    the line-to-continuum ratio are assumed    to be $3\%$ and $10\%$, and    the corresponding relative uncertainties of the  99.023 GHz continuum is $3\%$. The ratios of the frequency-integrated intensity ($\int I_{\nu,\textrm{L}}d\nu$) to the intensity calculated under the LTE assumption ($\int I_{\nu,\textrm{L}}^{\textrm{LTE}}d\nu$) for the H40$\alpha$, H80$\alpha$, and H120$\alpha$ lines are given in the three columns below the heading $\int I_{\nu,\textrm{L}}d\nu/\int I_{\nu,\textrm{L}}^{\textrm{LTE}}d\nu$. The LTE temperatures estimated by using a single line-to-continuum ratio of one of the three Hn$\alpha$ lines to the continuum emission at corresponding frequency are given in the three right-most columns.
\end{tablenotes}
\end{threeparttable}
\end{table*}

\section{Summary and conclusions}\label{sec:conclusion}


In    this work the departure coefficients $b_n$ and the amplification coefficient $\beta_{n,n+1}$ are calculated by using an nl-model with    the effects of line and continuum radiation. The effects of radiation fields, electron temperature, and density, and the velocity field on    the departure coefficients, the amplification coefficients, and    the Hn$\alpha$ lines are investigated. The results show   that    the population inversion ($\beta_{n,n+1}<0$) often occurs in  hydrogen atoms in the ionized gas that emits hydrogen RRLs. The effect of stimulated emission on the Hn$\alpha$ line widths is also studied, and we find the powerful stimulation effect can significantly narrow the line widths of the Hn$\alpha$ lines, even down    to less    than 10 km s$^{-1}$ under certain conditions, although  this effect is not important for the line widths when the EM$<3\times10^7$ cm$^{-6}$pc. We assess    the reliability of    the method of using multiple hydrogen line-to-continuum ratios to estimate  $T_e$, $n_e$, and EM of    the ionized gas. We find    this method is helpful    to obtain    the accurate values of    $T_e$ and EM. The accuracy of    the estimated $n_e$ is determined by    the stimulation effect on    the Hn$\alpha$ lines. The other details of our conclusions are summarized as follows:

1. The saturation intensity $J_s$ introduced by \citet{str96} is a rough criterion    for evaluating    the influence of    the radiation fields on $b_n$ and $\beta_{n,n+1}$, but may underestimate    the influence especially for    the high energy level n.  Not only    could the saturated higher frequency masing lines   cause adjacent lower frequency lines    to exhibit masing, but    the saturated lower frequency masing lines may also lead    to adjacent higher frequency masing lines.

2. The Hn$\alpha$ line emissions could be overestimated considerably if the radiation fields are not considered in calculating $b_n$ and $\beta_{n,n+1}$. Strong radiation fields could broaden the range of    the population inversion $\beta_{n,n+1}<0$ in the principal quantum number $n$.

3. The stimulation effect of the hydrogen recombination lines can be weakened by violent velocity fields. Steep velocity gradients can lead to considerable broadening of the line profile that reduces the absolute value of the negative optical depth. The narrower line profile function due to lower electron temperatures can increase the stimulation effect.

4. The maser effect of the Hn$\alpha$ lines is mainly determined by the line optical depth, and can occur with considerably negative line optical depth even if    the    total optical depth $\tau_{\nu}>0$. In addition, the overheating and inversion of the populations can also produce amplification of the Hn$\alpha$ lines, although    the    total optical depth and    the continuum optical depth are both very thick.

5. For uniform H II regions without strong background radio emission, most of the centimeter Hn$\alpha$ lines are appropriate for the LTE approximation when EM$<10^6$ cm$^{-6}$pc. However, when there is strong background emission (e.g., a hyper- or ultra-compact H II region), stimulated emission could also be important for the centimeter Hn$\alpha$ lines even if EM$\leq10^5$ cm$^{-6}$ pc.

6. The electron    temperature  estimated using a hydrogen RRL to free-free continuum ratio under the LTE assumption can be more than 50\% different from the actual electron    temperature.

In these conclusions, the influence of powerful radiation fields on level populations and hydrogen RRLs, the variation of stimulated emission with line width, and the deviation from the LTE approximation for RRLs are shown. They are helpful to estimate the significance of non-LTE conditions in observations of RRLs.


\section*{Acknowledgements}

The authors thank the referee, Dr. A. Roy, and an anonymous referee for their constructive comments and detailed inspections to significantly improve the quality of the manuscript. The work is supported by the National Key R$\&$D Program of China (No. 2017YFA0402604), the National Science Foundation of China No. 12003055, No. U1931104, and No. 11590782, China Postdoctoral Science Foundation No. 2020M671267, and Shanghai Post-doctoral Excellence Program No. 2018261. This work is also supported by the international partnership program of Chinese Academy of Sciences through Grant No. 114231KYSB20200009, the Strategic Priority Research Program of Chinese Academy of Sciences, Grant No. XDB41000000, and Key Research Project of Zhejiang Lab (No. 2021PE0AC03).

\clearpage

\begin{appendix}

\section{Comparison of  calculated departure coefficients with results in the literature} \label{sec:com}

Our nl-model introduced in \citet{zhu19} with the Case B assumption contains the transition processes including radiative recombination \citep{bur65}, spontaneous emission \citep{bro71}, collisional excitations and de-excitations, collisional ionization, three-body recombination \citep{vri80}, and angular momentum changing collisions \citep{pen64,guz16}. The methods of calculating the rate coefficients of these processes are similar to those used in \citet{pro18}. In the previous work, we used the iterative method to calculate the departure coefficients. The n-model was solved first under the assumption of $b_{nl}=b_{n}$, and the nl-model was used to calculate $b_{n}$ and $b_{nl}$ with $n$ less than the critical level $n_{crit}$ \citep{hum87,sto95,sal17}. However, it was found that our results are closer to those in \citet{sto95} than to the values in \citet{pro18}, although the rate coefficients of bound-bound collisional transitions used in \citet{sto95} are different. In this work we replaced the iterative method with a direct solver,  as done by \citet{pro18} with error estimates $\epsilon \sim 3\times10^{-5}$, but we find our results are still consistent with those in \citet{sto95} after the stopping criterion was improved.

\begin{figure}
\begin{center}
\includegraphics[scale=0.4]{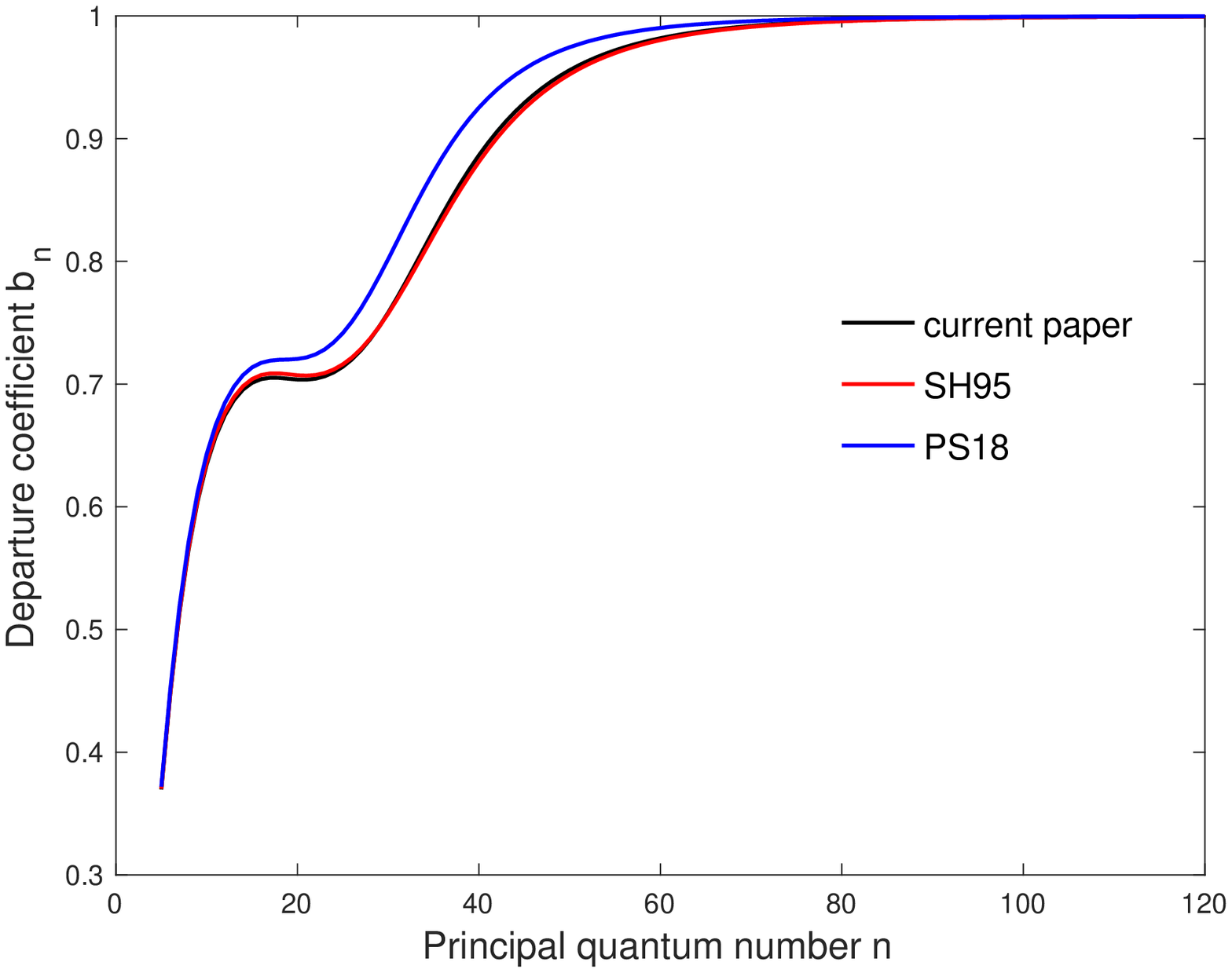}
\includegraphics[scale=0.4]{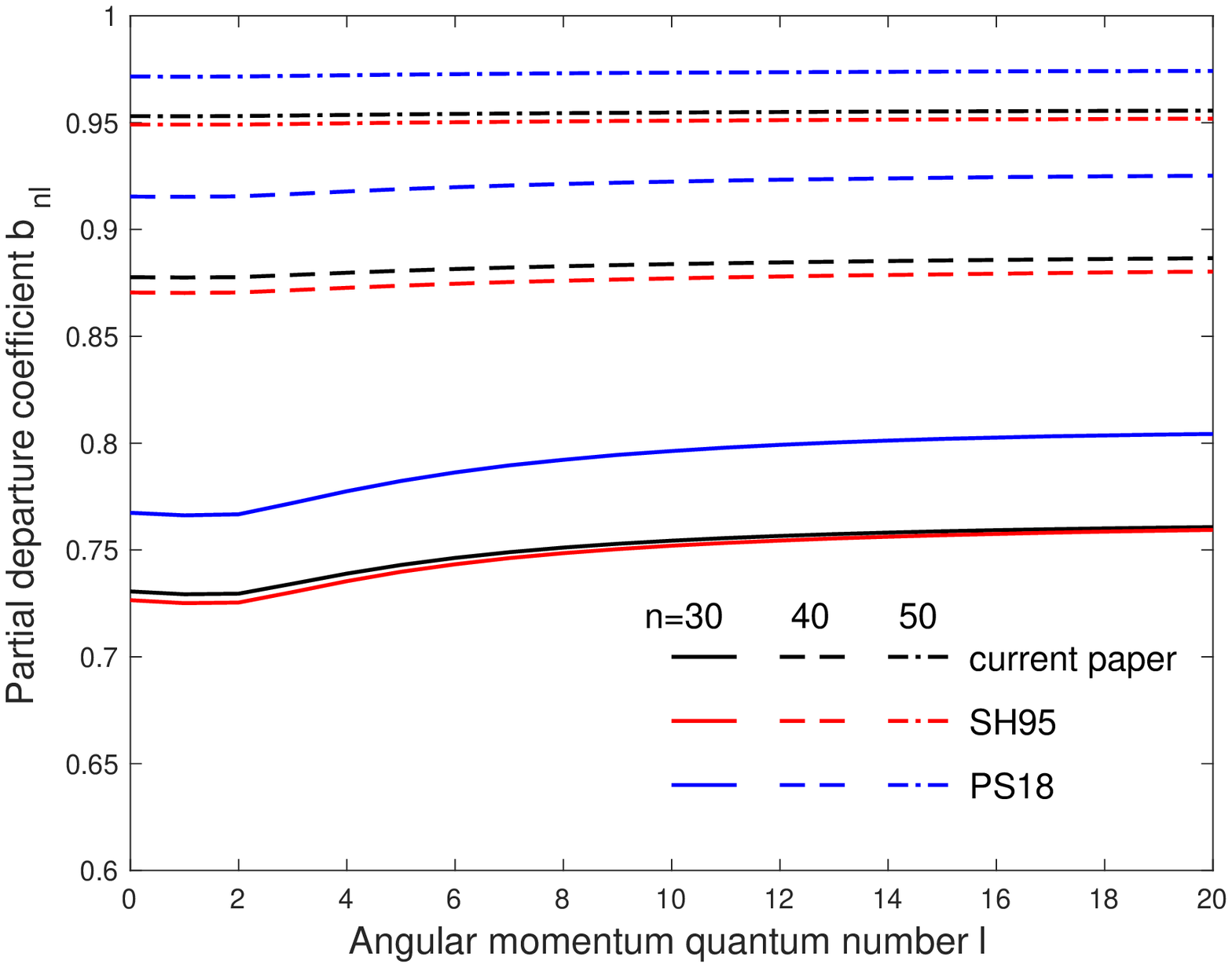}
\includegraphics[scale=0.4]{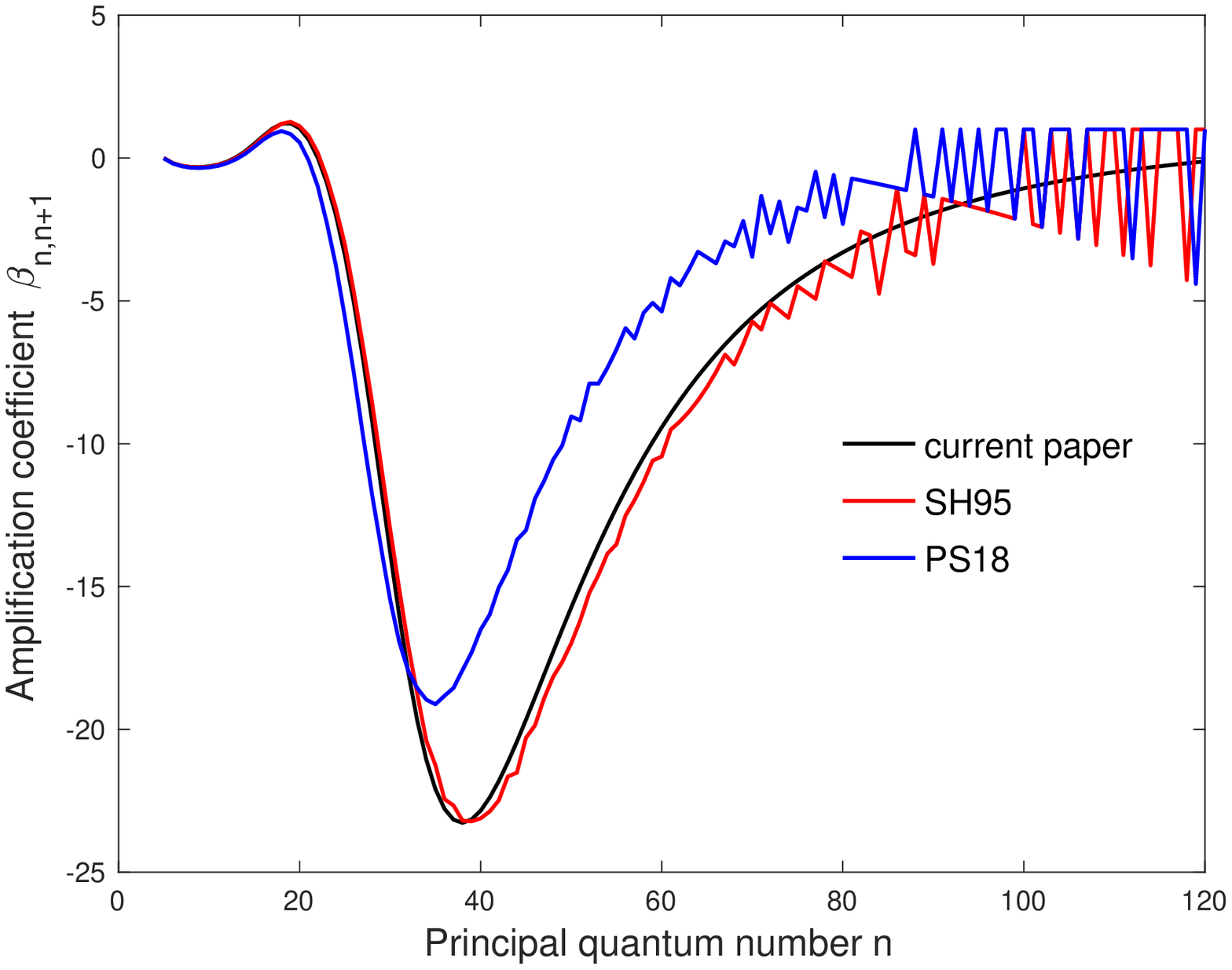}
\caption{Departure coefficients $b_n$, partial departure coefficients $b_{nl}$, and amplification coefficients $\beta_{n,n+1}$ calculated by our model (current paper), \citet{sto95} (SH95), and \citet{pro18} (PS18) are compared. The electron temperature $T_e$ is $10^4$ K, and the electron number density is $n_e=10^5$ cm$^{-3}$.}\label{fig:com}
\end{center}
\end{figure}

As an example, the values of $b_n$, $b_{nl}$, and $\beta_{n,n+1}$ for $T_e=10^4$ K and $n_e=10^5$ cm$^{-3}$ calculated by our model are compared with those provided by \citet{sto95} and \citet{pro18} in Fig. \ref{fig:com}. The values of the departure coefficients in the tables provided by \citet{sto95} and \citet{pro18} have four significant digits. This leads to the zigzags in the $\beta_{n,n+1}$ lines in the bottom panel. The differences in amplification coefficients $\beta_{n,n+1}$ and partial departure coefficients between \citet{sto95} and \citet{pro18} are not non-neglibile. It is clear that our results are consistent with those in \citet{sto95} since the small difference can be explained by the different rate coefficients of collisional excitations and de-excitations. Although the cause leading to the difference from the values of \citet{pro18} is still unknown, we suggest that the departure coefficients calculated by us and \citet{sto95} are more accurate. And our nl-model should be reliable.

\section{Tables of total optical depths, line optical depths, and line-to-continuum ratios}

The values of $\tau_{\nu}$ and $\tau_{\nu,\textrm{L}}$ at the centers of the hydrogen recombination lines for different temperatures, densities, and EMs are listed in Tables \ref{table:tauT8e3}- \ref{table:taulT1p2e4}. The line-to-continuum ratios are given in Tables \ref{table:fluxtoIcT8e3}-\ref{table:fluxtoIcT1p2e4}.

\begin{table*}\tiny
\centering
\caption{Total optical depth $\tau_{\nu}$ at the center of the Hn$\alpha$ line as functions of electron density $n_e$ and EM.} \label{table:tauT8e3}
\begin{threeparttable}
\begin{tabular}{|c|cccc|cccc|cccc|}
\hline
 & \multicolumn{4}{c|}{$n_e=1.0\times10^5$ cm$^{-3}$}  & \multicolumn{4}{c|}{$n_e=5.0\times10^5$ cm$^{-3}$} & \multicolumn{4}{c|}{$n_e=1.0\times10^6$ cm$^{-3}$} \\
\hline
  & \multicolumn{4}{c|}{EM [cm$^{-6}$pc]} & \multicolumn{4}{c|}{EM [cm$^{-6}$pc]} & \multicolumn{4}{c|}{EM [cm$^{-6}$pc]} \\
n & $1\times10^{9}$ & $5\times10^{9}$ & $1\times10^{10}$ & $5\times10^{10}$ & $1\times10^{9}$ & $5\times10^{9}$ & $1\times10^{10}$ & $5\times10^{10}$ & $1\times10^{9}$ & $5\times10^{9}$ & $1\times10^{10}$ & $5\times10^{10}$ \\
\hline
30 & -0.22 & -1.26 & ... & ... & -0.27 & -1.27 & -2.21 & ... & -0.23 & -1.11 & -2.00 & ...  \\
33 & -0.41 & -1.91 & ... & ... & -0.36 & -1.62 & -2.67 & ... & -0.27 & -1.29 & -2.31 & ...  \\
35 & -0.55 & -2.28 & ... & ... & -0.40 & -1.80 & -2.89 & ... & -0.28 & -1.35 & -2.43 & ...  \\
37 & -0.69 & -2.59 & ... & ... & -0.42 & -1.92 & -3.07 & ... & -0.28 & -1.37 & -2.50 & ...  \\
40 & -0.88 & -2.95 & ... & ... & -0.43 & -2.01 & -3.23 & ... & -0.27 & -1.33 & -2.49 & ...  \\
43 & -1.02 & -3.22 & ... & ... & -0.42 & -2.01 & -3.30 & ... & -0.24 & -1.20 & -2.36 & ... \\
45 & -1.08 & -3.35 & ... & ... & -0.41 & -1.96 & -3.29 & ... & -0.21 & -1.08 & -2.18 & ... \\
47 & -1.12 & -3.45 & ... & ... & -0.38 & -1.86 & -3.22 & ... & -0.18 & -0.92 & -1.93 & ... \\
50 & -1.15 & -3.54 & ... & ... & -0.32 & -1.62 & -3.01 & ... & -0.12 & -0.61 & -1.38 & ... \\
53 & -1.14 & -3.57 & ... & ... & -0.24 & -1.26 & -2.62 & ... & -0.04 & -0.21 & -0.62 & ... \\
55 & -1.11 & -3.55 & ... & ... & -0.17 & -0.95 & -2.21 & ... & 0.03 & 0.11 & 0.02 & ...  \\
57 & -1.06 & -3.51 & ... & ... & -0.10 & -0.58 & -1.64 & ... & 0.10 & 0.47 & 0.77 & ...  \\
60 & -0.95 & -3.37 & ... & ... & 0.04 & 0.11 & -0.39 & ... & 0.23 & 1.11 & 2.08 & ...  \\
63 & -0.79 & -3.14 & ... & ... & 0.21 & 0.97 & 1.36 & ... & 0.38 & 1.88 & 3.65 & ...  \\
65 & -0.66 & -2.91 & ... & ... & 0.35 & 1.63 & 2.75 & ... & 0.50 & 2.48 & 4.85 & ...  \\
67 & -0.50 & -2.60 & ... & ... & 0.50 & 2.39 & 4.32 & ... & 0.63 & 3.15 & 6.21 & ... \\
70 & -0.21 & -1.91 & ... & ... & 0.76 & 3.70 & 7.03 & ... & 0.87 & 4.32 & 8.57 & ... \\
\hline
 & \multicolumn{4}{c|}{$n_e=5.0\times10^6$ cm$^{-3}$}  & \multicolumn{4}{c|}{$n_e=1.0\times10^7$ cm$^{-3}$} & \multicolumn{4}{c|}{$n_e=5.0\times10^7$ cm$^{-3}$} \\
\hline
  & \multicolumn{4}{c|}{EM [cm$^{-6}$pc]} & \multicolumn{4}{c|}{EM [cm$^{-6}$pc]} & \multicolumn{4}{c|}{EM [cm$^{-6}$pc]} \\
n & $1\times10^{9}$ & $5\times10^{9}$ & $1\times10^{10}$ & $5\times10^{10}$ & $5\times10^{9}$ & $1\times10^{10}$ & $5\times10^{10}$ & $1\times10^{11}$ & $5\times10^{9}$ & $1\times10^{10}$ & $5\times10^{10}$ & $1\times10^{11}$ \\
\hline
30 & -0.10 & -0.50 & -1.00 & -3.57 & -0.29 & -0.58 & -2.79 & -4.06 & -0.01 & -0.02 & -0.18 & -0.57 \\
33 & -0.09 & -0.46 & -0.92 & -3.67 & -0.22 & -0.45 & -2.51 & -4.05 & 0.04 & 0.08 & 0.38 & 0.59 \\
35 & -0.08 & -0.40 & -0.82 & -3.63 & -0.17 & -0.34 & -2.12 & -3.89 & 0.08 & 0.16 & 0.76 & 1.40 \\
37 & -0.07 & -0.34 & -0.69 & -3.49 & -0.10 & -0.20 & -1.54 & -3.55 & 0.12 & 0.24 & 1.16 & 2.21 \\
40 & -0.04 & -0.21 & -0.43 & -3.00 & 0.02 & 0.03 & -0.32 & -2.37 & 0.18 & 0.36 & 1.77 & 3.49 \\
43 & -0.01 & -0.04 & -0.10 & -1.90 & 0.16 & 0.31 & 1.18 & 0.57 & 0.25 & 0.50 & 2.49 & 4.95 \\
45 & 0.02 & 0.08 & 0.16 & -0.64 & 0.26 & 0.52 & 2.30 & 3.19 & 0.32 & 0.63 & 3.12 & 6.22 \\
47 & 0.05 & 0.23 & 0.44 & 0.99 & 0.38 & 0.75 & 3.52 & 5.95 & 0.40 & 0.79 & 3.92 & 7.82 \\
50 & 0.10 & 0.48 & 0.94 & 3.78 & 0.58 & 1.15 & 5.59 & 10.44 & 0.55 & 1.11 & 5.53 & 11.04 \\
53 & 0.15 & 0.77 & 1.53 & 6.97 & 0.82 & 1.65 & 8.11 & 15.72 & 0.78 & 1.56 & 7.78 & 15.55 \\
55 & 0.20 & 1.00 & 1.98 & 9.39 & 1.03 & 2.05 & 10.15 & 19.93 & 0.97 & 1.95 & 9.73 & 19.46 \\
57 & 0.25 & 1.26 & 2.51 & 12.12 & 1.27 & 2.53 & 12.59 & 24.89 & 1.21 & 2.42 & 12.12 & 24.23 \\
60 & 0.35 & 1.73 & 3.46 & 17.00 & 1.72 & 3.43 & 17.13 & 34.07 & 1.67 & 3.33 & 16.67 & 33.33 \\
63 & 0.47 & 2.34 & 4.67 & 23.13 & 2.31 & 4.62 & 23.07 & 46.01 & 2.26 & 4.53 & 22.63 & 45.26 \\
65 & 0.57 & 2.83 & 5.65 & 28.11 & 2.80 & 5.60 & 27.97 & 55.84 & 2.76 & 5.51 & 27.57 & 55.14 \\
67 & 0.68 & 3.41 & 6.82 & 33.95 & 3.38 & 6.75 & 33.75 & 67.42 & 3.34 & 6.68 & 33.41 & 66.82 \\
70 & 0.89 & 4.47 & 8.94 & 44.60 & 4.44 & 8.87 & 44.36 & 88.66 & 4.41 & 8.82 & 44.10 & 88.20 \\
\hline
\end{tabular}
\begin{tablenotes}
\item \textbf{Notes.} The electron temperature is 8000 K.
\end{tablenotes}
\end{threeparttable}
\end{table*}

\begin{table*}\tiny
\centering
\caption{Total optical depth $\tau_\nu$ at the center of the Hn$\alpha$ line as functions of electron density $n_e$ and EM.}\label{table:tauT1e4}
\begin{threeparttable}
\begin{tabular}{|c|cccc|cccc|cccc|}
\hline
 & \multicolumn{4}{c|}{$n_e=1.0\times10^5$ cm$^{-3}$}  & \multicolumn{4}{c|}{$n_e=5.0\times10^5$ cm$^{-3}$} & \multicolumn{4}{c|}{$n_e=1.0\times10^6$ cm$^{-3}$} \\
\hline
  & \multicolumn{4}{c|}{EM [cm$^{-6}$pc]} & \multicolumn{4}{c|}{EM [cm$^{-6}$pc]} & \multicolumn{4}{c|}{EM [cm$^{-6}$pc]} \\
n & $1\times10^{9}$ & $5\times10^{9}$ & $1\times10^{10}$ & $5\times10^{10}$ & $1\times10^{9}$ & $5\times10^{9}$ & $1\times10^{10}$ & $5\times10^{10}$ & $1\times10^{9}$ & $5\times10^{9}$ & $1\times10^{10}$ & $5\times10^{10}$ \\
\hline
30 & -0.12 & -0.66 & -1.37 & ... & -0.15 & -0.74 & -1.39 & ... & -0.13 & -0.64 & -1.22 & -3.33 \\
33 & -0.23 & -1.13 & -1.97 & ... & -0.20 & -0.96 & -1.75 & ... & -0.15 & -0.75 & -1.42 & -3.65 \\
35 & -0.31 & -1.43 & -2.30 & ... & -0.23 & -1.07 & -1.93 & ... & -0.16 & -0.78 & -1.50 & -3.80 \\
37 & -0.39 & -1.69 & -2.58 & ... & -0.24 & -1.15 & -2.06 & ... & -0.16 & -0.79 & -1.53 & -3.90 \\
40 & -0.50 & -2.02 & -2.90 & ... & -0.25 & -1.20 & -2.16 & ... & -0.15 & -0.76 & -1.49 & -3.98 \\
43 & -0.58 & -2.26 & -3.12 & ... & -0.24 & -1.18 & -2.17 & ... & -0.14 & -0.68 & -1.36 & -3.95 \\
45 & -0.62 & -2.37 & -3.22 & ... & -0.23 & -1.13 & -2.12 & ... & -0.12 & -0.61 & -1.22 & -3.86 \\
47 & -0.65 & -2.45 & -3.30 & ... & -0.21 & -1.06 & -2.01 & ... & -0.10 & -0.50 & -1.03 & -3.70 \\
50 & -0.66 & -2.51 & -3.35 & ... & -0.18 & -0.88 & -1.75 & ... & -0.06 & -0.30 & -0.64 & -3.30 \\
53 & -0.65 & -2.50 & -3.35 & ... & -0.13 & -0.64 & -1.34 & ... & -0.01 & 0.04 & -0.13 & -2.53 \\
55 & -0.63 & -2.45 & -3.31 & ... & -0.08 & -0.43 & -0.96 & ... & 0.04 & 0.17 & 0.30 & -1.58 \\
57 & -0.59 & -2.37 & -3.23 & ... & -0.03 & -0.18 & -0.48 & ... & 0.09 & 0.42 & 0.79 & 0.19 \\
60 & -0.52 & -2.17 & -3.05 & ... & 0.06 & 0.29 & 0.43 & ... & 0.18 & 0.87 & 1.69 & 5.03 \\
63 & -0.41 & -1.85 & -2.77 & ... & 0.18 & 0.87 & 1.60 & ... & 0.28 & 1.41 & 2.79 & 11.20 \\
65 & -0.32 & -1.55 & -2.50 & ... & 0.27 & 1.34 & 2.53 & ... & 0.37 & 1.84 & 3.65 & 15.91 \\
67 & -0.21 & -1.17 & -2.13 & ... & 0.38 & 1.87 & 3.61 & ... & 0.47 & 2.33 & 4.63 & 21.17 \\
70 & -0.01 & -0.38 & -1.30 & ... & 0.57 & 2.80 & 5.49 & ... & 0.64 & 3.19 & 6.36 & 30.27 \\
\hline
 & \multicolumn{4}{c|}{$n_e=5.0\times10^6$ cm$^{-3}$}  & \multicolumn{4}{c|}{$n_e=1.0\times10^7$ cm$^{-3}$} & \multicolumn{4}{c|}{$n_e=5.0\times10^7$ cm$^{-3}$} \\
\hline
  & \multicolumn{4}{c|}{EM [cm$^{-6}$pc]} & \multicolumn{4}{c|}{EM [cm$^{-6}$pc]} & \multicolumn{4}{c|}{EM [cm$^{-6}$pc]} \\
n & $5\times10^{9}$ & $1\times10^{10}$ & $5\times10^{10}$ & $1\times10^{11}$ & $5\times10^{9}$ & $1\times10^{10}$ & $5\times10^{10}$ & $1\times10^{11}$ & $5\times10^{9}$ & $1\times10^{10}$ & $5\times10^{10}$ & $1\times10^{11}$ \\
\hline
30 & -0.29 & -0.58 & -2.52 & -3.63 & -0.17 & -0.34 & -1.69 & -3.01 & -0.01 & -0.02 & -0.12 & -0.29 \\
33 & -0.26 & -0.53 & -2.50 & -3.73 & -0.13 & -0.26 & -1.41 & -2.81 & 0.02 & 0.05 & 0.22 & 0.39 \\
35 & -0.24 & -0.47 & -2.37 & -3.71 & -0.10 & -0.20 & -1.11 & -2.47 & 0.05 & 0.10 & 0.47 & 0.89 \\
37 & -0.20 & -0.39 & -2.13 & -3.59 & -0.06 & -0.12 & -0.72 & -1.92 & 0.07 & 0.15 & 0.74 & 1.44 \\
40 & -0.12 & -0.24 & -1.51 & -3.17 & 0.02 & 0.03 & 0.03 & -0.53 & 0.12 & 0.24 & 1.20 & 2.37 \\
43 & -0.01 & -0.03 & -0.54 & -2.18 & 0.11 & 0.22 & 0.97 & 1.45 & 0.18 & 0.35 & 1.77 & 3.52 \\
45 & 0.07 & 0.13 & 0.31 & -0.87 & 0.18 & 0.36 & 1.71 & 3.01 & 0.23 & 0.45 & 2.26 & 4.51 \\
47 & 0.17 & 0.33 & 1.32 & 1.19 & 0.26 & 0.53 & 2.55 & 4.77 & 0.29 & 0.58 & 2.87 & 5.74 \\
50 & 0.34 & 0.67 & 3.11 & 5.13 & 0.41 & 0.82 & 4.04 & 7.86 & 0.41 & 0.82 & 4.08 & 8.16 \\
53 & 0.55 & 1.10 & 5.30 & 9.79 & 0.60 & 1.19 & 5.92 & 11.67 & 0.58 & 1.16 & 5.78 & 11.57 \\
55 & 0.72 & 1.44 & 7.02 & 13.39 & 0.75 & 1.50 & 7.45 & 14.78 & 0.73 & 1.45 & 7.25 & 14.50 \\
57 & 0.91 & 1.83 & 9.00 & 17.49 & 0.93 & 1.86 & 9.28 & 18.46 & 0.90 & 1.81 & 9.04 & 18.09 \\
60 & 1.27 & 2.54 & 12.61 & 24.85 & 1.27 & 2.54 & 12.69 & 25.32 & 1.25 & 2.49 & 12.45 & 24.90 \\
63 & 1.72 & 3.45 & 17.17 & 34.09 & 1.72 & 3.43 & 17.15 & 34.25 & 1.69 & 3.38 & 16.91 & 33.82 \\
65 & 2.09 & 4.19 & 20.89 & 41.57 & 2.08 & 4.16 & 20.81 & 41.60 & 2.06 & 4.12 & 20.60 & 41.19 \\
67 & 2.53 & 5.06 & 25.24 & 50.32 & 2.51 & 5.03 & 25.14 & 50.25 & 2.50 & 4.99 & 24.95 & 49.90 \\
70 & 3.32 & 6.64 & 33.19 & 66.26 & 3.31 & 6.61 & 33.06 & 66.09 & 3.29 & 6.58 & 32.91 & 65.83 \\
\hline
\end{tabular}
\begin{tablenotes}
\item \textbf{Notes.} The electron temperature is 10000 K.
\end{tablenotes}
\end{threeparttable}
\end{table*}

\begin{table*}\tiny
\centering
\caption{Total optical depth $\tau_\nu$ at the center of the  Hn$\alpha$ line as functions of electron density $n_e$ and EM.}
\label{table:tauT1p2e4}
\begin{threeparttable}
\begin{tabular}{|c|cccc|cccc|cccc|}
\hline
 & \multicolumn{4}{c|}{$n_e=1.0\times10^5$ cm$^{-3}$}  & \multicolumn{4}{c|}{$n_e=5.0\times10^5$ cm$^{-3}$} & \multicolumn{4}{c|}{$n_e=1.0\times10^6$ cm$^{-3}$} \\
\hline
  & \multicolumn{4}{c|}{EM [cm$^{-6}$pc]} & \multicolumn{4}{c|}{EM [cm$^{-6}$pc]} & \multicolumn{4}{c|}{EM [cm$^{-6}$pc]} \\
n & $1\times10^{9}$ & $5\times10^{9}$ & $1\times10^{10}$ & $5\times10^{10}$ & $1\times10^{9}$ & $5\times10^{9}$ & $1\times10^{10}$ & $5\times10^{10}$ & $1\times10^{9}$ & $5\times10^{9}$ & $1\times10^{10}$ & $5\times10^{10}$ \\
\hline
30 & -0.07 & -0.39 & -0.83 & ... & -0.10 & -0.46 & -0.89 & ... & -0.08 & -0.40 & -0.78 & -2.63  \\
33 & -0.14 & -0.70 & -1.32 & ... & -0.13 & -0.61 & -1.15 & ... & -0.10 & -0.47 & -0.92 & -2.93  \\
35 & -0.19 & -0.91 & -1.61 & ... & -0.14 & -0.68 & -1.28 & ... & -0.10 & -0.49 & -0.96 & -3.06  \\
37 & -0.24 & -1.11 & -1.86 & ... & -0.15 & -0.73 & -1.37 & ... & -0.10 & -0.50 & -0.97 & -3.14  \\
40 & -0.31 & -1.36 & -2.16 & ... & -0.15 & -0.75 & -1.42 & ... & -0.10 & -0.48 & -0.94 & -3.16  \\
43 & -0.36 & -1.54 & -2.37 & ... & -0.15 & -0.74 & -1.41 & ... & -0.08 & -0.42 & -0.84 & -3.06 \\
45 & -0.39 & -1.63 & -2.46 & ... & -0.14 & -0.70 & -1.35 & ... & -0.07 & -0.37 & -0.73 & -2.91  \\
47 & -0.41 & -1.69 & -2.53 & ... & -0.13 & -0.64 & -1.25 & ... & -0.06 & -0.29 & -0.59 & -2.66 \\
50 & -0.41 & -1.72 & -2.55 & ... & -0.11 & -0.52 & -1.03 & ... & -0.03 & -0.15 & -0.31 & -2.04 \\
53 & -0.40 & -1.68 & -2.52 & ... & -0.07 & -0.34 & -0.69 & ... & 0.01 & 0.04 & 0.06 & -0.89 \\
55 & -0.38 & -1.62 & -2.45 & ... & -0.04 & -0.19 & -0.40 & ... & 0.04 & 0.19 & 0.37 & 0.40 \\
57 & -0.36 & -1.53 & -2.35 & ... & 0.00 & -0.00 & -0.04 & ... & 0.08 & 0.38 & 0.73 & 2.20 \\
60 & -0.30 & -1.32 & -2.10 & ... & 0.07 & 0.34 & 0.63 & ... & 0.14 & 0.71 & 1.40 & 5.72 \\
63 & -0.22 & -1.01 & -1.73 & ... & 0.16 & 0.77 & 1.49 & ... & 0.23 & 1.12 & 2.23 & 10.09 \\
65 & -0.15 & -0.73 & -1.37 & ... & 0.23 & 1.12 & 2.19 & ... & 0.29 & 1.45 & 2.89 & 13.52 \\
67 & -0.07 & -0.38 & -0.89 & ... & 0.31 & 1.52 & 2.99 & ... & 0.37 & 1.83 & 3.65 & 17.43 \\
70 & 0.07 & 0.28 & 0.13 & ... & 0.45 & 2.23 & 4.42 & ... & 0.50 & 2.50 & 4.99 & 24.32 \\
\hline
 & \multicolumn{4}{c|}{$n_e=5.0\times10^6$ cm$^{-3}$}  & \multicolumn{4}{c|}{$n_e=1.0\times10^7$ cm$^{-3}$} & \multicolumn{4}{c|}{$n_e=5.0\times10^7$ cm$^{-3}$} \\
\hline
  & \multicolumn{4}{c|}{EM [cm$^{-6}$pc]} & \multicolumn{4}{c|}{EM [cm$^{-6}$pc]} & \multicolumn{4}{c|}{EM [cm$^{-6}$pc]} \\
n & $5\times10^{9}$ & $1\times10^{10}$ & $5\times10^{10}$ & $1\times10^{11}$ & $5\times10^{9}$ & $1\times10^{10}$ & $5\times10^{10}$ & $1\times10^{11}$ & $5\times10^{9}$ & $1\times10^{10}$ & $5\times10^{10}$ & $1\times10^{11}$ \\
\hline
30 & -0.18 & -0.36 & -1.70 & -2.81 & -0.11 & -0.21 & -1.07 & -2.06 & -0.01 & -0.01 & -0.08 & -0.17  \\
33 & -0.17 & -0.33 & -1.63 & -2.83 & -0.08 & -0.17 & -0.86 & -1.78 & 0.02 & 0.03 & 0.15 & 0.28 \\
35 & -0.15 & -0.30 & -1.49 & -2.73 & -0.06 & -0.12 & -0.65 & -1.44 & 0.03 & 0.07 & 0.33 & 0.64  \\
37 & -0.12 & -0.24 & -1.28 & -2.51 & -0.03 & -0.07 & -0.39 & -0.95 & 0.05 & 0.11 & 0.53 & 1.04 \\
40 & -0.07 & -0.14 & -0.79 & -1.89 & 0.02 & 0.03 & 0.12 & 0.06 & 0.09 & 0.18 & 0.88 & 1.76 \\
43 & 0.00 & 0.01 & -0.09 & -0.71 & 0.08 & 0.17 & 0.79 & 1.42 & 0.13 & 0.27 & 1.34 & 2.68 \\
45 & 0.06 & 0.13 & 0.50 & 0.46 & 0.14 & 0.27 & 1.33 & 2.53 & 0.17 & 0.35 & 1.73 & 3.46 \\
47 & 0.13 & 0.27 & 1.22 & 1.93 & 0.20 & 0.40 & 1.97 & 3.82 & 0.22 & 0.45 & 2.23 & 4.45 \\
50 & 0.26 & 0.52 & 2.53 & 4.65 & 0.31 & 0.63 & 3.12 & 6.15 & 0.32 & 0.64 & 3.19 & 6.38 \\
53 & 0.43 & 0.85 & 4.17 & 8.04 & 0.46 & 0.92 & 4.59 & 9.11 & 0.45 & 0.91 & 4.54 & 9.08  \\
55 & 0.56 & 1.11 & 5.49 & 10.74 & 0.58 & 1.16 & 5.80 & 11.55 & 0.57 & 1.14 & 5.70 & 11.40 \\
57 & 0.71 & 1.42 & 7.04 & 13.87 & 0.73 & 1.45 & 7.24 & 14.44 & 0.71 & 1.42 & 7.12 & 14.23 \\
60 & 0.99 & 1.98 & 9.86 & 19.57 & 0.99 & 1.99 & 9.94 & 19.86 & 0.98 & 1.96 & 9.80 & 19.60 \\
63 & 1.35 & 2.70 & 13.45 & 26.79 & 1.35 & 2.69 & 13.45 & 26.89 & 1.33 & 2.66 & 13.31 & 26.63  \\
65 & 1.64 & 3.28 & 16.37 & 32.66 & 1.63 & 3.27 & 16.34 & 32.67 & 1.62 & 3.24 & 16.21 & 32.41  \\
67 & 1.98 & 3.96 & 19.80 & 39.53 & 1.97 & 3.95 & 19.75 & 39.48 & 1.96 & 3.93 & 19.63 & 39.25 \\
70 & 2.61 & 5.21 & 26.05 & 52.04 & 2.60 & 5.20 & 25.97 & 51.94 & 2.59 & 5.18 & 25.88 & 51.76 \\
\hline
\end{tabular}
\begin{tablenotes}
\item \textbf{Notes.} The electron temperature is 12000 K.
\end{tablenotes}
\end{threeparttable}
\end{table*}

\begin{table*}\tiny
\centering
\caption{Line optical depth $\tau_{\nu,\textrm{L}}$ at the center of the Hn$\alpha$ line as functions of electron density $n_e$ and EM.} \label{table:taulT8e3}
\begin{threeparttable}
\begin{tabular}{|c|cccc|cccc|cccc|}
\hline
 & \multicolumn{4}{c|}{$n_e=1.0\times10^5$ cm$^{-3}$}  & \multicolumn{4}{c|}{$n_e=5.0\times10^5$ cm$^{-3}$} & \multicolumn{4}{c|}{$n_e=1.0\times10^6$ cm$^{-3}$} \\
\hline
  & \multicolumn{4}{c|}{EM [cm$^{-6}$pc]} & \multicolumn{4}{c|}{EM [cm$^{-6}$pc]} & \multicolumn{4}{c|}{EM [cm$^{-6}$pc]} \\
n & $1\times10^{9}$ & $5\times10^{9}$ & $1\times10^{10}$ & $5\times10^{10}$ & $1\times10^{9}$ & $5\times10^{9}$ & $1\times10^{10}$ & $5\times10^{10}$ & $1\times10^{9}$ & $5\times10^{9}$ & $1\times10^{10}$ & $5\times10^{10}$ \\
\hline
30 & -0.22 & -1.28 & ... & ... & -0.28 & -1.29 & -2.25 & ... & -0.24 & -1.13 & -2.04 & ...  \\
33 & -0.42 & -1.95 & ... & ... & -0.36 & -1.66 & -2.74 & ... & -0.28 & -1.33 & -2.38 & ...  \\
35 & -0.56 & -2.33 & ... & ... & -0.41 & -1.85 & -3.00 & ... & -0.29 & -1.40 & -2.53 & ...  \\
37 & -0.71 & -2.66 & ... & ... & -0.44 & -2.00 & -3.21 & ... & -0.29 & -1.44 & -2.65 & ...  \\
40 & -0.90 & -3.08 & ... & ... & -0.46 & -2.14 & -3.48 & ... & -0.29 & -1.45 & -2.74 & ...  \\
43 & -1.05 & -3.41 & ... & ... & -0.46 & -2.21 & -3.69 & ... & -0.28 & -1.40 & -2.75 & ...  \\
45 & -1.13 & -3.61 & ... & ... & -0.46 & -2.22 & -3.81 & ... & -0.27 & -1.34 & -2.70 & ...  \\
47 & -1.19 & -3.79 & ... & ... & -0.45 & -2.20 & -3.91 & ... & -0.25 & -1.27 & -2.62 & ...  \\
50 & -1.25 & -4.05 & ... & ... & -0.42 & -2.13 & -4.04 & ... & -0.22 & -1.13 & -2.41 & ...  \\
53 & -1.29 & -4.32 & ... & ... & -0.39 & -2.01 & -4.11 & ... & -0.18 & -0.96 & -2.11 & ...  \\
55 & -1.30 & -4.50 & ... & ... & -0.36 & -1.90 & -4.10 & ... & -0.16 & -0.84 & -1.87 & ...  \\
57 & -1.30 & -4.70 & ... & ... & -0.33 & -1.76 & -4.01 & ... & -0.14 & -0.72 & -1.61 & ...  \\
60 & -1.28 & -5.02 & ... & ... & -0.28 & -1.53 & -3.68 & ... & -0.10 & -0.53 & -1.21 & ...  \\
63 & -1.24 & -5.39 & ... & ... & -0.23 & -1.28 & -3.13 & ... & -0.07 & -0.36 & -0.84 & ...  \\
65 & -1.21 & -5.65 & ... & ... & -0.20 & -1.10 & -2.72 & ... & -0.05 & -0.26 & -0.62 & ...  \\
67 & -1.16 & -5.92 & ... & ... & -0.17 & -0.93 & -2.32 & ... & -0.03 & -0.17 & -0.43 & ...  \\
70 & -1.09 & -6.29 & ... & ... & -0.12 & -0.69 & -1.74 & ... & -0.01 & -0.07 & -0.20 & ...  \\
\hline
 & \multicolumn{4}{c|}{$n_e=5.0\times10^6$ cm$^{-3}$}  & \multicolumn{4}{c|}{$n_e=1.0\times10^7$ cm$^{-3}$} & \multicolumn{4}{c|}{$n_e=5.0\times10^7$ cm$^{-3}$} \\
\hline
  & \multicolumn{4}{c|}{EM [cm$^{-6}$pc]} & \multicolumn{4}{c|}{EM [cm$^{-6}$pc]} & \multicolumn{4}{c|}{EM [cm$^{-6}$pc]} \\
n & $1\times10^{9}$ & $5\times10^{9}$ & $1\times10^{10}$ & $5\times10^{10}$ & $5\times10^{9}$ & $1\times10^{10}$ & $5\times10^{10}$ & $1\times10^{11}$ & $5\times10^{9}$ & $1\times10^{10}$ & $5\times10^{10}$ & $1\times10^{11}$ \\
\hline
30 & -0.10 & -0.52 & -1.03 & -3.76 & -0.31 & -0.62 & -2.98 & -4.45 & -0.03 & -0.06 & -0.37 & -0.95 \\
33 & -0.10 & -0.49 & -0.99 & -4.03 & -0.26 & -0.52 & -2.86 & -4.76 & 0.01 & 0.01 & 0.02 & -0.12 \\
35 & -0.09 & -0.46 & -0.92 & -4.15 & -0.22 & -0.44 & -2.64 & -4.93 & 0.03 & 0.06 & 0.24 & 0.36 \\
37 & -0.08 & -0.41 & -0.84 & -4.24 & -0.17 & -0.35 & -2.28 & -5.04 & 0.04 & 0.09 & 0.41 & 0.73 \\
40 & -0.07 & -0.33 & -0.68 & -4.23 & -0.10 & -0.21 & -1.55 & -4.83 & 0.06 & 0.11 & 0.55 & 1.03 \\
43 & -0.05 & -0.24 & -0.49 & -3.85 & -0.04 & -0.08 & -0.77 & -3.34 & 0.06 & 0.11 & 0.56 & 1.08 \\
45 & -0.03 & -0.18 & -0.37 & -3.26 & 0.00 & 0.00 & -0.32 & -2.04 & 0.05 & 0.11 & 0.53 & 1.03 \\
47 & -0.02 & -0.12 & -0.25 & -2.47 & 0.03 & 0.06 & 0.06 & -0.97 & 0.05 & 0.10 & 0.49 & 0.96 \\
50 & -0.01 & -0.04 & -0.09 & -1.36 & 0.06 & 0.13 & 0.45 & 0.16 & 0.04 & 0.08 & 0.39 & 0.96 \\
53 & 0.00 & 0.02 & 0.04 & -0.48 & 0.08 & 0.15 & 0.65 & 0.81 & 0.03 & 0.06 & 0.32 & 0.64 \\
55 & 0.01 & 0.05 & 0.10 & -0.06 & 0.08 & 0.16 & 0.71 & 1.04 & 0.03 & 0.06 & 0.28 & 0.56 \\
57 & 0.01 & 0.07 & 0.14 & 0.26 & 0.08 & 0.16 & 0.72 & 1.16 & 0.03 & 0.05 & 0.25 & 0.50 \\
60 & 0.02 & 0.09 & 0.17 & 0.55 & 0.07 & 0.14 & 0.67 & 1.16 & 0.02 & 0.04 & 0.21 & 0.42 \\
63 & 0.02 & 0.09 & 0.18 & 0.69 & 0.07 & 0.13 & 0.62 & 1.11 & 0.02 & 0.04 & 0.18 & 0.36 \\
65 & 0.02 & 0.09 & 0.18 & 0.73 & 0.06 & 0.12 & 0.58 & 1.06 & 0.02 & 0.04 & 0.15 & 0.30 \\
67 & 0.02 & 0.09 & 0.17 & 0.74 & 0.06 & 0.11 & 0.54 & 1.01 & 0.01 & 0.03 & 0.14 & 0.27 \\
70 & 0.02 & 0.08 & 0.16 & 0.73 & 0.05 & 0.10 & 0.49 & 0.92 & 0.01 & 0.02 & 0.12 & 0.24 \\
\hline
\end{tabular}
\begin{tablenotes}
\item \textbf{Notes.} The electron temperature is 8000 K.
\end{tablenotes}
\end{threeparttable}
\end{table*}

\begin{table*}\tiny
\centering
\caption{Line optical depth $\tau_{\nu,\textrm{L}}$ at the center of the  Hn$\alpha$ line as functions of electron density $n_e$ and EM.} \label{table:taulT1e4}
\begin{threeparttable}
\begin{tabular}{|c|cccc|cccc|cccc|}
\hline
 & \multicolumn{4}{c|}{$n_e=1.0\times10^5$ cm$^{-3}$}  & \multicolumn{4}{c|}{$n_e=5.0\times10^5$ cm$^{-3}$} & \multicolumn{4}{c|}{$n_e=1.0\times10^6$ cm$^{-3}$} \\
\hline
  & \multicolumn{4}{c|}{EM [cm$^{-6}$pc]} & \multicolumn{4}{c|}{EM [cm$^{-6}$pc]} & \multicolumn{4}{c|}{EM [cm$^{-6}$pc]} \\
n & $1\times10^{9}$ & $5\times10^{9}$ & $1\times10^{10}$ & $5\times10^{10}$ & $1\times10^{9}$ & $5\times10^{9}$ & $1\times10^{10}$ & $5\times10^{10}$ & $1\times10^{9}$ & $5\times10^{9}$ & $1\times10^{10}$ & $5\times10^{10}$ \\
\hline
30 & -0.12 & -0.68 & -1.40 & ... & -0.16 & -0.76 & -1.42 & ... & -0.13 & -0.66 & -1.25 & -3.48 \\
33 & -0.23 & -1.15 & -2.03 & ... & -0.21 & -0.99 & -1.80 & ... & -0.16 & -0.77 & -1.48 & -3.92 \\
35 & -0.32 & -1.47 & -2.38 & ... & -0.23 & -1.11 & -2.01 & ... & -0.17 & -0.82 & -1.58 & -4.20 \\
37 & -0.40& -1.75 & -2.69 & ... & -0.25 & -1.21 & -2.18 & ... & -0.17 & -0.85 & -1.64 & -4.47 \\
40 & -0.52 & -2.11 & -3.08 & ... & -0.27 & -1.29 & -2.35 & ... & -0.17 & -0.86 & -1.68 & -4.91 \\
43 & -0.61 & -2.40 & -3.41 & ... & -0.27 & -1.33 & -2.46 & ... & -0.17 & -0.83 & -1.65 & -5.42 \\
45 & -0.66 & -2.57 & -3.62 & ... & -0.27 & -1.33 & -2.51 & ... & -0.16 & -0.80 & -1.61 & -5.83 \\
47 & -0.70 & -2.71 & -3.82 & ... & -0.27 & -1.32 & -2.53 & ... & -0.15 & -0.76 & -1.55 & -6.31 \\
50 & -0.74 & -2.90 & -4.13 & ... & -0.25 & -1.27 & -2.52 & ... & -0.14 & -0.69 & -1.42 & -7.16 \\
53 & -0.76 & -3.06 & -4.47 & ... & -0.24 & -1.20 & -2.46 & ... & -0.12 & -0.60 & -1.25 & -8.13 \\
55 & -0.77 & -3.16 & -4.72 & ... & -0.22 & -1.14 & -2.37 & ... & -0.11 & -0.54 & -1.12 & -8.67 \\
57 & -0.77 & -3.26 & -5.01 & ... & -0.21 & -1.07 & -2.26 & ... & -0.09 & -0.47 & -0.99 & -8.71 \\
60 & -0.76 & -3.40 & -5.52 & ... & -0.18 & -0.94 & -2.04 & ... & -0.07 & -0.37 & -0.78 & -7.30 \\
63 & -0.74 & -3.53 & -6.13 & ... & -0.16 & -0.81 & -1.76 & ... & -0.05 & -0.27 & -0.57 & -5.60 \\
65 & -0.73 & -3.60 & -6.60 & ... & -0.14 & -0.71 & -1.56 & ... & -0.04 & -0.21 & -0.45 & -4.58 \\
67 & -0.71 & -3.65 & -7.10 & ... & -0.12 & -0.62 & -1.36 & ... & -0.03 & -0.15 & -0.33 & -3.67 \\
70 & -0.67 & -3.66 & -7.86 & ... & -0.09 & -0.48 & -1.06 & ... & -0.02 & -0.09 & -0.19 & -2.51 \\
\hline
 & \multicolumn{4}{c|}{$n_e=5.0\times10^6$ cm$^{-3}$}  & \multicolumn{4}{c|}{$n_e=1.0\times10^7$ cm$^{-3}$} & \multicolumn{4}{c|}{$n_e=5.0\times10^7$ cm$^{-3}$} \\
\hline
  & \multicolumn{4}{c|}{EM [cm$^{-6}$pc]} & \multicolumn{4}{c|}{EM [cm$^{-6}$pc]} & \multicolumn{4}{c|}{EM [cm$^{-6}$pc]} \\
n & $5\times10^{9}$ & $1\times10^{10}$ & $5\times10^{10}$ & $1\times10^{11}$ & $5\times10^{9}$ & $1\times10^{10}$ & $5\times10^{10}$ & $1\times10^{11}$ & $5\times10^{9}$ & $1\times10^{10}$ & $5\times10^{10}$ & $1\times10^{11}$ \\
\hline
30 & -0.30 & -0.60 & -2.67 & -3.92 & -0.18 & -0.37 & -1.84 & -3.30 & -0.02 & -0.05 & -0.26 & -0.58 \\
33 & -0.29 & -0.58 & -2.77 & -4.27 & -0.16 & -0.32 & -1.68 & -3.35 & -0.00 & -0.01 & -0.05 & -0.15 \\
35 & -0.27 & -0.55 & -2.77 & -4.50 & -0.14 & -0.28 & -1.50 & -3.26 & 0.01 & 0.02 & 0.07 & 0.11 \\
37 & -0.25 & -0.51 & -2.69 & -4.72 & -0.11 & -0.23 & -1.28 & -3.04 & 0.02 & 0.04 & 0.17 & 0.32 \\
40 & -0.21 & -0.42 & -2.44 & -5.02 & -0.08 & -0.16 & -0.90 & -2.39 & 0.03 & 0.05 & 0.27 & 0.51 \\
43 & -0.16 & -0.33 & -2.01 & -5.13 & -0.04 & -0.08 & -0.51 & -1.50 & 0.03 & 0.06 & 0.29 & 0.57 \\
45 & -0.13 & -0.26 & -1.66 & -4.82 & -0.02 & -0.03 & -0.26 & -0.93 & 0.03 & 0.06 & 0.29 & 0.56 \\
47 & -0.09 & -0.19 & -1.29 & -4.02 & 0.00 & 0.00 & -0.05 & -0.44 & 0.03 & 0.05 & 0.27 & 0.53 \\
50 & -0.05 & -0.10 & -0.75 & -2.60 & 0.02 & 0.05 & 0.18 & 0.13 & 0.02 & 0.04 & 0.22 & 0.43 \\
53 & -0.01 & -0.02 & -0.31 & -1.41 & 0.04 & 0.07 & 0.31 & 0.47 & 0.02 & 0.04 & 0.18 & 0.36 \\
55 & 0.01 & 0.02 & -0.07 & -0.79 & 0.04 & 0.08 & 0.36 & 0.60 & 0.02 & 0.03 & 0.16 & 0.32 \\
57 & 0.02 & 0.05 & 0.10 & -0.31 & 0.04 & 0.08 & 0.38 & 0.66 & 0.01 & 0.03 & 0.14 & 0.29 \\
60 & 0.04 & 0.07 & 0.28 & 0.19 & 0.04 & 0.08 & 0.36 & 0.67 & 0.01 & 0.02 & 0.12 & 0.24 \\
63 & 0.04 & 0.09 & 0.37 & 0.48 & 0.04 & 0.07 & 0.34 & 0.64 & 0.01 & 0.02 & 0.10 & 0.21 \\
65 & 0.05 & 0.09 & 0.39 & 0.58 & 0.03 & 0.07 & 0.32 & 0.61 & 0.01 & 0.02 & 0.09 & 0.18 \\
67 & 0.05 & 0.09 & 0.41 & 0.65 & 0.03 & 0.06 & 0.30& 0.58 & 0.01 & 0.02 & 0.08 & 0.16 \\
70 & 0.04 & 0.09 & 0.41 & 0.69 & 0.03 & 0.06 & 0.28 & 0.53 & 0.01 & 0.01 & 0.07 & 0.14 \\
\hline
\end{tabular}
\begin{tablenotes}
\item \textbf{Notes.} The electron temperature is 10000 K.
\end{tablenotes}
\end{threeparttable}
\end{table*}

\begin{table*}\tiny
\centering
\caption{Line optical depth $\tau_{\nu,\textrm{L}}$ at the center of the Hn$\alpha$ line as functions of electron density $n_e$ and EM.} \label{table:taulT1p2e4}
\begin{threeparttable}
\begin{tabular}{|c|cccc|cccc|cccc|}
\hline
 & \multicolumn{4}{c|}{$n_e=1.0\times10^5$ cm$^{-3}$}  & \multicolumn{4}{c|}{$n_e=5.0\times10^5$ cm$^{-3}$} & \multicolumn{4}{c|}{$n_e=1.0\times10^6$ cm$^{-3}$} \\
\hline
  & \multicolumn{4}{c|}{EM [cm$^{-6}$pc]} & \multicolumn{4}{c|}{EM [cm$^{-6}$pc]} & \multicolumn{4}{c|}{EM [cm$^{-6}$pc]} \\
n & $1\times10^{9}$ & $5\times10^{9}$ & $1\times10^{10}$ & $5\times10^{10}$ & $1\times10^{9}$ & $5\times10^{9}$ & $1\times10^{10}$ & $5\times10^{10}$ & $1\times10^{9}$ & $5\times10^{9}$ & $1\times10^{10}$ & $5\times10^{10}$ \\
\hline
30 & -0.07 & -0.40 & -0.85 & ... & -0.10 & -0.48 & -0.91 & ... & -0.08 & -0.41 & -0.80 & -2.74  \\
33 & -0.14 & -0.72 & -1.36 & ... & -0.13 & -0.63 & -1.19 & ... & -0.10 & -0.49 & -0.96 & -3.15 \\
35 & -0.20 & -0.94 & -1.67 & ... & -0.15 & -0.71 & -1.34 & ... & -0.11 & -0.53 & -1.03 & -3.38  \\
37 & -0.25 & -1.15 & -1.95 & ... & -0.16 & -0.77 & -1.46 & ... & -0.11 & -0.54 & -1.06 & -3.59  \\
40 & -0.33 & -1.43 & -2.31 & ... & -0.17 & -0.83 & -1.57 & ... & -0.11 & -0.55 & -1.08 & -3.90  \\
43 & -0.39 & -1.66 & -2.60 & ... & -0.17 & -0.85 & -1.64 & ... & -0.11 & -0.54 & -1.07 & -4.23  \\
45 & -0.42 & -1.79 & -2.78 & ... & -0.17 & -0.86 & -1.66 & ... & -0.10 & -0.52 & -1.04 & -4.47  \\
47 & -0.45 & -1.90 & -2.94 & ... & -0.17 & -0.85 & -1.66 & ... & -0.10 & -0.50 & -1.00 & -4.72  \\
50 & -0.47 & -2.02 & -3.17 & ... & -0.17 & -0.83 & -1.64 & ... & -0.09 & -0.46 & -0.93 & -5.10  \\
53 & -0.49 & -2.13 & -3.40 & ... & -0.16 & -0.78 & -1.58 & ... & -0.08 & -0.41 & -0.83 & -5.32 \\
55 & -0.49 & -2.18 & -3.57 & ... & -0.15 & -0.75 & -1.52 & ... & -0.07 & -0.37 & -0.75 & -5.20  \\
57 & -0.50 & -2.23 & -3.75 & ... & -0.14 & -0.71 & -1.45 & ... & -0.06 & -0.33 & -0.67 & -4.82  \\
60 & -0.49 & -2.29 & -4.05 & ... & -0.12 & -0.63 & -1.31 & ... & -0.05 & -0.26 & -0.54 & -4.01  \\
63 & -0.49 & -2.33 & -4.38 & ... & -0.11 &.-0.55 & -1.15 & ... & -0.04 & -0.20 & -0.42 & -3.16  \\
65 & -0.48 & -2.34 & -4.60 & ... & -0.10 & -0.49 & -1.04 & ... & -0.03 & -0.16 & -0.34 & -2.63  \\
67 & -0.46 & -2.34 & -4.81 & ... & -0.08 & -0.43 & -0.92 & ... & -0.02 & -0.13 & -0.26 & -2.13  \\
70 & -0.44 & -2.30 & -5.03 & ... & -0.07 & -0.35 & -0.74 & ... & -0.02 & -0.08 & -0.17 & -1.49  \\
\hline
 & \multicolumn{4}{c|}{$n_e=5.0\times10^6$ cm$^{-3}$}  & \multicolumn{4}{c|}{$n_e=1.0\times10^7$ cm$^{-3}$} & \multicolumn{4}{c|}{$n_e=5.0\times10^7$ cm$^{-3}$} \\
\hline
  & \multicolumn{4}{c|}{EM [cm$^{-6}$pc]} & \multicolumn{4}{c|}{EM [cm$^{-6}$pc]} & \multicolumn{4}{c|}{EM [cm$^{-6}$pc]} \\
n & $5\times10^{9}$ & $1\times10^{10}$ & $5\times10^{10}$ & $1\times10^{11}$ & $5\times10^{9}$ & $1\times10^{10}$ & $5\times10^{10}$ & $1\times10^{11}$ & $5\times10^{9}$ & $1\times10^{10}$ & $5\times10^{10}$ & $1\times10^{11}$ \\
\hline
30 & -0.19 & -0.38 & -1.82 & -3.05 & -0.12 & -0.24 & -1.18 & -2.30 & -0.02 & -0.04 & -0.19 & -0.40  \\
33 & -0.19 & -0.38 & -1.85 & -3.26 & -0.10 & -0.21 & -1.08 & -2.21 & -0.01 & -0.01 & -0.07 & -0.15 \\
35 & -0.18 & -0.36 & -1.81 & -3.36 & -0.09 & -0.19 & -0.97 & -2.06 & 0.00 & 0.00 & 0.01 & 0.01 \\
37 & -0.17 & -0.33 & -1.72 & -3.41 & -0.08 & -0.16 & -0.84 & -1.85 & 0.01 & 0.02 & 0.08 & 0.14  \\
40 & -0.14 & -0.29 & -1.53 & -3.36 & -0.06 & -0.11 & -0.61 & -1.41 & 0.01 & 0.03 & 0.14 & 0.28  \\
43 & -0.11 & -0.23 & -1.26 & -3.05 & -0.03 & -0.07 & -0.38 & -0.92 & 0.02 & 0.03 & 0.17 & 0.33 \\
45 & -0.09 & -0.19 & -1.06 & -2.67 & -0.02 & -0.04 & -0.23 & -0.60 & 0.02 & 0.03 & 0.17 & 0.34  \\
47 & -0.07 & -0.15 & -0.84 & -2.19 & -0.01 & -0.01 & -0.10 & -0.31 & 0.02 & 0.03 & 0.16 & 0.32 \\
50 & -0.04 & -0.09 & -0.53 & -1.46 & 0.01 & 0.02 & 0.06 & 0.04 & 0.01 & 0.03 & 0.13 & 0.27  \\
53 & -0.02 & -0.04 & -0.26 & -0.82 & 0.02 & 0.04 & 0.16 & 0.26 & 0.01 & 0.02 & 0.11 & 0.23  \\
55 & -0.00 & -0.01 & -0.11 & -0.46 & 0.02 & 0.04 & 0.20 & 0.34 & 0.01 & 0.02 & 0.10 & 0.20 \\
57 & 0.01 & 0.01 & 0.01 & -0.18 & 0.02 & 0.05 & 0.22 & 0.39 & 0.01 & 0.02 & 0.09 & 0.17 \\
60 & 0.02 & 0.03 & 0.13 & 0.12 & 0.02 & 0.04 & 0.22 & 0.41 & 0.01 & 0.02 & 0.08 & 0.14  \\
63 & 0.02 & 0.05 & 0.20 & 0.30 & 0.02 & 0.04 & 0.21 & 0.40 & 0.01 & 0.01 & 0.07 & 0.12  \\
65 & 0.03 & 0.05 & 0.23 & 0.37 & 0.02 & 0.04 & 0.20 & 0.38 & 0.01 & 0.01 & 0.06 & 0.11  \\
67 & 0.03 & 0.05 & 0.24 & 0.41 & 0.02 & 0.04 & 0.19 & 0.36 & 0.00 & 0.01 & 0.05 & 0.10  \\
70 & 0.03 & 0.05 & 0.24 & 0.44 & 0.02 & 0.03 & 0.17 & 0.33 & 0.00 & 0.01 & 0.04 & 0.09  \\
\hline
\end{tabular}
\begin{tablenotes}
\item \textbf{Notes.} The electron temperature is 12000 K.
\end{tablenotes}
\end{threeparttable}
\end{table*}

\begin{table*}\tiny
\centering
\caption{Line-to-continuum ratios $\int I_{\nu,\textrm{L}}d\nu/I_{\nu,\textrm{C}}$ [Hz] of the Hn$\alpha$ lines as functions of electron density $n_e$ and EM.}
\label{table:fluxtoIcT8e3}
\begin{threeparttable}
\begin{tabular}{|c|cccc|cccc|cccc|}
\hline
 & \multicolumn{4}{c|}{$n_e=1.0\times10^5$ cm$^{-3}$}  & \multicolumn{4}{c|}{$n_e=5.0\times10^5$ cm$^{-3}$} & \multicolumn{4}{c|}{$n_e=1.0\times10^6$ cm$^{-3}$} \\
\hline
  & \multicolumn{4}{c|}{EM [cm$^{-6}$pc]} & \multicolumn{4}{c|}{EM [cm$^{-6}$pc]} & \multicolumn{4}{c|}{EM [cm$^{-6}$pc]} \\
n & $1\times10^{9}$ & $5\times10^{9}$ & $1\times10^{10}$ & $5\times10^{10}$ & $1\times10^{9}$ & $5\times10^{9}$ & $1\times10^{10}$ & $5\times10^{10}$ & $1\times10^{9}$ & $5\times10^{9}$ & $1\times10^{10}$ & $5\times10^{10}$ \\
\hline
30 & 7.67e7 & 1.37e8 & ... & ... & 8.87e7 & 1.46e8 & 2.48e8 & ... & 9.13e7 & 1.40e8 & 2.26e8 & ...  \\
33 & 4.88e7 & 1.15e8 & ... & ... & 5.29e7 & 1.04e8 & 1.91e8 & ... & 5.25e7 & 8.92e7 & 1.58e8 & ...  \\
35 & 3.78e7 & 1.04e8 & ... & ... & 3.82e7 & 8.27e7 & 1.61e8 & ... & 3.70e7 & 6.61e7 & 1.24e8 & ...  \\
37 & 2.99e7 & 9.40e7 & ... & ... & 2.80e7 & 6.60e7 & 1.35e8 & ... & 2.65e7 & 4.92e7 & 9.67e7 & ...  \\
40 & 2.17e7 & 8.00e7 & ... & ... & 1.78e7 & 4.66e7 & 1.03e8 & ... & 1.65e7 & 3.17e7 & 6.61e7 & ...  \\
43 & 1.61e7 & 6.78e7 & ... & ... & 1.17e7 & 3.29e7 & 7.88e7 & ... & 1.05e7 & 2.06e7 & 4.43e7 & ...  \\
45 & 1.33e7 & 6.08e7 & ... & ... & 8.93e6 & 2.60e7 & 6.58e7 & ... & 7.93e6 & 1.55e7 & 3.35e7 & ...  \\
47 & 1.10e7 & 5.44e7 & ... & ... & 6.88e6 & 2.05e7 & 5.47e7 & ... & 6.03e6 & 1.17e7 & 2.49e7 & ...  \\
50 & 8.33e6 & 4.64e7 & ... & ... & 4.74e6 & 1.41e7 & 4.06e7 & ... & 4.07e6 & 7.64e6 & 1.54e7 & ...  \\
53 & 6.36e6 & 3.99e7 & ... & ... & 3.31e6 & 9.61e6 & 2.87e7 & ... & 2.80e6 & 5.01e6 & 9.16e6 & ...  \\
55 & 5.33e6 & 3.62e7 & ... & ... & 2.63e6 & 7.36e6 & 2.17e7 & ... & 2.20e6 & 3.78e6 & 6.37e6 & ...  \\
57 & 4.47e6 & 3.28e7 & ... & ... & 2.11e6 & 5.58e6 & 1.55e7 & ... & 1.74e6 & 2.86e6 & 4.40e6 & ...  \\
60 & 3.45e6 & 2.83e7 & ... & ... & 1.52e6 & 3.64e6 & 8.27e6 & ... & 1.24e6 & 1.87e6 & 2.53e6 & ...  \\
63 & 2.66e6 & 2.39e7 & ... & ... & 1.10e6 & 2.34e6 & 4.08e6 & ... & 8.86e5 & 1.22e6 & 1.49e5 & ...  \\
65 & 2.24e6 & 2.08e7 & ... & ... & 8.96e5 & 1.73e6 & 2.59e6 & ... & 7.12e5 & 9.22e5 & 1.06e6 & ...  \\
67 & 1.88e6 & 1.75e7 & ... & ... & 7.29e5 & 1.28e6 & 1.71e6 & ... & 5.73e5 & 6.94e5 & 7.68e5 & ...  \\
70 & 1.45e6 & 1.21e6 & ... & ... & 5.36e5 & 8.20e5 & 9.90e5 & ... & 4.13e5 & 4.54e5 & 4.87e5 & ...  \\
\hline
 & \multicolumn{4}{c|}{$n_e=5.0\times10^6$ cm$^{-3}$}  & \multicolumn{4}{c|}{$n_e=1.0\times10^7$ cm$^{-3}$} & \multicolumn{4}{c|}{$n_e=5.0\times10^7$ cm$^{-3}$} \\
\hline
  & \multicolumn{4}{c|}{EM [cm$^{-6}$pc]} & \multicolumn{4}{c|}{EM [cm$^{-6}$pc]} & \multicolumn{4}{c|}{EM [cm$^{-6}$pc]} \\
n & $1\times10^{9}$ & $5\times10^{9}$ & $1\times10^{10}$ & $5\times10^{10}$ & $5\times10^{9}$ & $1\times10^{10}$ & $5\times10^{10}$ & $1\times10^{11}$ & $5\times10^{9}$ & $1\times10^{10}$ & $5\times10^{10}$ & $1\times10^{11}$ \\
\hline
30 & 9.25e7 & 1.12e8 & 1.42e8 & 6.34e8 & 1.03e8 & 1.19e8 & 3.98e8 & 9.46e8 & 9.28e7 & 9.40e7 & 1.07e8 & 1.37e8 \\
33 & 5.06e7 & 6.11e7 & 7.80e7 & 4.15e8 & 5.52e7 & 6.26e7 & 2.06e8 & 6.06e8 & 4.92e7 & 4.91e7 & 4.89e7 & 5.19e7 \\
35 & 3.47e7 & 4.18e7 & 5.29e7 & 3.10e8 & 3.73e7 & 4.17e7 & 1.26e8 & 4.39e8 & 3.32e7 & 3.28e7 & 2.99e7 & 2.88e7 \\
37 & 2.43e7 & 2.90e7 & 3.61e7 & 2.27e8 & 2.57e7 & 2.82e7 & 7.27e7 & 3.00e8 & 2.29e7 & 2.23e7 & 1.87e7 & 1.65e7 \\
40 & 1.47e7 & 1.72e7 & 2.08e7 & 1.30e8 & 1.51e7 & 1.61e7 & 3.04e7 & 1.27e8 & 1.36e7 & 1.29e7 & 9.36e6 & 7.43e6 \\
43 & 9.19e6 & 1.04e7 & 1.21e7 & 5.91e7 & 9.16e6 & 9.41e6 & 1.32e7 & 2.98e7 & 8.27e6 & 7.66e6 & 4.55e6 & 3.40e6 \\
45 & 6.83e6 & 7.58e6 & 8.58e6 & 2.96e7 & 6.65e6 & 6.66e6 & 7.87e6 & 1.26e7 & 6.00e6 & 5.41e6 & 2.73e6 & 2.08e6 \\
47 & 5.14e6 & 5.55e6 & 6.09e6 & 1.45e7 & 4.87e6 & 4.74e6 & 4.88e6 & 6.72e6 & 4.38e6 & 3.80e6 & 1.61e6 & 1.33e6 \\
50 & 3.42e6 & 3.53e6 & 3.68e6 & 5.82e6 & 3.08e6 & 2.85e6 & 2.53e6 & 3.22e6 & 2.73e6 & 2.20e6 & 7.56e5 & 7.20e5 \\
53 & 2.32e6 & 2.26e6 & 2.23e6 & 2.84e6 & 1.95e6 & 1.69e6 & 1.40e6 & 1.74e6 & 1.68e6 & 1.21e6 & 3.91e5 & 4.04e5 \\
55 & 1.80e6 & 1.68e6 & 1.59e6 & 1.89e6 & 1.43e6 & 1.17e6 & 9.65e5 & 1.19e6 & 1.20e6 & 7.83e5 & 2.66e5 & 2.79e5 \\
57 & 1.41e6 & 1.24e6 & 1.13e6 & 1.30e6 & 1.04e6 & 8.04e5 & 6.77e5 & 8.30e5 & 8.47e5 & 4.90e5 & 1.85e5 & 1.95e5 \\
60 & 9.85e5 & 7.80e5 & 6.69e5 & 7.67e5 & 6.30e5 & 4.49e5 & 4.08e5 & 4.96e5 & 4.79e5 & 2.27e5 & 1.11e5 & 1.17e5 \\
63 & 6.89e5 & 4.79e5 & 3.99e5 & 4.70e5 & 3.67e5 & 2.52e5 & 2.52e5 & 3.04e5 & 2.53e5 & 1.02e5 & 6.78e4 & 7.13e4 \\
65 & 5.44e5 & 3.42e5 & 2.87e5 & 3.44e5 & 2.51e5 & 1.76e5 & 1.84e5 & 2.22e5 & 1.59e5 & 6.14e4 & 4.95e4 & 5.21e4 \\
67 & 4.28e5 & 2.43e5 & 2.10e5 & 2.54e5 & 1.70e5 & 1.26e5 & 1.36e5 & 1.64e5 & 9.59e4 & 3.96e4 & 3.65e4 & 3.84e4 \\
70 & 2.96e5 & 1.47e5 & 1.35e5 & 1.64e5 & 9.56e4 & 7.94e4 & 8.81e4 & 1.06e5 & 4.33e4 & 2.33e4 & 2.35e4 & 2.47e4 \\
\hline
\end{tabular}
\begin{tablenotes}
\item \textbf{Notes.} The electron temperature is assumed to be 8000 K. AeB means A$\times10^\textrm{B}$.
\end{tablenotes}
\end{threeparttable}
\end{table*}

\begin{table*}\tiny
\centering
\caption{Line-to-continuum ratios $\int I_{\nu,\textrm{L}}d\nu/I_{\nu,\textrm{C}}$ [Hz] of the Hn$\alpha$ lines as functions of electron density $n_e$ and EM.}
\label{table:fluxtoIcT1e4}
\begin{threeparttable}
\begin{tabular}{|c|cccc|cccc|cccc|}
\hline
 & \multicolumn{4}{c|}{$n_e=1.0\times10^5$ cm$^{-3}$}  & \multicolumn{4}{c|}{$n_e=5.0\times10^5$ cm$^{-3}$} & \multicolumn{4}{c|}{$n_e=1.0\times10^6$ cm$^{-3}$} \\
\hline
  & \multicolumn{4}{c|}{EM [cm$^{-6}$pc]} & \multicolumn{4}{c|}{EM [cm$^{-6}$pc]} & \multicolumn{4}{c|}{EM [cm$^{-6}$pc]} \\
n & $1\times10^{9}$ & $5\times10^{9}$ & $1\times10^{10}$ & $5\times10^{10}$ & $1\times10^{9}$ & $5\times10^{9}$ & $1\times10^{10}$ & $5\times10^{10}$ & $1\times10^{9}$ & $5\times10^{9}$ & $1\times10^{10}$ & $5\times10^{10}$ \\
\hline
30 & 5.69e7 & 7.82e7 & 1.19e8 & ... & 6.43e7 & 8.76e7 & 1.25e8 & ... & 6.63e7 & 8.63e7 & 1.18e8 & ...  \\
33 & 3.47e7 & 5.96e7 & 1.00e8 & ... & 3.76e7 & 5.79e7 & 9.14e7 & ... & 3.78e7 & 5.28e7 & 7.76e7 & ...  \\
35 & 2.60e7 & 5.20e7 & 9.07e7 & ... & 2.69e7 & 4.46e7 & 7.47e7 & ... & 2.66e7 & 3.86e7 & 5.92e7 & ...  \\
37 & 2.00e7 & 4.59e7 & 8.18e7 & ... & 1.96e7 & 3.47e7 & 6.12e7 & ... & 1.90e7 & 2.86e7 & 4.52e7 & ...  \\
40 & 1.40e7 & 3.84e7 & 6.99e7 & ... & 1.24e7 & 2.39e7 & 4.52e7 & ... & 1.18e7 & 1.85e7 & 3.03e7 & ...  \\
43 & 1.00e7 & 3.22e7 & 5.98e7 & ... & 8.14e6 & 1.67e7 & 3.34e7 & ... & 7.60e6 & 1.22e7 & 2.04e7 & ...  \\
45 & 8.14e6 & 2.87e7 & 5.41e7 & ... & 6.23e6 & 1.32e7 & 2.72e7 & ... & 5.74e6 & 9.33e6 & 1.56e7 & ...  \\
47 & 6.66e6 & 2.55e7 & 4.91e7 & ... & 4.82e6 & 1.05e7 & 2.20e7 & ... & 4.38e6 & 7.18e6 & 1.19e7 & ...  \\
50 & 5.00e6 & 2.13e7 & 4.30e7 & ... & 3.33e6 & 7.44e6 & 1.58e7 & ... & 2.98e6 & 4.89e6 & 7.94e6 & ...  \\
53 & 3.80e6 & 1.79e7 & 3.80e7 & ... & 2.35e6 & 5.28e6 & 1.11e7 & ... & 2.07e6 & 3.36e6 & 5.22e6 & ...  \\
55 & 3.18e6 & 1.58e7 & 3.52e7 & ... & 1.88e6 & 4.20e6 & 8.51e6 & ... & 1.64e6 & 2.62e6 & 3.92e6 & ...  \\
57 & 2.68e6 & 1.40e7 & 3.26e7 & ... & 1.51e6 & 3.33e6 & 6.42e6 & ... & 1.31e6 & 2.04e6 & 2.93e6 & ...  \\
60 & 2.09e6 & 1.15e7 & 2.88e7 & ... & 1.11e6 & 2.34e6 & 4.05e6 & ... & 9.38e5 & 1.41e6 & 1.87e6 & ...  \\
63 & 1.64e6 & 9.21e6 & 2.47e7 & ... & 8.16e5 & 1.63e6 & 2.47e6 & ... & 6.81e5 & 9.67e5 & 1.19e6 & ...  \\
65 & 1.39e6 & 7.75e6 & 2.15e7 & ... & 6.70e5 & 1.27e6 & 1.76e6 & ... & 5.53e5 & 7.50e5 & 8.75e5 & ...  \\
67 & 1.19e6 & 6.33e6 & 1.79e7 & ... & 5.51e5 & 9.85e5 & 1.26e6 & ... & 4.49e5 & 5.80e5 & 6.48e5 & ...  \\
70 & 9.37e5 & 4.37e6 & 1.20e7 & ... & 4.13e5 & 6.68e5 & 7.80e5 & ... & 3.31e5 & 3.92e5 & 4.18e5 & ...  \\
\hline
 & \multicolumn{4}{c|}{$n_e=5.0\times10^6$ cm$^{-3}$}  & \multicolumn{4}{c|}{$n_e=1.0\times10^7$ cm$^{-3}$} & \multicolumn{4}{c|}{$n_e=5.0\times10^7$ cm$^{-3}$} \\
\hline
  & \multicolumn{4}{c|}{EM [cm$^{-6}$pc]} & \multicolumn{4}{c|}{EM [cm$^{-6}$pc]} & \multicolumn{4}{c|}{EM [cm$^{-6}$pc]} \\
n & $5\times10^{9}$ & $1\times10^{10}$ & $5\times10^{10}$ & $1\times10^{11}$ & $5\times10^{9}$ & $1\times10^{10}$ & $5\times10^{10}$ & $1\times10^{11}$ & $5\times10^{9}$ & $1\times10^{10}$ & $5\times10^{10}$ & $1\times10^{11}$ \\
\hline
30 & 7.68e7 & 8.91e7 & 2.66e8 & 5.53e8 & 7.35e7 & 8.03e7 & 1.69e8 & 3.79e8 & 6.92e7 & 7.01e7 & 7.75e7 & 8.97e7  \\
33 & 4.24e7 & 4.96e7 & 1.62e8 & 3.84e8 & 3.99e7 & 4.34e7 & 8.82e7 & 2.17e8 & 3.71e7 & 3.72e7 & 3.81e7 & 4.00e7  \\
35 & 2.92e7 & 3.42e7 & 1.15e8 & 3.02e8 & 2.73e7 & 2.95e7 & 5.66e7 & 1.42e8 & 2.52e7 & 2.51e7 & 2.43e7 & 2.39e7 \\
37 & 2.05e7 & 2.39e7 & 7.92e7 & 2.36e8 & 1.90e7 & 2.04e7 & 3.62e7 & 8.67e7 & 1.75e7 & 1.73e7 & 1.57e7 & 1.45e7  \\
40 & 1.24e7 & 1.43e7 & 4.30e7 & 1.53e8 & 1.13e7 & 1.20e7 & 1.85e7 & 3.57e7 & 1.04e7 & 1.02e7 & 8.26e6 & 6.89e6  \\
43 & 7.70e6 & 8.71e6 & 2.16e7 & 7.99e7 & 7.00e6 & 7.24e6 & 9.49e6 & 1.40e7 & 6.43e6 & 6.11e6 & 4.26e6 & 3.22e6  \\
45 & 5.68e6 & 6.34e6 & 1.34e7 & 4.01e7 & 5.14e6 & 5.23e6 & 6.14e6 & 7.91e6 & 4.71e6 & 4.38e6 & 2.66e6 & 1.94e6  \\
47 & 4.23e6 & 4.63e6 & 8.28e6 & 1.72e7 & 3.81e6 & 3.80e6 & 4.00e6 & 4.74e6 & 3.47e6 & 3.14e6 & 1.62e6 & 1.19e6  \\
50 & 2.75e6 & 2.92e6 & 4.14e6 & 5.97e6 & 2.46e6 & 2.36e6 & 2.14e6 & 2.41e6 & 2.21e6 & 1.89e6 & 7.56e5 & 6.24e5  \\
53 & 1.81e6 & 1.84e6 & 2.19e6 & 2.81e6 & 1.60e6 & 1.46e6 & 1.18e6 & 1.32e6 & 1.41e6 & 1.11e6 & 3.67e5 & 3.48e5  \\
55 & 1.37e6 & 1.35e6 & 1.48e6 & 1.84e6 & 1.20e6 & 1.05e6 & 8.16e5 & 9.13e5 & 1.03e6 & 7.58e5 & 2.41e5 & 2.41e5  \\
57 & 1.04e6 & 9.86e5 & 1.03e6 & 1.26e6 & 8.93e5 & 7.43e5 & 5.72e5 & 6.40e5 & 7.53e5 & 5.04e5 & 1.65e5 & 1.69e5 \\
60 & 6.76e5 & 6.07e5 & 6.13e5 & 7.40e5 & 5.67e5 & 4.34e5 & 3.45e5 & 3.85e5 & 4.56e5 & 2.58e5 & 9.83e4 & 1.01e5 \\
63 & 4.33e5 & 3.69e5 & 3.78e5 & 4.52e5 & 3.50e5 & 2.48e5 & 2.13e5 & 2.37e5 & 2.64e5 & 1.23e5 & 6.04e4 & 6.20e4 \\
65 & 3.18e5 & 2.65e5 & 2.77e5 & 3.30e5 & 2.49e5 & 1.71e5 & 1.57e5 & 1.74e5 & 1.77e5 & 7.38e4 & 4.41e4 & 4.54e4 \\
67 & 2.32e5 & 1.92e5 & 2.05e5 & 2.44e5 & 1.74e5 & 1.20e5 & 1.16e5 & 1.29e5 & 1.15e5 & 4.48e4 & 3.26e4 & 3.35e4 \\
70 & 1.42e5 & 1.22e5 & 1.33e5 & 1.58e5 & 1.00e5 & 7.32e4 & 7.52e4 & 8.33e4 & 5.96e4 & 2.33e4 & 2.10e4 & 2.16e4 \\
\hline
\end{tabular}
\begin{tablenotes}
\item \textbf{Notes.} The electron temperature is assumed to be 10000 K. AeB means A$\times10^\textrm{B}$.
\end{tablenotes}
\end{threeparttable}
\end{table*}

\begin{table*}\tiny
\centering
\caption{Line-to-continuum ratios $\int I_{\nu,\textrm{L}}d\nu/I_{\nu,\textrm{C}}$ [Hz] of the Hn$\alpha$ lines as functions of electron density $n_e$ and EM.}
\label{table:fluxtoIcT1p2e4}
\begin{threeparttable}
\begin{tabular}{|c|cccc|cccc|cccc|}
\hline
 & \multicolumn{4}{c|}{$n_e=1.0\times10^5$ cm$^{-3}$}  & \multicolumn{4}{c|}{$n_e=5.0\times10^5$ cm$^{-3}$} & \multicolumn{4}{c|}{$n_e=1.0\times10^6$ cm$^{-3}$} \\
\hline
  & \multicolumn{4}{c|}{EM [cm$^{-6}$pc]} & \multicolumn{4}{c|}{EM [cm$^{-6}$pc]} & \multicolumn{4}{c|}{EM [cm$^{-6}$pc]} \\
n & $1\times10^{9}$ & $5\times10^{9}$ & $1\times10^{10}$ & $5\times10^{10}$ & $1\times10^{9}$ & $5\times10^{9}$ & $1\times10^{10}$ & $5\times10^{10}$ & $1\times10^{9}$ & $5\times10^{9}$ & $1\times10^{10}$ & $5\times10^{10}$ \\
\hline
30 & 4.54e7 & 5.55e7 & 7.32e7 & ... & 5.03e7 & 6.22e7 & 7.95e7 & ... & 5.19e7 & 6.22e7 & 7.70e7 & ...  \\
33 & 2.70e7 & 3.91e7 & 5.84e7 & ... & 2.91e7 & 3.95e7 & 5.50e7 & ... & 2.94e7 & 3.73e7 & 4.90e7 & ...  \\
35 & 1.99e7 & 3.26e7 & 5.18e7 & ... & 2.07e7 & 2.98e7 & 4.38e7 & ... & 2.07e7 & 2.71e7 & 3.68e7 & ...  \\
37 & 1.50e7 & 2.79e7 & 4.64e7 & ... & 1.50e7 & 2.28e7 & 3.50e7 & ... & 1.48e7 & 1.99e7 & 2.77e7 & ...  \\
40 & 1.02e7 & 2.25e7 & 3.95e7 & ... & 9.50e6 & 1.55e7 & 2.52e7 & ... & 9.18e6 & 1.29e7 & 1.85e7 & ...  \\
43 & 7.21e6 & 1.85e7 & 3.37e7 & ... & 6.21e6 & 1.08e7 & 1.84e7 & ... & 5.90e6 & 8.53e6 & 1.26e7 & ...  \\
45 & 5.80e6 & 1.62e7 & 3.04e7 & ... & 4.75e6 & 8.53e6 & 1.49e7 & ... & 4.46e6 & 6.57e6 & 9.75e6 & ...  \\
47 & 4.71e6 & 1.43e7 & 2.74e7 & ... & 3.67e6 & 6.80e6 & 1.20e7 & ... & 3.42e6 & 5.10e6 & 7.59e6 & ...  \\
50 & 3.50e6 & 1.18e7 & 2.35e7 & ... & 2.55e6 & 4.89e6 & 8.74e6 & ... & 2.33e6 & 3.53e6 & 5.23e6 & ...  \\
53 & 2.65e6 & 9.68e6 & 2.03e7 & ... & 1.80e6 & 3.54e6 & 6.28e6 & ... & 1.63e6 & 2.48e6 & 3.60e6 & ...  \\
55 & 2.22e6 & 8.50e6 & 1.84e7 & ... & 1.45e6 & 2.87e6 & 5.00e6 & ... & 1.29e6 & 1.96e6 & 2.80e6 & ...  \\
57 & 1.87e6 & 7.43e6 & 1.67e7 & ... & 1.17e6 & 2.32e6 & 3.94e6 & ... & 1.03e6 & 1.56e6 & 2.17e6 & ...  \\
60 & 1.46e6 & 6.03e6 & 1.42e7 & ... & 8.57e5 & 1.69e6 & 2.70e6 & ... & 7.46e5 & 1.10e6 & 1.46e6 & ...  \\
63 & 1.15e6 & 4.80e6 & 1.16e7 & ... & 6.37e5 & 1.22e6 & 1.80e6 & ... & 5.46e5 & 7.81e5 & 9.77e5 & ...  \\
65 & 9.82e5 & 4.06e6 & 9.74e6 & ... & 5.26e5 & 9.79e5 & 1.36e6 & ... & 4.45e5 & 6.18e5 & 7.41e5 & ...  \\
67 & 8.43e5 & 3.37e6 & 7.85e6 & ... & 4.35e5 & 7.82e5 & 1.02e6 & ... & 3.65e5 & 4.88e5 & 5.61e5 & ...  \\
70 & 6.73e5 & 2.47e6 & 5.10e6 & ... & 3.30e5 & 5.53e5 & 6.59e5 & ... & 2.71e5 & 3.39e5 & 3.69e5 & ...  \\
\hline
 & \multicolumn{4}{c|}{$n_e=5.0\times10^6$ cm$^{-3}$}  & \multicolumn{4}{c|}{$n_e=1.0\times10^7$ cm$^{-3}$} & \multicolumn{4}{c|}{$n_e=5.0\times10^7$ cm$^{-3}$} \\
\hline
  & \multicolumn{4}{c|}{EM [cm$^{-6}$pc]} & \multicolumn{4}{c|}{EM [cm$^{-6}$pc]} & \multicolumn{4}{c|}{EM [cm$^{-6}$pc]} \\
n & $5\times10^{9}$ & $1\times10^{10}$ & $5\times10^{10}$ & $1\times10^{11}$ & $5\times10^{9}$ & $1\times10^{10}$ & $5\times10^{10}$ & $1\times10^{11}$ & $5\times10^{9}$ & $1\times10^{10}$ & $5\times10^{10}$ & $1\times10^{11}$ \\
\hline
30 & 5.83e7 & 6.46e7 & 1.40e8 & 2.79e8 & 5.67e7 & 6.04e7 & 9.96e7 & 1.81e8 & 5.47e7 & 5.52e7 & 5.99e7 & 6.67e7  \\
33 & 3.23e7 & 3.61e7 & 8.28e7 & 1.83e8 & 3.10e7 & 3.30e7 & 5.38e7 & 9.94e7 & 2.95e7 & 2.96e7 & 3.06e7 & 3.21e7 \\
35 & 2.23e7 & 2.50e7 & 5.82e7 & 1.37e8 & 2.13e7 & 2.26e7 & 3.60e7 & 6.49e7 & 2.01e7 & 2.01e7 & 2.00e7 & 2.00e7  \\
37 & 1.57e7 & 1.77e7 & 4.07e7 & 1.01e8 & 1.49e7 & 1.58e7 & 2.42e7 & 4.13e7 & 1.40e7 & 1.39e7 & 1.32e7 & 1.26e7  \\
40 & 9.60e6 & 1.07e7 & 2.35e7 & 5.89e7 & 8.99e6 & 9.43e6 & 1.34e7 & 2.03e7 & 8.41e6 & 8.25e6 & 7.19e6 & 6.29e6  \\
43 & 6.03e6 & 6.70e6 & 1.33e7 & 2.92e7 & 5.60e6 & 5.80e6 & 7.45e6 & 9.78e6 & 5.21e6 & 5.02e6 & 3.87e6 & 3.06e6  \\
45 & 4.49e6 & 4.95e6 & 8.97e6 & 1.67e7 & 4.14e6 & 4.24e6 & 5.04e6 & 6.06e6 & 3.84e6 & 3.64e6 & 2.50e6 & 1.86e6  \\
47 & 3.37e6 & 3.67e6 & 6.04e6 & 9.38e6 & 3.09e6 & 3.12e6 & 3.41e6 & 3.83e6 & 2.85e6 & 2.65e6 & 1.58e6 & 1.13e6  \\
50 & 2.22e6 & 2.37e6 & 3.32e6 & 4.25e6 & 2.02e6 & 1.99e6 & 1.89e6 & 2.00e6 & 1.84e6 & 1.64e6 & 7.62e5 & 5.69e5  \\
53 & 1.48e6 & 1.54e6 & 1.85e6 & 2.16e6 & 1.33e6 & 1.26e6 & 1.06e6 & 1.11e6 & 1.19e6 & 9.96e5 & 3.62e5 & 3.11e5  \\
55 & 1.14e6 & 1.15e6 & 1.27e6 & 1.45e6 & 1.01e6 & 9.25e5 & 7.27e5 & 7.68e5 & 8.92e5 & 7.04e5 & 2.29e5 & 2.15e5  \\
57 & 8.73e5 & 8.58e5 & 8.84e5 & 1.00e6 & 7.68e5 & 6.72e5 & 5.09e5 & 5.40e5 & 6.63e5 & 4.88e5 & 1.53e5 & 1.51e5  \\
60 & 5.84e5 & 5.45e5 & 5.30e5 & 5.95e5 & 5.02e5 & 4.08e5 & 3.06e5 & 3.26e5 & 4.19e5 & 2.69e5 & 8.92e4 & 9.04e4  \\
63 & 3.86e5 & 3.40e5 & 3.27e5 & 3.66e5 & 3.22e5 & 2.41e5 & 1.89e5 & 2.01e5 & 2.56e5 & 1.39e5 & 5.47e4 & 5.56e4  \\
65 & 2.90e5 & 2.47e5 & 2.41e5 & 2.69e5 & 2.36e5 & 1.68e5 & 1.39e5 & 1.48e5 & 1.80e5 & 8.61e4 & 4.01e4 & 4.08e4  \\
67 & 2.16e5 & 1.80e5 & 1.79e5 & 1.99e5 & 1.71e5 & 1.17e5 & 1.03e5 & 1.10e5 & 1.23e5 & 5.26e4 & 2.97e4 & 3.02e4 \\
70 & 1.37e5 & 1.13e5 & 1.16e5 & 1.29e5 & 1.02e5 & 6.99e4 & 6.71e4 & 7.12e4 & 6.65e4 & 2.56e4 & 1.92e4 & 1.95e4 \\
\hline
\end{tabular}
\begin{tablenotes}
\item \textbf{Notes.} The electron temperature is assumed to be 12000 K. AeB means A$\times10^\textrm{B}$.
\end{tablenotes}
\end{threeparttable}
\end{table*}

\end{appendix}


\begin{thebibliography}{99}
\bibitem[Afflerbach et al.(1994)]{aff94} Afflerbach, A., Churchwell, E., Hofner, P., \& Kurtz, S. 1994, \apj, 437, 697
\bibitem[Aleman et al.(2018)]{ale18} Aleman, I., Exter, K., Ueta, T., Walton, S., Tielens, A. G. G. M., et al. 2018, \mnras, 477, 4499
\bibitem[Altenhoff et al.(1960)]{alt60} Altenhoff, W., Mezger, P. G., Wendker, H., \& Westerhout, G. 1960, Ver\"{o}ff. Sternwarte Bonn, 59, 48
\bibitem[Anderson et al.(2011)]{and11} Anderson, L. D., Bania, T. M., Balser, D. S., \& Rood, R. T. 2011, \apjs, 194, 32
\bibitem[Arthur \& Hoare(2006)]{art06} Arthur, S. J. \& Hoare, M. G. 2006, \apjs, 165, 283
\bibitem[B\'{a}ez-Rubio et al.(2013)]{bae13} B\'{a}ez-Rubio, A., Mart\'{i}n-Pintado, J., Thum, C., \& Planesas, P. 2013, \aap, 553, A45
\bibitem[B\'{a}ez-Rubio et al.(2018)]{bae18} B\'{a}ez-Rubio, A., Mart\'{i}n-Pintado, J., Rico-Villas, F., \& Jim\'{e}nez-Serra, I. 2018, \apjl, 867, L6
\bibitem[Bodenheimer et al.(1979)]{bod79} Bodenheimer, P.,    tenorio-Tagle, G., \& York, H. W. 1979, \apj, 233, 85
\bibitem[Brocklehurst(1970)]{bro70} Brocklehurst, M. 1970, \mnras, 148, 417
\bibitem[Brocklehurst(1971)]{bro71} Brocklehurst, M. 1971, \mnras, 153, 471
\bibitem[Burgess(1965)]{bur65} Burgess, A. 1965, MmRAS, 69, 1
\bibitem[Chen et al.(2020)]{che20} Chen, H.-Y., Chen, X., Wang, J.-Z., Shen, Z.-Q., \& Yang, K. 2020 \apjs, 248, 3
\bibitem[Churchwell \& Walmsley(1975)]{chu75} Churchwell, E., \& Walmsley, C. M. 1975, \aap, 38, 451
\bibitem[Cox et al.(1995)]{cox95} Cox, P., Mart\'{i}n-Pintado, J., Bachiller, R., Bronfman, L., Cernicharo, J., Nyman, L.-A., Roelfsema, P. R. 1995, \aap, 295, L39
\bibitem[de la Fuente et al.(2020a)]{fue20a} de la Fuente, E., Porras, A., Trinidad, M. A., Kurtz, S. E., Kemp, S. N., et al. 2020, \mnras, 492, 895
\bibitem[de la Fuente et al.(2020b)]{fue20b} de la Fuente, E., Tafoya, D., Trinidad, M. A., Porras, A., Nigoche-Netro, A. 2020, \mnras, 497, 4436
\bibitem[De Pree et al.(2000)]{pre00} De Pree, C. G., Wilner, D. J., Goss, W. M., Welch, W. J., \& McGrath, E. 2000, \apj, 540,308
\bibitem[De Pree et al.(2020)]{pre20} De Pree, C. G., Wilner, D. J., Kristensen, L. E., Galv\'{a}n-Madrid, R., Goss, W. M., et al. 2020, \aj, 160, 234
\bibitem[Diaz-Miller et al.(1998)]{dia98} Diaz-Miller, R. I., Franco, J., \& Shore, S. N. 1998, \apj, 501, 192
\bibitem[Ferland et al.(2017)]{fer17} Ferland, G. J., Chatzikos, M., Guzm\'{a}n, F., Lykins, M. L., van Hoof, P. A. M., et al. 2017, Rev. Mex. Astron. Astrofis., 53, 385
\bibitem[Gaetz \& Salpeter(1983)]{gae83} Gaetz,    t. J., \& Salpeter, E. E. 1983, \apjs, 52, 155
\bibitem[Goldberg(1966)]{gol66} Goldberg, L. 1966, \apj, 144, 1225
\bibitem[Gordon \& Sorochenko(2002)]{gor02} Gordon, M. A., \& Sorochenko, R. L. 2002, Radio Recombination Lines (Berlin: Springer)
\bibitem[Gordon \& Walmsley(1990)]{gor90} Gordon, M. A., \& Walmsley, C. M. 1990, \apj, 365, 606
\bibitem[Guzm\'{a}n et al.(2016)]{guz16} Guzm\'{a}n, F., Badnell, N. R., Williams, R. J. R., van Hoof, P. A. M., Chatzikos, M., et al. 2016, \mnras, 459, 3498
\bibitem[Guzm\'{a}n et al.(2019)]{guz19} Guzm\'{a}n, F., Chatzikos, M., van Hoof, P. A. M., Balser, D. S., Dehghanian, M., et al. 2019, \mnras, 486, 1003
\bibitem[H\"{o}glund \& Mezger(1965)]{hog65} H\"{o}glund, B. \& Mezger, P. G. 1965, Science, 150, 339
\bibitem[Hummer \& Storey(1987)]{hum87} Hummer, D. G. \& Storey, P. J. 1987, \mnras, 224, 801
\bibitem[Jim\'{e}nez-Serra et al.(2011)]{jim11} Jim\'{e}nez-Serra, I., Mart\'{i}n-Pintado, J., B\'{a}ez-Rubio, A., Patel, N.,    thum, C. 2011, \apj, 732, L27
\bibitem[Kegel(1979)]{keg79} Kegel, W. H. 1979, \aaps, 38, 131
\bibitem[Keto et al.(2008)]{ket08} Keto, E. R., Zhang, Q., \& Kurtz, S. 2008, \apj, 672, 423
\bibitem[Kim et al.(2017)]{kim17} Kim, W.-J., Wyrowski, F., Urquhart, J. S., Menten, K. M., \& Csengeri,    t. 2017, \aap, 602, A37
\bibitem[Koeppen \& Kegel(1980)]{koe80} Koeppen, J., Kegel, W. H. 1980, \aaps, 42, 59
\bibitem[Kurtz et al.(1994)]{kur94} Kurtz, S., Churchwell, E., \& Wood, D. O. S. 1994, \apjs, 91, 659
\bibitem[Lockman (1989)]{loc89} Lockman, F. J. 1989, \apjs, 71, 469
\bibitem[Liu et al.(2021)]{liu21} Liu, H.-L., Liu,    t., Evans, N. J., Wang, K., Garay, G., et al. 2021, \mnras, 505, 2801
\bibitem[Mac Low et al.(1991)]{mac91} Mac Low, M.-M., van Buren, D., Wood, D. O. S., \& Churchwell, E. 1991, \apj, 369, 395
\bibitem[Mart\'{i}n-Pintado et al.(1989)]{mar89} Mart\'{i}n-Pintado, J., Bachiller, R.,    thum,, C., \& Walmsley, C. M. 1989, \aap, 215, L13
\bibitem[Murchikova et al.(2020)]{mur20} Murchikova, L., Murphy, E. J., Lis, D. C., Armus, L., de Mink, S., et al. 2020, \apj, 903, 29
\bibitem[Nguyen-Luong et al.(2017)]{ngu17} Nguyen-Luong, Q., Anderson, L. D., Motte, F., Kim, K.-T., Schilke, P., et al. 2017, \apjl, 844, 25
\bibitem[Oster(1961)]{ost61} Oster, L. 1961, RvMP, 33, 525
\bibitem[Peimbert(1979)]{pei79} Peimber, M. 1979, in IAU Symp. 84,    the Large-Scale Characteristics of    the Galaxy, ed. W. B. Burton (Dordrecht: Reidel), 307
\bibitem[Pengelly \& Seaton(1964)]{pen64} Pengelly, R. M. \& Seaton, M. F. 1964, \mnras, 127, 165
\bibitem[Peters et al.(2012)]{pet12} Peters,    t., Longmore, S. N., \& Dullemond, C. P. 2012 \mnras, 425, 2352
\bibitem[Planesas et al.(1991)]{pla91} Planesas, P., G\'{o}mez-Gonz\'{a}lez, J., Rodr\'{i}guez, L. F., \& Cant\'{o}, J. 1991, \rmxaa, 22, 19
\bibitem[Prozesky \& Smits(2018)]{pro18} Prozesky, A., \& Smits, D. P. 2018, \mnras, 478, 2766
\bibitem[Prozesky \& Smits(2020)]{pro20} Prozesky, A., \& Smits, D. P. 2020, \mnras, 491, 2536
\bibitem[R\"{o}llig et al.(1999)]{rol99} R\"{o}llig, M., Kegel, W. H., Mauersberger, R., \& Doerr, C. 1999, \aap, 343, 939
\bibitem[Salgado et al.(2017)]{sal17} Salgado, F., Morabito, L. K., Oonk, J. B. R., Salas, P., Toribio, M. C., et al. 2017, \apj, 837, 141
\bibitem[Sejnowski \& Hjellming(1969)]{sej69} Sejnowski,    t. J. \& Hjellming, R. M. 1969, \apj, 156, 915
\bibitem[Scoville \& Murchikova(2013)]{sco13} Scoville, N. \& Murchikova, L. 2013, \apj, 779, 75
\bibitem[Shaver(1970)]{sha70} Shaver, P. A. 1970, ApL, 5, 167
\bibitem[Shaver et al.(1979)]{sha79} Shaver, P. A., McGee, R. X., \& Pottasch, S. R. 1979, \nat, 280, 476
\bibitem[Shaver et al.(1983)]{sha83} Shaver, P. A., McGee, R. X., Newton, L. M., Danks, A. C., \& Pottasch, S. R. 1983, \mnras, 204, 53
\bibitem[Sorochenko \& Borodzich(1965)]{sor65} Sorochenko, R. L., \& Borodzich, E. V. 1965, Dokl, Akad. Nauk SSSR, 163, 603. English    translation: 1966, Sov. Phys.- Dokl., 10, 588
\bibitem[Spitzer(1978)]{spi78} Spitzer, L. 1978, Physical Processes in    the Interstellar Medium (New York: Wiley-Interscience), 333
\bibitem[Storey \& Hummer(1995)]{sto95} Storey, P. J. \& Hummer, D. G. 1995, \mnras, 272, 41
\bibitem[Strelnitski et al.(1996)]{str96} Strelnitski, V. S., Ponomarev, V., \& Smith, H. A. 1996, \apj, 470, 1118
\bibitem[Tenorio-Tagle(1979)]{ten79} Tenorio-Tagle, G. 1979, \aap, 71, 59
\bibitem[Thum et al.(1995)]{thu95} Thum, C., Strelnitski, V. S., Mart\'{i}n-Pintado, J., Matthews, H. E., \& Smith, H. A. 1995, \aap, 300, 843
\bibitem[Thum et al.(2013)]{thu13} Thum, C., Neri, R., B\'{a}ez-Bubio, \& Krips, M. 2013, \aap, 556, 129
\bibitem[Thum et al.(1998)]{thu98} Thum, C., Mart\'{n}-Pintado, J., Quirrenbach, A., \& Matthews, H. E. 1998, \aap, 333, L63
\bibitem[Walmsley(1990)]{wal90} Walmsley, C. M. 1990, \aaps, 82, 201
\bibitem[Wilson et al.(2015)]{wil15} Wilson, T. L., Bania, T. M., \& Balser, D. S. 2015, \apj, 812, 45
\bibitem[Wood \& Churchwell(1989)]{woo89} Wood, D. O. S., \& Churchwell, E. 1989, \apjs, 69, 831
\bibitem[Vriens \& Smeets(1980)]{vri80} Vriens, L. \& Smeets, A. H. M. 1980, Physical Review A, 22, 940
\bibitem[Zhu et al.(2019)]{zhu19} Zhu, F.-Y., Zhu, Q.-F., Wang, J.-Z., \& Zhang, J.-S. 2019, \apj, 881, 14

\end{thebibliography}
\end{document}